%% file: main.tex
\documentclass[a4paper, fontsize = 12pt, numbers = noenddot, parskip = no]{scrartcl}
\input{preamble}

\newcommand{\csmc}{\text{\upshape\gls{CSMC}}}

\newcommand{\particlerwm}{\text{\upshape\gls{PARTICLERWM}}}

\newcommand{\particlemala}{\text{\upshape\gls{PARTICLEMALA}}}
\newcommand{\particleamala}{\text{\upshape\gls{PARTICLEAMALA}}}

\newcommand{\particlemgrad}{\text{\upshape\gls{PARTICLEMGRAD}}}
\newcommand{\particleagrad}{\text{\upshape\gls{PARTICLEAGRAD}}}
\newcommand{\particleagradplus}{\text{\upshape\gls{PARTICLEAGRADPLUS}}}

\newcommand{\particleapcnl}{\text{\upshape\gls{PARTICLEAPCNL}}}
\newcommand{\particlepcnl}{\text{\upshape\gls{PARTICLEPCNL}}}
\newcommand{\particleapcnlplus}{\text{\upshape\gls{PARTICLEAPCNLPLUS}}}

\title{
  \Large{Particle-MALA and Particle-mGRAD:\\Gradient-based MCMC methods for high-dimensional state-space models}
}
\author[1]{Adrien Corenflos}
\author[2]{Axel Finke}
\affil[1]{\large{Department of Electrical Engineering and Automation,\protect\\Aalto University, Finland\protect\\\href{mailto:adrien.corenflos@aalto.fi}{adrien.corenflos@aalto.fi}}}
\affil[2]{\large{Department of Mathematical Sciences,\protect\\Loughborough University, UK\protect\\\href{mailto:a.finke@lboro.ac.uk}{a.finke@lboro.ac.uk}}}

\date{\large{\today}}

\usepackage{authblk}
\makeatletter
\renewcommand\AB@affilsepx{, \protect\Affilfont}
\makeatother

\begin{document}

\glsunset{MCMC}
\glsunset{MALA}
\glsunset{MGRAD}
\glsunset{AGRAD}
\glsunset{PARTICLEMALA}
\glsunset{PARTICLEMGRAD}
\maketitle
\vspace{-1cm}

\begin{abstract}
\noindent{}State-of-the-art methods for Bayesian inference in state-space models are (a) \emph{\gls{CSMC}} algorithms; (b) sophisticated `classical' \gls{MCMC} algorithms like \emph{\gls{MALA},} or \emph{\gls{MGRAD}} from \citet{titsias2018auxiliary}. The former propose $N$ particles at each time step to exploit the model's `decorrelation-over-time' property and thus scale favourably with the time horizon, $T$, but break down if the dimension of the latent states, $D$, is large. The latter leverage gradient-/prior-informed local proposals to scale favourably with $D$ but exhibit sub-optimal scalability with $T$ due to a lack of model-structure exploitation. We introduce methods which combine the strengths of both approaches. The first, \emph{\gls{PARTICLEMALA},} spreads $N$ particles locally around the current state using gradient information, thus extending \gls{MALA} to $T > 1$ time steps and $N > 1$ proposals. The second, \emph{\gls{PARTICLEMGRAD}}, additionally incorporates (conditionally) Gaussian prior dynamics into the proposal, thus extending the \gls{MGRAD} algorithm to $T > 1$ time steps and $N > 1$ proposals. We prove that \gls{PARTICLEMGRAD} interpolates between \gls{CSMC} and \gls{PARTICLEMALA}, resolving the `tuning problem' of choosing between \gls{CSMC} (superior for highly informative prior dynamics) and \gls{PARTICLEMALA} (superior for weakly informative prior dynamics). We similarly extend other `classical' \gls{MCMC} approaches like \emph{\glsdesc{AMALA}}, \emph{\gls{AGRAD}}, and \emph{\gls{PCNL}} to $T > 1$ time steps and $N > 1$ proposals. In experiments, for both highly and weakly informative prior dynamics, our methods substantially improve upon both \gls{CSMC} and sophisticated `classical' \gls{MCMC} approaches. 
\end{abstract}

\section{Introduction}
\label{sec:introduction}
\input{01_introduction}

\section{Existing methodology}
\label{sec:background}
\input{02_background.tex}

\section{Particle extensions of MALA and aMALA}
\label{sec:particlemala}

\input{03_particlemala}

\section{Particle extensions of mGRAD and aGRAD}
\label{sec:particlemgrad} 
\input{04_particlemgrad}

\section{Experimental validation and comparison}
\label{sec:simulations}
\input{05_simulations}

\section{Conclusion}
\label{sec:conclusions}
\input{06_conclusion}

\subsection*{Author contributions}
A.C.\ and A.F.\ jointly developed the methodology, writing was primarily done by A.F., A.C.\ implemented and conducted the experiments, after which both A.C.\ and A.F.\ edited and reviewed the final manuscript.

\renewcommand*{\bibfont}{\footnotesize}
\setlength{\bibsep}{3pt plus 0.3ex}
\bibliography{input/literature}

\appendix

\input{appendix.tex}

\end{document}

%% file: preamble.tex
\usepackage[left=2cm, right=2cm, top=2cm, bottom=3cm]{geometry}
\usepackage{authblk}
\usepackage[T1]{fontenc}
\usepackage[utf8]{inputenc}
\usepackage{lmodern}
\usepackage{lscape}
\usepackage{float}
\usepackage[normalem]{ulem}
\usepackage{verbatim}
\usepackage{siunitx}

\graphicspath{{fig/arxiv/}}

\setkomafont{sectioning}{\normalfont \rmfamily \bfseries}%
\setkomafont{paragraph}{\normalfont \rmfamily \bfseries}%

\usepackage{silence}
\usepackage{url}
\usepackage{hyperref}%
\hypersetup{%
  breaklinks=true,%
  linktocpage=true,%
  colorlinks=true,%
  linkcolor=blue,%
  citecolor=purple,%
  filecolor=blue,%
  urlcolor=blue,%
  pdftitle={},%
  pdfsubject={},%
  pdfauthor={},%
  pdfkeywords={},%
  pdfproducer={}%
  }%

\usepackage{mathtools} %
\mathtoolsset{showonlyrefs=true}%
\usepackage[ntheorem, framemethod=TikZ]{mdframed} %
\usepackage{framed} %
\usepackage[amsmath, hyperref, framed]{ntheorem}%
\usepackage{amssymb} %
\usepackage{bm}
\usepackage{wasysym}

\input{input/mathematics.tex}

\usepackage{accents}

\usepackage{pifont}%

\usepackage{multirow, multicol, booktabs, ctable, makecell}
\usepackage{tabularx, bigdelim}
\usepackage{nicematrix}
\usepackage{threeparttable}
\usepackage{tablefootnote}

\usepackage{siunitx}

\usepackage{graphicx}
\DeclareGraphicsExtensions{.pdf, .png} %
\usepackage{float}

\usepackage{caption, subcaption}
\usepackage{wrapfig}
\usepackage{xcolor} %

\usepackage{tikz}

\qedsymbol{\ensuremath{\Box}}
\theoremstyle{plain}%

\theoremheaderfont{\normalfont\bfseries\rmfamily} 
\theorembodyfont{\normalfont\slshape}%
\theoremseparator{.}
\newtheorem{proposition}{Proposition}%
\newtheorem{lemma}{Lemma}%
\newtheorem{remark}{Remark}%
\newtheorem{example}{Example}%

\theorembodyfont{\rmfamily}%
\newmdtheoremenv[
 ntheorem=true,
 skipbelow = .5\baselineskip plus 1ex minus 1ex,
 skipabove = .5\baselineskip plus 1ex minus 1ex,
 innerleftmargin = 0pt,
 innerrightmargin = 0pt,
 leftline = false,
 rightline = false,
 needspace = 0ex %
]{framedAlgorithm}[theorem]{Algorithm}

\AtBeginEnvironment{framedAlgorithm}{\begin{minipage}{\textwidth}}
\AtEndEnvironment{framedAlgorithm}{\end{minipage}}
\newmdtheoremenv[
 ntheorem=true,
 skipbelow = .5\baselineskip plus 1ex minus 1ex,
 skipabove = .5\baselineskip plus 1ex minus 1ex,
 innerleftmargin = .4\baselineskip plus 1ex minus 1ex,
 innerrightmargin = .4\baselineskip plus 1ex minus 1ex,
 leftline = true,
 rightline = true,
 needspace = 0ex %
]{framedExample}[theorem]{Application to state-space models}

\theoremstyle{empty}%
\theoremheaderfont{} 

\theoremstyle{nonumberplain}%
\theorembodyfont{\upshape \rmfamily}%
\theoremheaderfont{\upshape \rmfamily\bfseries}%
\theoremsymbol{\ensuremath{_\Box}}%
\newtheorem{namedProof}{Proof}%

\usepackage[textwidth=2.5cm, textsize=scriptsize]{todonotes}

\makeatletter
\newcommand{\myitem}[1]{%
\item[#1.]\protected@edef\@currentlabel{#1}%
}
\makeatother

\usepackage{ifthen}%
\usepackage{mfirstuc}
\usepackage{xkeyval}%
\usepackage{xfor}%
\usepackage{amsgen}%
\usepackage{etoolbox}%
\usepackage{datatool-base}%

\usepackage[%
  xindy,%
  style=long,%
  toc=true,%
  nonumberlist,%
  acronym,%
  acronymlists={abbreviations},%
  hyperfirst=true%
  ]{glossaries}%
  
\input{input/abbreviations.tex}

\usepackage{csquotes}%
\usepackage{natbib}
\setcitestyle{authoryear}

%% file: input/mathematics.tex
\newcommand{\myBlockMat}[1]{\mathcal{M}_{#1}}

\newcommand{\qedwhite}{\hfill \ensuremath{\Box}}

\DeclareMathOperator{\myVec}{vec}

\newcommand{\placeholder}{a}%
\newcommand{\placeholderAlt}{a}
\newcommand{\placeholderAltAlt}{b}

\newcommand{\gradientIndicator}{\kappa}
\newcommand{\sta}{\mathbf{x}} %
\newcommand{\staAlt}{\tilde{\mathbf{x}}} %
\newcommand{\obs}{\mathbf{y}} %
\newcommand{\aux}{\mathbf{u}} %
\newcommand{\target}{\pi}
\newcommand{\proposalAlt}{q}
\newcommand{\Mutation}{M}

\newcommand{\Potential}{G}
\newcommand{\qsemigroup}{Q}
\newcommand{\genericVec}{\mathbf{v}}

\newcommand{\genericGrad}{\boldsymbol{\phi}}
\newcommand{\genericGradAlt}{\boldsymbol{\psi}}
\newcommand{\genericGradAltAlt}{\boldsymbol{\varphi}}
\newcommand{\priorMean}{\mathbf{m}}
\newcommand{\priorVar}{\mathbf{C}}
\newcommand{\preconVar}{\smash{\widetilde{\priorVar}}}
\newcommand{\priorIntercept}{\mathbf{b}}
\newcommand{\priorFactor}{\mathbf{F}}
\newcommand{\marginalMean}{\boldsymbol{\mu}}
\newcommand{\marginalVar}{\boldsymbol{\Sigma}}
\newcommand{\kalmanMean}{\boldsymbol{\mu}}
\newcommand{\kalmanVar}{\boldsymbol{\Sigma}}
\newcommand{\kalmanGain}{\mathbf{A}}
\newcommand{\targetAlt}{\target'}
\newcommand{\MutationAlt}{\Mutation'}

\newcommand{\PotentialAlt}{\Potential'}
\newcommand{\qsemigroupAlt}{\qsemigroup'}
\newcommand{\priorMeanAlt}{\priorMean'}
\newcommand{\priorVarAlt}{\priorVar'}
\newcommand{\priorInterceptAlt}{\priorIntercept'}
\newcommand{\priorFactorAlt}{\priorFactor'}
\newcommand{\priorMeanAltAlt}{\boldsymbol{\nu}}
\newcommand{\priorVarAltAlt}{\boldsymbol{\Upsilon}}

\newcommand{\mgradProposalVar}{\mathbf{B}}
\newcommand{\eMat}{\mathbf{E}}
\newcommand{\gMat}{\mathbf{G}}
\newcommand{\hMat}{\mathbf{H}}
\newcommand{\dMat}{\mathbf{D}}
\newcommand{\vMat}{\mathbf{V}}
\newcommand{\aMat}{\mathbf{A}}
\newcommand{\bMat}{\mathbf{B}}
\newcommand{\fMat}{\mathbf{F}}

\DeclareMathOperator{\dN}{N} %

\newcommand{\diff}{\mathrm{d}} %
\newcommand{\intDiff}{\,\mathrm{d}} %
\DeclareMathOperator{\Prob}{\mathbb{P}} %
\DeclareMathOperator{\E}{\mathbb{E}} %
\DeclareMathOperator{\ind}{\mathbb{I}} %
\newcommand{\ccdot}{\,\cdot\,} %
\newcommand{\T}{\mathrm{T}} %
\newcommand{\compl}{\mathrm{c}} %
\DeclareMathOperator{\bo}{\mathrm{O}}%
\newcommand{\lip}{\textnormal{\textsc{lip}}}%
\newcommand{\tv}{\textnormal{\textsc{tv}}}
\DeclareMathOperator{\diag}{diag}%
\DeclareMathOperator{\kl}{KL}%
\newcommand{\iMat}{\mathbf{I}}

\newcommand{\unitMat}{\bm{1}}
\newcommand{\zeroMat}{\bm{0}}

\newcommand{\calH}{\mathcal{H}}%

\newcommand{\reals}{\mathbb{R}}%
\newcommand{\naturals}{\mathbb{N}}%
\newcommand{\spaceX}{\mathcal{X}}%
\newcommand{\spaceY}{\mathcal{Y}}%

\DeclareFontFamily{U}{mathx}{\hyphenchar\font45}
\DeclareFontShape{U}{mathx}{m}{n}{<-> mathx10}{}
\DeclareSymbolFont{mathx}{U}{mathx}{m}{n}
\DeclareMathAccent{\widebar}{0}{mathx}{"73}

\newcommand{\prodSubstackAligned}[6]{
  \prod_{\substack{\mathllap{#1} #2 \mathrlap{#3}\\
                   \mathllap{#4} #5 \mathrlap{#6}}}}%

%% file: input/abbreviations.tex
\newacronym{IID}{IID}{independent and identically distributed}
\newacronym{WP}{w.p.\@}{with probabilitity}
\newacronym{TV}{TV}{total variation}
\newacronym{ESS}{ESS}{effective sample size}
\newacronym{MH}{MH}{Metropolis--Hastings}
\newacronym{SMC}{SMC}{sequential Monte Carlo}
\newacronym{RWM}{RWM}{random-walk Metropolis}
\newacronym{IMH}{IMH}{independent Metropolis--Hastings}
\newacronym{MCMC}{MCMC}{Markov chain Monte Carlo}

\newacronym{AMALA}{aMALA}{auxiliary MALA}
\newacronym{MALA}{MALA}{Metropolis-adjusted Langevin algorithm}

\newacronym{AGRAD}{aGRAD}{auxiliary gradient} 
\newacronym{MGRAD}{mGRAD}{marginal gradient}

\newacronym{CN}{CN}{Crank--Nicolson}
\newacronym{CNL}{CNL}{Crank--Nicolson--Langevin}
\newacronym{PCNL}{PCNL}{preconditioned Crank--Nicolson--Langevin}
\newacronym{PCN}{PCN}{preconditioned Crank--Nicolson}

\newacronym{ACN}{aCN}{auxiliary Crank--Nicolson}
\newacronym{APCN}{aPCN}{auxiliary preconditioned Crank--Nicolson}
\newacronym{ACNL}{aCNL}{auxiliary Crank--Nicolson--Langevin}
\newacronym{APCNL}{aPCNL}{auxiliary preconditioned Crank--Nicolson--Langevin}

\newacronym{CSMC}{CSMC}{conditional sequential Monte Carlo}

\newacronym{PARTICLERWM}{Particle-RWM}{particle random-walk Metropolis}

\newacronym{PARTICLEMALA}{Particle-MALA}{particle MALA}
\newacronym{PARTICLEAMALA}{Particle-aMALA}{particle auxiliary MALA} 
\newacronym{PARTICLEAMALAPLUS}{Particle-aMALA+}{particle lookahead auxiliary MALA} 

\newacronym{PARTICLEMGRAD}{Particle-mGRAD}{particle marginal gradient}
\newacronym{PARTICLEAGRAD}{Particle-aGRAD}{particle auxiliary gradient}
\newacronym{PARTICLEAGRADPLUS}{Particle-aGRAD+}{particle lookahead auxiliary gradient}

\newacronym{PARTICLEPCN}{Particle-PCN}{particle PCN}
\newacronym{PARTICLEAPCN}{Particle-aCNL}{particle auxiliary PCN}
\newacronym{PARTICLEAPCNPLUS}{Particle-aCNL+}{particle lookahead auxiliary PCN}

\newacronym{PARTICLECNL}{Particle-CNL}{particle CNL}
\newacronym{PARTICLEACNL}{Particle-aCNL}{particle auxiliary CNL}
\newacronym{PARTICLEACNLPLUS}{Particle-aCNL+}{particle lookahead auxiliary CNL}

\newacronym{PARTICLEPCNL}{Particle-PCNL}{particle PCNL}
\newacronym{PARTICLEAPCNL}{Particle-aPCNL}{particle auxiliary PCNL}
\newacronym{PARTICLEAPCNLPLUS}{Particle-aPCNL+}{particle lookahead auxiliary PCNL}

%% file: 01_introduction.tex
\subsection{Feynman--Kac models}
\label{subsubsec:feynman--kac_models}

\glsreset{MH}
\glsreset{CSMC}
\glsreset{IMH}
\glsreset{RWM}
\glsunset{WP}
\glsreset{PCNL}
\glsreset{MCMC}

\glsunset{PARTICLERWM}

\glsreset{MALA}
\glsreset{AMALA}
\glsunset{PARTICLEMALA}
\glsunset{PARTICLEAMALA}
\glsunset{PARTICLEAMALAPLUS}

\glsreset{MGRAD}
\glsreset{AGRAD}
\glsunset{PARTICLEMGRAD}
\glsunset{PARTICLEAGRAD}
\glsunset{PARTICLEAGRADPLUS}

\glsunset{PARTICLEPCNL}
\glsunset{PARTICLEAPCNL}
\glsunset{PARTICLEAPCNLPLUS}

The aim of this work is to construct efficient \emph{\gls{MCMC}} updates for sampling from a continuous \emph{joint smoothing} distribution $\target_T(\sta_{1:T})$ on $\spaceX^T$, where $\spaceX \coloneqq \reals^D$ and where for any $t \leq T$, we have defined the following distributions (termed \emph{filters}):
\begin{align}\label{eq:gamma_def}
 \pi_t(\sta_{1:t}) \propto \prod_{s=1}^t \qsemigroup_s(\sta_{s-1:s}).
\end{align}
Here, $\qsemigroup_t(\sta_{t-1:t}) > 0$ is differentiable and can be evaluated point-wise. Throughout this work, we use the convention that quantities with `time' subscripts $t \leq 0$ or $t > T$ should be ignored, so that, e.g., $\qsemigroup_1(x_{0:1}) \equiv \qsemigroup_1(\sta_1)$ and $\qsemigroup_{T+1}(\sta_{T:T+1}) \equiv 1$. We will frequently work with some decomposition
\begin{align}
 \qsemigroup_t(\sta_{t-1:t}) = \Mutation_t(\sta_t|\sta_{t-1}) \Potential_t(\sta_{t-1:t}),
\end{align}
such that
\begin{itemize}
  \item $\Mutation_t(\ccdot|\sta_{t-1})$ is a density (w.r.t.\ a suitable version of the Lebesgue measure) and also defines a Markov transition kernel called \emph{mutation kernel}; %
  \item $\Potential_t(\sta_{t-1:t}) > 0$ is called \emph{potential function}. %
\end{itemize}
We assume that these densities and potential functions are differentiable and that they (as well as their gradients) can be evaluated point-wise. Motivated by the following example, we will sometimes refer to $\smash{\Mutation_{1:T}(\sta_{1:T}) \coloneqq \prod_{t=1}^T \Mutation_t(\sta_t|\sta_{t-1})}$ as the \emph{prior dynamics} of the \emph{latent states} $\sta_{1:T}$ and $\smash{\Potential_{1:T}(\sta_{1:T}) \coloneqq \prod_{t=1}^T G_t(\sta_{t-1:t})}$ as the \emph{likelihood}.

\begin{example}[state-space model] \label{ex:ssm:special_case_of_feynman--kac_model}
  One important special case of Feynman--Kac models are \emph{state-space models}. A state-space model is a bivariate Markov chain $(\sta_t, \obs_t)_{t \geq 1}$ on $\spaceX \times \spaceY$, where $\spaceX \coloneqq \reals^D$ and $\spaceY \coloneqq \reals^{D'}$, with initial density $p(\sta_1, \obs_1) = f_1(\sta_1) g_1(\obs_1|\sta_1)$ and transition densities $p(\sta_t, \obs_t| \sta_{t-1}) = f_t(\sta_t|\sta_{t-1}) g_t(\obs_t|\sta_t)$ (w.r.t.\ a suitable version of the Lebesgue measure). State-space models assume that only the measurements $\obs_{1:T}$ can be observed whilst the Markov chain $(\sta_t)_{t \geq 1}$ (often representing the evolution of the phenomenon of interest) is latent. The joint smoothing distribution then encodes our knowledge of the latent states $\sta_{1:T}$ given the available data $\obs_{1:T}$:
    \begin{align}
    \target_T(\sta_{1:T}) \coloneqq p(\sta_{1:T} | \obs_{1:T}) 
    \propto %
    \prod_{t=1}^T f_t(\sta_t|\sta_{t-1}) g_t(\obs_t|\sta_t). \label{eq:joint_smoothing_distribution}
    \end{align}
  One possible way of casting such a state-space model as a Feynman--Kac model (there are others) is then to take $\Mutation_t(\sta_t|\sta_{t-1}) = f_t(\sta_t|\sta_{t-1})$ and $\Potential_t(\sta_{t-1:t}) = g_t(\obs_t|\sta_t)$. In this case, $\Mutation_{1:T}(\sta_{1:T}) = p(\sta_{1:T})$, $\Potential_{1:T}(\sta_{1:T}) = p(\obs_{1:T}|\sta_{1:T})$, and $\pi_t(\sta_{1:t}) = p(\sta_{1:t}|\obs_{1:t})$, for $t \leq T$.%
\end{example}

\subsection{Sampling the latent states}

Performing inference about the latent states $\sta_{1:T}$ requires calculating expectations of the form $\E_{\sta_{1:T} \sim \pi_T}[\varphi(\sta_{1:T})]$, for some integrable test function $\varphi\colon \spaceX^T \to \reals$. Unfortunately, such expectations do not admit closed-form expressions in most realistic problems and must be approximated by some Monte Carlo estimate $\smash{\frac{1}{I}\sum_{i=1}^I \varphi(\sta_{1:T}[i])}$ using samples $(\sta_{1:T}[i])_{i=1}^I$ (approximately) distributed according to $\pi_T$. These often come from some \gls{MCMC} algorithm targeting $\pi_T$.

\begin{figure}
  \vspace{-3ex}
  \centering
  \hspace{-4.8em}
  \begin{subfigure}{0.405\textwidth}
      \centering\captionsetup{width = 0.9\textwidth}%
      \includegraphics[]{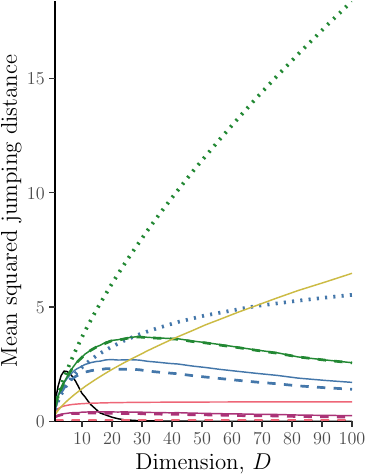}
      \caption{Empirical scaling with $D$ for fixed time horizon $T = 25$.}
      \label{fig:lgssm_dimension}
  \end{subfigure}
  \hspace{-0.8em}
  \begin{subfigure}{0.37\textwidth}
      \centering\captionsetup{width = \textwidth}%
      \includegraphics[]{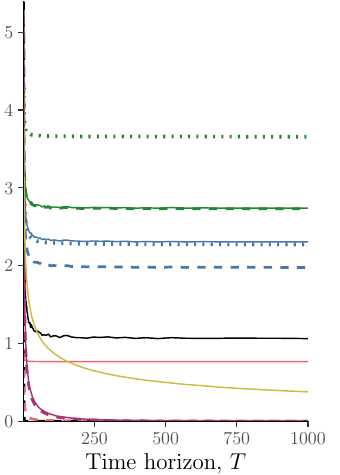}
      \caption{Empirical scaling with $T$ for fixed state dimension $D = 10$.}
      \label{fig:lgssm_time_horizon}
  \end{subfigure}\hspace{-1em}
    \begin{subfigure}{0.15\textwidth}
      \centering\captionsetup{width = \textwidth, labelformat = empty}%
      \includegraphics[]{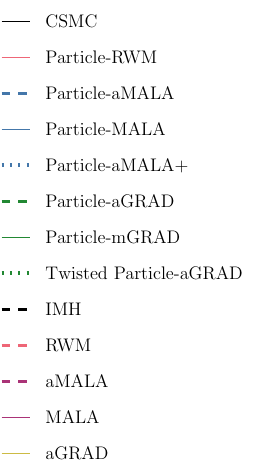}\hspace{-20em}
      \vspace{6ex}
      \label{fig:lgssm_legend}
  \end{subfigure}
  \caption{Toy linear-Gaussian state-space model with $\Mutation_t(\sta_t|\sta_{t-1}) = \dN(\sta_t; \sta_{t-1}, \lambda \iMat)$ and $\Potential_t(\sta_{t-1:t}) = \dN(\obs_t; \sta_t, \iMat)$, where $\iMat$ is the ($D \times D$)-identity matrix and $\lambda = 1$. For a fair comparison, all methods use $N + 1 = 32$ particles. The step sizes are: $(TD)^{-1}$ for (multi-proposal) \gls{RWM}, $D^{-1}$ for \gls{PARTICLERWM}, $(TD)^{-1/3}$ for (multi-proposal) \gls{AMALA}/\gls{MALA}/\gls{AGRAD}, and $D^{-{1/3}}$ for the remaining (i.e., new) methods. Panel~\subref{fig:lgssm_dimension} illustrates that as $D$ increases, some `classical' \gls{MCMC} algorithms (\gls{RWM}, \gls{MALA}, \gls{AMALA} and \gls{AGRAD}) are stable but the \gls{CSMC} algorithm breaks down. Conversely, Panel~\subref{fig:lgssm_time_horizon} illustrates that as $T$ increases, the \gls{CSMC} algorithm is stable in $T$ but all `classical' \gls{MCMC} algorithms (\gls{IMH}, \gls{RWM}, \gls{MALA}, \gls{AMALA} and \gls{AGRAD}) break down.}
  \label{fig:lgssm_scaling}
\end{figure}

\paragraph{`Classical' MCMC methods.} Unfortunately, simple \gls{MCMC} approaches like the \emph{\gls{IMH}} algorithm \citep{hastings1970monte} perform poorly if the problem size: $D \times T$, is large due the difficulty of constructing efficient \emph{global} (a.k.a.\ \emph{independent}) proposal distributions in high dimensions. To circumvent this difficulty, \gls{MCMC} algorithms with \emph{local} moves like the \emph{\gls{RWM}} algorithm \citep{metropolis1953equation}, propose a new state of the Markov chain near the current state. By decreasing the proposal scale at a suitable rate with the problem size, the \gls{RWM} algorithm can circumvent this curse of dimension \citep{roberts1997weak}. Further improved performance can be achieved by exploiting %
\begin{itemize}
  \item \emph{gradient information,} i.e.\ by including gradients of the log-likelihood or log-target density into the proposal as in the \emph{\gls{MALA}} from \citet{besag1994representations} %
  and in the \emph{\gls{AMALA}} from \citet{titsias2018auxiliary}; and \emph{additionally}
  \item \emph{prior information,} i.e.\ by explicitly incorporating the prior dependence structure into the proposal as in the \emph{\gls{PCNL}} and related algorithms \citep[see, e.g.,][and references therein]{cotter2013crank} or in the \emph{\gls{MGRAD}} and \emph{\gls{AGRAD}}\footnote{Throughout this work, `\gls{AGRAD}' refers more specifically to the `aGrad-z' algorithm from \citet{titsias2018auxiliary}.} algorithms from \citet{titsias2011riemann, titsias2018auxiliary}.
\end{itemize}
Figure~\ref{fig:lgssm_dimension} illustrates that `classical' \gls{MCMC} algorithms can scale favourably with $D$.

However, `classical' \gls{MCMC} algorithms are agnostic to the `decorrelation-over-time' structure of the target distribution $\pi_T(\sta_{1:T})$, i.e., to the fact that, for suitably regular models, the correlation of $x_s$ and $\sta_t$ under $\pi(\sta_{1:T})$ decays with $\lvert t - s\rvert$. For example, for the simple \gls{RWM} algorithm and \gls{MALA}, the \emph{step size} $\delta > 0$ (i.e., proposal variance) would need to decrease at a suitable rate with $T$ ($\delta \in \bo((DT)^{-1})$ and $\delta \in \bo((DT)^{-1/3})$, respectively) even if the model was completely independent across time steps \citep{roberts2001optimal}. Therefore, it stands to reason that the scaling of `classical' \gls{MCMC} methods like \gls{MALA}, \gls{PCNL} or \gls{MGRAD}/\gls{AGRAD} with the time horizon $T$ could be improved by empowering them to exploit this model structure.

\glsreset{CSMC}
\paragraph{CSMC methods.}
Another popular $\pi_T$-invariant \gls{MCMC}-kernel, $P_\csmc$,  is induced by running the \emph{\gls{CSMC}} algorithm proposed in the seminal work \citet{andrieu2010particle, whiteley2010particle}. Given the current state $\sta_{1:T} \in \spaceX^T$ of the Markov chain (then called the \emph{reference path}) this algorithm generates $\staAlt_{1:T} \sim P_\csmc(\ccdot |\sta_{1:T})$ as follows, where we write $[n] \coloneqq \{1, \dotsc, n\}$ and $[n]_0 \coloneqq [n] \cup \{0\}$:
\begin{enumerate}
  \item \label{enum:csmc_motivation:1} For $t = 1, \dotsc, T$, sample some index $k_t$ from a uniform distribution on $[N]_0$; set $\sta_t^{k_t} \coloneqq \sta_t$ and sample the remaining \emph{particles} $\smash{\sta_t^{-k_t} \coloneqq (\sta_t^0, \dotsc, \sta_t^{k_t-1}, \sta_t^{k_t + 1}, \dotsc, \sta_t^N)}$ conditionally independently such that for $n \neq k_t$, 
  \begin{align}
    \sta_t^n \sim \Mutation_t(\ccdot |\sta_{t-1}^{a_{t-1}^n}), \label{eq:standard_mutation}
  \end{align}
  for \emph{ancestor indices} $a_{t-1}^n \in [N]_0$ whose r\^ole is explained later.
  
  \item \label{enum:csmc_motivation:2} Return $\smash{\staAlt_{1:T} \coloneqq (\sta_1^{l_1}, \dotsc, \sta_t^{l_T})}$, for indices $l_1, \dotsc, l_T \in [N]_0$ sampled from an appropriate distribution.
\end{enumerate}
Informally, the \gls{CSMC} algorithm can be interpreted as employing $T$ separate accept--reject steps (one at each time point) which allows it to exploit the `decorrelation-over-time' property of $\pi_T(\sta_{1:T})$ (akin to a `classical' \gls{MCMC} algorithm with blocking in the `time' direction as noted by \citealt{singh2017blocking}). For suitably regular problems, the \gls{CSMC} algorithm therefore scales more favourably with $T$ than `classical' \gls{MCMC} approaches as illustrated in Figure~\ref{fig:lgssm_time_horizon}.

Unfortunately, as shown in \citet{finke2023conditional}, the \gls{CSMC} algorithm suffers from a curse of dimension in the state dimension $D$: as $D$ increases, it becomes increasingly likely that $\staAlt_{1:T}$ coincides exactly with $\sta_{1:T}$, i.e., the induced \gls{MCMC} chain gets stuck. This is unsurprising because the \gls{CSMC} algorithm generalises the \gls{IMH} algorithm to $T > 1$ time steps and $N > 1$ proposals. Indeed, note that \eqref{eq:standard_mutation} is again an independent (i.e.\ global) proposal in the sense that it does not depend on the time-$t$ component of the current state of the Markov chain, $\sta_t$; and such proposals are known to scale poorly with dimension (due to the difficulty of finding efficient global proposals in high dimensions). The only potential remedy: increasing $N$ exponentially with $D$, is prohibitively costly.

\paragraph{Existing combinations of `classical' MCMC and CSMC.} To circumvent this problem, \citet{finke2023conditional} introduced the \emph{\gls{PARTICLERWM}}\footnote{Referred to as `random-walk \gls{CSMC}' therein.} algorithm which scatters the particles locally around the reference path \citep[see also][for related approaches]{shestopaloff2018sampling, malory2021bayesian}. That is, conditional on the reference path $\sta_{1:T}$, the remaining particles $\sta_t^{-k_t}$ are proposed from a joint distribution under which 
\begin{align}
  \sta_t^n \sim \dN(\sta_t, \delta_t \iMat), \label{eq:particlerwm_mutation}
\end{align}
for $n \neq k_t$, where $\iMat$ is the $(D \times D)$-identity matrix. As noted in \citet{tjelmeland2004using}, sampling from this joint proposal distribution can be achieved by first sampling an auxiliary variable $\smash{\aux_t \sim \dN(\sta_t, \tfrac{\delta_t}{2}\iMat)}$ and then $\smash{\sta_t^n \sim \dN(\aux_t, \tfrac{\delta_t}{2} \iMat)}$, for $n \neq k_t$. 
\citet{finke2023conditional} also showed that scaling the step size as $\delta_t \in \bo(D^{-1})$ (independently of $T$) guarantees stability in high dimensions. This is again unsurprising because the \gls{PARTICLERWM} algorithm generalises the \gls{RWM} algorithm with Gaussian proposals (and proposal variance $\delta_1$) to $T > 1$ time steps and $N > 1$ proposals. Recently, \citet{corenflos2023auxiliary} showed that the \gls{PARTICLERWM} algorithm can be viewed as a Gibbs-sampling step for the auxiliary variables $\aux_t$ followed by a \gls{CSMC} update which targets a modified Feynman--Kac model which depends on $\aux_{1:T}$, allowing for greater flexibility in the choice proposals. Including related `pseudo observations' $\aux_t$ into \gls{CSMC} updates had previously been suggested by \citet{murray2013disturbance, fearnhead2016augmentation, karppinen2021conditional} but primarily aimed at overcoming the problem that the \gls{CSMC} algorithm mixes poorly if the initial distribution $\Mutation_1(\sta_1)$ is diffuse (and potentially also for improving mixing in the presence of `static' model parameters).

\subsection{Contributions}
\glsreset{AMALA}

Recall that in the `classical' \gls{MCMC} setting, improved performance can often be achieved by enhancing the proposal distribution using gradient or prior information. Thus, in this work, we introduce a methodology which combines the strength of \gls{CSMC} methods (i.e., exploitation of the `decorrelation-over-time' property of the target distribution) with the strengths of sophisticated `classical' \gls{MCMC} approaches (i.e., gradient-enhanced local proposals). 

In the remainder of this section, we detail the contributions of this paper (Table~\ref{tab:overview} summarises our proposed methodology). 

In Section~\ref{sec:particlemala}, we introduce the following \gls{CSMC} type methods which propose particles locally around the reference path guided by gradient information:
  \begin{itemize}
   \item \textbf{\gls{PARTICLEAMALA}.}\ In Section~\ref{subsec:particleamala}, we extend the \gls{PARTICLERWM} algorithm to incorporate gradient information into the proposals. That is, conditional on the reference path $\sta_{1:T}$, the remaining particles $\sta_t^{-k_t}$ are proposed from a joint distribution under which 
    \begin{align}
      \sta_t^n \sim \dN(\sta_t + \tfrac{\delta_t}{2} \nabla_{\sta_t} \log \pi_t(\sta_{1:T}), \delta_t \iMat), \label{eq:particleamala_mutation}
    \end{align}
    for $n \neq k_t$. Sampling from this joint proposal can be achieved by first sampling an auxiliary variable $\smash{\aux_t \sim \dN(\sta_t + \gradientIndicator \tfrac{\delta_t}{2} \nabla_{\sta_t} \log \pi_t(\sta_{1:t}), \tfrac{\delta_t}{2}\iMat)}$ and then $\sta_t^n \sim \dN(\aux_t, \tfrac{\delta_t}{2} \iMat)$, for $n \neq k_t$. We call this method \emph{\gls{PARTICLEAMALA}} because the auxiliary variables $\aux_t$ are explicitly included in the space, i.e.\ they appear in the particle weights, and because the algorithm generalises a version of \emph{\gls{AMALA}} from \citet{titsias2018auxiliary} to $T > 1$ time steps and $N > 1$ proposals.

     \item\textbf{\gls{PARTICLEMALA}.} In Section~\ref{subsec:particlemala}, we improve \gls{PARTICLEAMALA} by marginalising out the auxiliary variables $\aux_t$. We call the resulting method \emph{\gls{PARTICLEMALA}} because it generalises \emph{\gls{MALA}} \citep{besag1994representations} to $T > 1$ time steps and $N > 1$ proposals.

     \item\textbf{\gls{PARTICLEAMALAPLUS}.} In Section~\ref{subsec:particleamalaplus}, we improve \gls{PARTICLEAMALA} differently by replacing the `filter' gradient $\nabla_{\sta_t} \log \pi_t(\sta_{1:t})$ in \eqref{eq:particleamala_mutation} with the `smoothing' gradient $\nabla_{\sta_t} \log \pi_T(\sta_{1:T})$ which is beneficial when future observations are informative about the current state. %
    We call the resulting method \emph{\gls{PARTICLEAMALAPLUS}}.
    \end{itemize}
In Section~\ref{sec:particlemgrad}, we consider the special case that the Feynman--Kac model has conditionally Gaussian mutation kernels: $\Mutation_t(\sta_t|\sta_{t-1}) = \dN(\sta_t; \priorMean_t(\sta_{t-1}), \priorVar_t(\sta_{t-1}))$. In this setting, we introduce the following methods which propose particles locally around the reference path guided by both gradient information and prior information:
    \begin{itemize}
     \item \textbf{\gls{PARTICLEAGRAD}.} In Section~\ref{subsec:particleagrad}, we propose an algorithm which, conditional on the reference path $\sta_{1:T}$, proposes the remaining particles $\sta_t^{-k_t}$ from a joint distribution under which 
     \begin{align}
     \sta_t^n & \sim \dN\bigl((\iMat - \kalmanGain_t(\sta_{t-1}^{a_{t-1}^n})) \priorMean_t(\smash{\sta_{t-1}^{a_{t-1}^n}}) + \kalmanGain_t(\sta_{t-1}^{a_{t-1}^n}) [\sta_t + \tfrac{\delta_t}{2} \nabla_{\sta_t} \log G_t(\sta_{t-1:t})],%
      \mgradProposalVar_t(\sta_{t-1}^{a_{t-1}^n}) \bigr),
        \label{eq:particleagrad_marginal_proposal}
    \end{align}
    for $n \neq k_t$, 
    where $\kalmanGain_t(\sta) \coloneqq (\priorVar_t(\sta) + \tfrac{\delta_t}{2}\iMat)^{-1} \priorVar_t(\sta)$ and $\mgradProposalVar_t(\sta) \coloneqq \tfrac{\delta_t}{2} \kalmanGain_t(\sta)^2 + \kalmanGain_t(\sta)$.
    Sampling from this joint proposal can be achieved by first sampling an auxiliary variable $\smash{\aux_t \sim \dN(\sta_t +  \tfrac{\delta_t}{2} \nabla_{\sta_t} \log \Potential_t(\sta_{t-1:t}), \tfrac{\delta_t}{2}\iMat)}$ and then $\smash{\sta_t^n \sim \MutationAlt_t(\sta_t|\sta_{t-1}^{a_{t-1}^n}; \aux_t)}$, for $n \neq k_t$, where $\MutationAlt_t(\sta_t|\sta_{t-1}; \aux_t) = p(\sta_t|\sta_{t-1}, \aux_t)$ is the fully-adapted auxiliary particle-filter proposal for the state-space model with Gaussian transitions $\sta_t|\sta_{t-1} \sim \Mutation_t(\sta_t|\sta_{t-1})$ and pseudo observations $\aux_t|\sta_t \sim \dN(\aux_t; \sta_t; \frac{\delta_t}{2} \iMat)$. We call this the \emph{\gls{PARTICLEAGRAD}} algorithm because the auxiliary variables $\aux_t$ again appear in the particle weights, and because it generalises the powerful \emph{\gls{AGRAD}} algorithm from \citet{titsias2018auxiliary} to $T > 1$ time steps and $N > 1$ proposals.

    \item \textbf{\gls{PARTICLEMGRAD}.} In Section~\ref{subsec:particlemgrad}, under the assumption that $\priorVar_t(\sta_{t-1}) = \priorVar_t$ and in analogy to Section~\ref{subsec:particlemala}, we improve \gls{PARTICLEAGRAD} by marginalising out the auxiliary variables $\aux_t$. We call the resulting method \emph{\gls{PARTICLEMGRAD}} because it generalises the powerful \emph{\gls{MGRAD}} algorithm from \citet{titsias2018auxiliary} to $T > 1$ time steps and $N > 1$ proposals.
        
    \item \textbf{\gls{PARTICLEAGRADPLUS}.} In Section~\ref{subsec:particleagradplus}, in analogy to Section~\ref{subsec:particleamalaplus}, we improve \gls{PARTICLEAGRAD} by replacing the `filter-potential' gradients $\nabla_{\sta_t} \log \Potential_t(\sta_{t-1:t})$ in \eqref{eq:particleagrad_marginal_proposal} with `smoothing-potential' gradients $\nabla_{\sta_t} \log \Potential_{1:T}(\sta_{1:T})$ which may be beneficial if $\Potential_t(\sta_{t-1:t})$ varies significantly in $\sta_{t-1}$. We call this method \emph{\gls{PARTICLEAGRADPLUS}}. 
    
    \item \textbf{Twisted \gls{PARTICLEAGRAD}(+).} In Section~\ref{subsec:twisted_particleagrad}, under the assumption that $\priorMean_t(\sta_{t-1}) = \priorFactor_t \sta_{t-1} + \priorIntercept_t$ and $\priorVar_t(\sta_{t-1}) = \priorVar_t$, we improve \gls{PARTICLEAGRAD} and \gls{PARTICLEAGRADPLUS} by instead using all future auxiliary variables $\aux_{t:T}$ to propose $\smash{\sta_t^n \sim \MutationAlt_t(\sta_t|\sta_{t-1}^{a_{t-1}^n}; \aux_{t:T})}$, for $n \neq k_t$, where $\MutationAlt_t(\sta_t|\sta_{t-1}; \aux_{t:T}) = p(\sta_t|\sta_{t-1}, \aux_{t:T})$ is the fully \emph{twisted} particle filter proposal for the state-space model with Gaussian transitions and pseudo observations $\aux_t$ mentioned above. We call the resulting methods \emph{twisted \gls{PARTICLEAGRAD}} and \emph{twisted \gls{PARTICLEAGRADPLUS}}.
   \end{itemize}
   In Section~\ref{subsec:interpolation}, we prove that \gls{PARTICLEAGRAD} and \gls{PARTICLEMGRAD} (and their smoothing-gradient/twisted variants) solve the `tuning' problem of having to choose between:
    \begin{enumerate}
        \item \label{enum:standard_csmc} the \gls{CSMC} algorithm (which proposes particles solely based on the prior dynamics);
        \item \label{enum:purely_local} the \gls{PARTICLEAMALA}, \gls{PARTICLEMALA} or \gls{PARTICLEAMALAPLUS} (which propose particles solely locally around the reference path). 
    \end{enumerate}
    \begin{figure}
      \centering
      \includegraphics{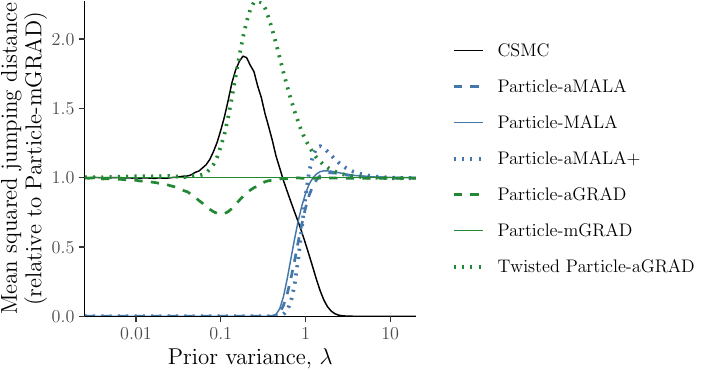}
      \caption{Empirical illustration of the `interpolation' from Propositions~\ref{prop:convergence_for_highly_informative_prior} and \ref{prop:convergence_for_weakly_informative_prior} in the toy linear-Gaussian state-space model from Figure~\ref{fig:lgssm_scaling} (with $D = T = 10$).} 
      \label{fig:lgssm_variance}
    \end{figure}
    This choice is not always clear: on the one hand, Choice~\ref{enum:purely_local} can exhibit superior performance in high dimensions. On the other hand, if the prior dynamics are highly informative then Choice~\ref{enum:standard_csmc} can outperform Choice~\ref{enum:purely_local}. Specifically, we prove that the following results hold in stationarity and under the simplifying assumption that the model factorises over time, i.e., if $\Potential_t$, $\priorMean_t$, $\priorVar_t$ (and hence $\kalmanGain_t$ and $\mgradProposalVar_t$ in \eqref{eq:particleagrad_marginal_proposal}) do not depend on the state at time $t-1$:
    \begin{itemize}
      \item \textbf{Proposition~\ref{prop:convergence_for_highly_informative_prior}.} \gls{PARTICLEAGRAD} and \gls{PARTICLEMGRAD} reduce to the \gls{CSMC} algorithm  as prior dynamics become \emph{more} informative. Informally, we then have $\kalmanGain_t \approx \zeroMat$ and $\mgradProposalVar_t \approx \priorVar_t$ so that \eqref{eq:particleagrad_marginal_proposal} reduces to \eqref{eq:standard_mutation}.
      \item \textbf{Proposition~\ref{prop:convergence_for_weakly_informative_prior}.} \gls{PARTICLEAGRAD} and \gls{PARTICLEMGRAD} reduce to \gls{PARTICLEAMALA} and \gls{PARTICLEMALA}, respectively, as prior dynamics become \emph{less} informative. Informally, we then have $\kalmanGain_t \approx \iMat$ and $\mgradProposalVar_t \approx \delta_t \iMat$ so that \eqref{eq:particleagrad_marginal_proposal} reduces to \eqref{eq:particleamala_mutation}.
    \end{itemize}
    Propositions~\ref{prop:convergence_for_highly_informative_prior} and \ref{prop:convergence_for_weakly_informative_prior} are illustrated in Figure~\ref{fig:lgssm_variance} for a model in which the independence across time-steps is not verified.
    As a by-product, these propositions show that the \gls{AGRAD}/\gls{MGRAD} algorithms from \citet{titsias2018auxiliary} can be viewed as automatically interpolating between the \gls{IMH} algorithm and \gls{AMALA}/\gls{MALA}, depending on the `informativeness' of the prior. To our knowledge, this has not been pointed out in the literature. As another by-product, the methodology presented in this section also addresses the `tuning problem' of having to choose whether to sample the initial latent state $\sta_1$ within the \gls{CSMC} scheme (which is preferable if the prior on the initial state is informative) or to treat it as a `static' parameter to be sampled separately~\citep[which is preferable if this prior is diffuse, see][]{murray2013disturbance, fearnhead2016augmentation, karppinen2021conditional}. %

In Section~\ref{sec:simulations}, we demonstrate the performance of our methodology on a high-dimensional multivariate stochastic volatility model, often used as a benchmark in the particle filtering literature. The different methods proposed in this article dramatically improve on existing \gls{CSMC} and related methods and also on `classical' \gls{MCMC} methods in terms of effective sample size for different levels of prior informativeness. %

All proofs (e.g., of the fact that the proposed methods leave $\pi_T(\sta_{1:T})$ invariant) are deferred to the appendix. Additionally, in Appendix~\ref{app:sec:particlepcnl}, we introduce \emph{\gls{PARTICLEPCNL}} methods which generalise the \glsreset{PCNL}\emph{\gls{PCNL}} algorithm from \citet{cotter2013crank} to $T > 1$ time steps and $N > 1$ proposals. The methods proposed in this work and their special cases if $N = T = 1$ are summarised in Table~\ref{tab:overview}. Note that for $T = 1$ but $N > 1$, our work implies novel \emph{multi-proposal} versions of `classical' \gls{MCMC} kernels like \gls{MALA}, \gls{AMALA}, \gls{MGRAD}, \gls{AGRAD} and \gls{PCNL}. These may be of independent interest because they can exploit parallel computing architectures for inference in non-dynamic models.

Importantly, and in keeping with existing \gls{CSMC} methodology, the computational cost of all our proposed algorithms is linear in both $T$ and $N$, in time and memory.

Finally, the Python code for reproducing our experiments is publicly available at \url{https://github.com/AdrienCorenflos/particle_mala}. 
It was written as a library and can be extended to accommodate other models than the ones considered here.

\glsunset{AGRAD}
\glsunset{MGRAD}
\begin{table}%
  \centering
   \setlength{\extrarowheight}{5pt}
   \caption{The methods mentioned in this work (new methods are in \textit{italic}).}
   \begin{threeparttable}
  \begin{tabular}{lrr}
    \toprule
    \textbf{Method} & \!\!\!\!\!\!\!\!\!\!\textbf{Section} & 
    \makecell[r]{\textbf{Special case}\\\textbf{if} $N = T = 1$}
    \\\midrule 
    \acrshort{CSMC}\tnote{\textdagger}    & \ref{subsec:csmc}     & \gls{IMH}\tabularnewline
    \acrshort{PARTICLERWM} & \ref{subsec:particlerwm}   & \gls{RWM} \tabularnewline
    \textit{\acrshort{PARTICLEAMALA}}     & \ref{subsec:particleamala}   & \gls{AMALA} \tabularnewline
    \textit{\acrshort{PARTICLEMALA}}     & \ref{subsec:particlemala}   & \gls{MALA} \tabularnewline
    \textit{\acrshort{PARTICLEAMALAPLUS}}    & \ref{subsec:particleamalaplus}  & \gls{AMALA} \tabularnewline
    \textit{\acrshort{PARTICLEAGRAD}}    & \ref{subsec:particleagrad}  & \gls{AGRAD}  \tabularnewline
    \textit{\acrshort{PARTICLEMGRAD}}    & \ref{subsec:particlemgrad}  & \gls{MGRAD}  \tabularnewline
    \textit{\acrshort{PARTICLEAGRADPLUS}}   & \ref{subsec:particleagradplus} & \gls{AGRAD} \tabularnewline
    \textit{Twisted \acrshort{PARTICLEAGRAD}(+)}  & \ref{subsec:twisted_particleagrad} & \gls{AGRAD} \tabularnewline
    \textit{\acrshort{PARTICLEPCNL}} \& more\tnote{\textdaggerdbl}   & Appendix~\ref{app:sec:particlepcnl}  & \acrshort{PCNL}
    \\\bottomrule
  \end{tabular}
  \begin{tablenotes}%
    \footnotesize
    \item[\textdagger] In our taxonomy, \gls{CSMC} could be called `Particle-\gls{IMH}'. However, the latter already refers to an entirely different algorithm in \citet{andrieu2010particle}.
    \item[\textdaggerdbl] We again also describe auxiliary-variable, smoothing-gradient (`+') and twisted versions.
    \end{tablenotes}
    \end{threeparttable}
  \label{tab:overview}
\end{table}

%% file: 02_background.tex
\subsection{CSMC (particle extension of IMH)}
\label{subsec:csmc}

\subsubsection{Algorithm}

Assume that we can generate \gls{IID} samples from the mutation kernels $\Mutation_t(\sta_t|\sta_{t-1})$. A method for constructing a $\pi_T$-invariant \gls{MCMC} kernel $P_\csmc(\staAlt_{1:T}|\sta_{1:T})$ is then given by the \emph{\gls{CSMC}} algorithm from \citet{andrieu2010particle} which proposes $N$ particles at each time step to build up an efficient proposal. Algorithm~\ref{alg:csmc} summarises the scheme, where `\gls{WP}' is short for `\glsdesc{WP}'. We also recursively define the $n$th surviving particle lineage at time $t$ as
\begin{align}
  \sta_{1:t}^{(n)} \coloneqq (\sta_{1:t-1}^{(a_{t-1}^n)}, \sta_t^n).
\end{align}
In particular, therefore, $\smash{\sta_{t-1:t}^{(n)} = (\sta_{t-1}^{a_{t-1}^n}, \sta_t^n)}$.
\begin{framedAlgorithm}[\gls{CSMC}] \label{alg:csmc}
Given $\sta_{1:T} \in \spaceX^T$:
\begin{enumerate}
    \item \label{alg:csmc:1} for $t = 1, \dotsc, T$, 
    \begin{enumerate}
        \item sample $k_t$ from a uniform distribution on $[N]_0$ and set $\sta_t^{k_t} \coloneqq \sta_t$,
        \item \label{alg:csmc:1:b} if $t > 1$, set $\smash{a_{t-1}^{k_t} \coloneqq k_{t-1}}$ and sample $a_{t-1}^n = i$ \gls{WP} $\smash{W_{t-1}^i}$, for $n \in [N]_0 \setminus \{k_t\}$,
        
        \item \label{alg:csmc:mutation} sample $\smash{\sta_t^n \sim \Mutation_t(\ccdot|\sta_{t-1}^{a_{t-1}^n})}$ for $n \in [N]_0 \setminus \{k_t\}$, 
        
        \item \label{alg:csmc:weight_calculation} for $n \in [N]_0$, set $\smash{w_t^n \propto G_t(\sta_{t-1:t}^{(n)})}$.
        
        \item \label{alg:csmc:weight_normalisation} for $n \in [N]_0$, set $\smash{W_t^n \coloneqq  w_t^n / \sum_{m = 0}^N w_t^m}$;%
    \end{enumerate}
    
    \item \label{alg:csmc:2} sample $i \in [N]_0 \setminus \{k_T\}$ \gls{WP} $\dfrac{W_T^i}{1 - W_T^{k_T}}$; set $l_T \coloneqq i$ \gls{WP} $\smash{1 \wedge \dfrac{1- W_T^{k_T}}{1 - W_T^i}}$; otherwise, set $l_T \coloneqq k_T$;

    \item \label{alg:csmc:backward_sampling} for $t = T-1, \dotsc, 1$, sample $l_t = i \in [N]_0$ \gls{WP}
          $
          \dfrac{W_t^{i} \qsemigroup_{t+1}(\sta_t^{i}, \sta_{t+1}^{l_{t+1}})}{\sum_{n=0}^N W_t^n \qsemigroup_{t+1}(\sta_t^{n}, \sta_{t+1}^{l_{t+1}})};
          $
    \item \label{alg:csmc:4} return $\staAlt_{1:T} \coloneqq (\sta_1^{l_1}, \dotsc, \sta_t^{l_T})$.
\end{enumerate}
\end{framedAlgorithm}
Algorithm~\ref{alg:csmc} includes two extensions to the original presentation of the \gls{CSMC} algorithm in \citet{andrieu2010particle}:
\begin{itemize}
  \item Step~\ref{alg:csmc:2} uses the so-called \emph{forced-move} extension for \gls{CSMC} algorithms which was proposed in \citet{chopin2013particle} \citep[see also][]{liu1996peskun}. The algorithm would still be valid if we instead sampled $l_T = i \in [N]_0$ with probability $W_T^i$.%
  \item Step~\ref{alg:csmc:backward_sampling} is the \emph{backward-sampling} extension from \citet{whiteley2010particle}.  The algorithm would still be valid if we instead set $\smash{l_t = a_t^{l_{t+1}}}$ (but typically much less efficient, especially if $T$ is large).
\end{itemize}
Importantly,  sampling $\staAlt_{1:T}$ given $\sta_{1:T}$ as described in Algorithm~\ref{alg:csmc} induces a Markov kernel $P_\csmc(\staAlt_{1:T}|\sta_{1:T})$ which leaves $\pi_T$ invariant. For sufficiently ergodic models, this \gls{MCMC} kernel can yield highly efficient updates of the sequence of latent states, even if the time horizon $T$ is large \citep{lee2020coupled, karjalainen2023mixing}.

\subsubsection{Relationship with `classical' MCMC algorithms} 
\label{subsubsec:csmc:relationship_with_classical_mcmc_algorithms}

Interestingly, the \gls{CSMC} algorithm generalises the classical \emph{\gls{IMH}} algorithm \citep{hastings1970monte} in the sense that the former reduces to the latter if $T = N = 1$. This can be seen as follows, where we suppress the `time' subscript $t = 1$ everywhere to simplify the notation. Given that the current state of the Markov chain is $\sta = \sta^0$ (we can assume that $k = 0$ without loss of generality), Step~\ref{alg:csmc:mutation} of Algorithm~\ref{alg:csmc} proposes $\sta^1 \sim  \Mutation$. The remaining steps return $\staAlt \coloneqq \sta^1$ as the new state with acceptance probability $1 \wedge \alpha_{\text{\upshape\gls{IMH}}}(\sta^0, \sta^1)$, where
\begin{align}
  \alpha_{\text{\upshape\gls{IMH}}}(\sta^0, \sta^1) 
  & \coloneqq \frac{1 - W^0}{1 - W^1}
  = \frac{G(\sta^1)}{G(\sta^0)}
  = \frac{\pi(\sta^1) \Mutation(\sta^0)}{\pi(\sta^0) \Mutation(\sta^1)}.
\end{align}
Otherwise, the old state $\staAlt \coloneqq \sta^0 = \sta$ is returned as the new state.

\subsubsection{Breakdown in high dimensions}
Unfortunately, as shown in \citet{finke2023conditional}, Algorithm~\ref{alg:csmc} suffers from a curse of dimension if $D$ is large (unless the number of proposed particles, $N$, grows exponentially in $D$ but that is prohibitive). This is not surprising since the \gls{IMH} algorithm is known to break down in high dimensions (due to the difficulty of finding an efficient global proposal distribution $\Mutation$ in high dimensions). %

\subsection{Particle-RWM}
\label{subsec:particlerwm}
\glsreset{PARTICLERWM}

\subsubsection{Algorithm}
To circumvent the curse of dimension, \citet{finke2023conditional} \citep[see also][for related methods]{shestopaloff2018sampling, malory2021bayesian} developed the \emph{\gls{PARTICLERWM}} algorithm which scatters the proposed particles locally around the reference path using Gaussian perturbations as outlined in Algorithm~\ref{alg:particlerwm}.

\begin{framedAlgorithm}[\gls{PARTICLERWM}] \label{alg:particlerwm}
Implement Algorithm~\ref{alg:csmc} but replace the particle proposal (Step~\ref{alg:csmc:mutation}) and the weight calculation (Step~\ref{alg:csmc:weight_calculation}) by
\begin{enumerate}
    \myitem{1c} \label{alg:particlerwm:mutation} sample $\smash{\aux_t \sim \dN(\sta_t, \tfrac{\delta_t}{2} \iMat)}$, and $\smash{\sta_t^n \sim \dN(\aux_t, \tfrac{\delta_t}{2}\iMat)}$, for $n \in [N]_0 \setminus \{k_t\}$, 
    
    \myitem{1d} \label{alg:particlerwm:weight_calculation} for $n \in [N]_0$, set $\smash{w_t^n \propto \qsemigroup_t(\sta_{t-1:t}^{(n)})}$.
\end{enumerate}
\end{framedAlgorithm}

Notably, Step~\ref{alg:particlerwm:mutation} marginally samples $\smash{\sta_t^n \sim \dN(\sta_t, \delta_t \iMat)}$, for $n \neq k_t$.

\subsubsection{Interpretation as a CSMC update on an extended space}
\label{subsubsection:particlerwm:extended_state-space_justification}

\citet{corenflos2023auxiliary} showed that Algorithm~\ref{alg:particlerwm} can be derived by including the auxiliary variables $\aux_t$ into the space and thus considering the extended distribution
\begin{align}
  \targetAlt_T(\sta_{1:T}, \aux_{1:T}) 
  & \coloneqq \target_T(\sta_{1:T}) \prod_{t=1}^T \dN(\aux_t; \sta_t, \tfrac{\delta_t}{2}\iMat),
\end{align}
which admits $\target_T(\sta_{1:T})$ as a marginal and which can be targeted by alternating the following two steps. Given $\sta_{1:T} \in \spaceX^T$,
\begin{enumerate}
    \item sample $\aux_t \sim \dN(\sta_t, \tfrac{\delta_t}{2} \iMat)$, for $t = 1,\dotsc, T$;
    \item run the \gls{CSMC} algorithm (Algorithm~\ref{alg:csmc}) but with $\Mutation_t(\sta_t|\sta_{t-1})$, $\Potential_t(\sta_{t-1:t})$, and $\qsemigroup_t(\sta_{t-1:t})$ replaced by $\smash{\MutationAlt_t(\sta_t|\sta_{t-1}; \aux_t) \coloneqq \dN(\sta_t; \aux_t, \tfrac{\delta_t}{2}\iMat)}$, $\smash{\PotentialAlt_t(\sta_{t-1:t}) \coloneqq \qsemigroup_t(\sta_{t-1:t})}$ and $\smash{\qsemigroupAlt_t(\sta_{t-1:t}; \aux_t)} \coloneqq \smash{\MutationAlt_t(\sta_t|\sta_{t-1}; \aux_t) \PotentialAlt_t(\sta_{t-1:t})}$.
\end{enumerate}
In particular, this shows that sampling $\staAlt_{1:T}$ given $\sta_{1:T}$ via Algorithm~\ref{alg:particlerwm} induces a Markov kernel $P_\particlerwm(\staAlt_{1:T}|\sta_{1:T})$ which leaves $\pi_T$ invariant.

\subsubsection{Relationship with `classical' MCMC algorithms}
\label{subsubsec:particlerwm:relationship_with_other_methods}

The \gls{PARTICLERWM} algorithm generalises the classical (Gaussian) \emph{\gls{RWM}} algorithm of \citet{metropolis1953equation} in the sense that the former reduces to the latter if $T = N = 1$. This can be seen as follows, where we again suppress the `time' subscript $t = 1$ everywhere to simplify the notation. Given that the current state of the Markov chain is $\sta = \sta^0$ (we can again assume that $k = 0$ without loss of generality), Step~\ref{alg:particlerwm:mutation} of Algorithm~\ref{alg:particlerwm} proposes $\sta^1 \sim  \dN(\sta^0, \delta \iMat)$. The remaining steps return $\staAlt \coloneqq \sta^1$ as the new state with acceptance probability $1 \wedge \alpha_{\text{\upshape\gls{RWM}}}(\sta^0, \sta^1)$, where
\begin{align}
  \alpha_{\text{\upshape\gls{RWM}}}(\sta^0, \sta^1) 
  & \coloneqq \frac{1 - W^0}{1 - W^1}
  = \frac{\pi(\sta^1)}{\pi(\sta^0)}.
\end{align}
Otherwise, the old state $\staAlt \coloneqq \sta^0 = \sta$ is returned as the new state.

\subsubsection{Stability in high dimensions}

\citet{finke2023conditional} proved that the \gls{PARTICLERWM} algorithm circumvents the curse of dimensionality if the proposal variance is scaled as $\delta_t \in \bo(D^{-1})$ \citep[see also][for a proof for non-Gaussian exchangeable proposals but in the case where the model factorises over time]{malory2021bayesian}. However, from the literature on classical \gls{MCMC} algorithms, it is well known that faster convergence rates can be achieved by incorporating gradient information into the proposal \citep{roberts1998optimal}. Thus, in the next section, we extend the \gls{PARTICLERWM} to allow for gradient-informed proposals.

%% file: 03_particlemala.tex
\subsection{Particle-aMALA}
\label{subsec:particleamala}

We now propose \emph{\gls{PARTICLEAMALA}}, a method which extends the \gls{PARTICLERWM} algorithm from \citet{finke2023conditional} by allowing for the use of gradient information in the proposal. For the moment, gradients are taken w.r.t.\ the filtering densities and we employ an indicator $\gradientIndicator \in \{0,1\}$ to permit switching off the use of gradient information.

We now write
\begin{align}
  \MutationAlt_t(\sta_t|\sta_{t-1}; \aux_t) & \coloneqq \dN(\sta_t; \aux_t, \tfrac{\delta_t}{2}\iMat), \label{eq:particleamala:mutation}\\
  \PotentialAlt_t(\sta_{t-1:t}; \aux_t) 
   &\coloneqq \qsemigroup_t(\sta_{t-1:t}) \frac{\dN(\aux_t; \sta_t + \gradientIndicator \tfrac{\delta_t}{2} 
   \nabla_{\sta_t} \log \pi_t(\sta_{1:t})%
   , \tfrac{\delta_t}{2} \iMat)}{\dN(\aux_t; \sta_t, \tfrac{\delta_t}{2}\iMat)}, \label{eq:particleamala:potential}
\end{align}
as well as $\qsemigroupAlt_t(\sta_{t-1:t}; \aux_t) \coloneqq \MutationAlt_t(\sta_t|\sta_{t-1}; \aux_t) \PotentialAlt_t(\sta_{t-1:t}; \aux_t)$, where we note that
\begin{align}
 \nabla_{\sta_t} \log \pi_t(\sta_{1:t}) = \nabla_{\sta_t} \log \qsemigroup_t(\sta_{t-1:t}).
\end{align}
A single iteration of the \gls{PARTICLEAMALA} is then as follows.

\begin{framedAlgorithm}[\gls{PARTICLEAMALA}] \label{alg:particleamala}
Implement Algorithm~\ref{alg:csmc} but replace the particle proposal (Step~\ref{alg:csmc:mutation}) and the weight calculation (Step~\ref{alg:csmc:weight_calculation}) by
\begin{enumerate}
    \myitem{1c}\label{alg:particleamala:mutation} sample $\smash{\aux_t \sim \dN(\sta_t + \gradientIndicator \tfrac{\delta_t}{2}
    \nabla_{\sta_t} \log \pi_t(\sta_{1:t})
    , \tfrac{\delta_t}{2} \iMat)}$, and $\smash{\sta_t^n \sim \dN(\aux_t, \tfrac{\delta_t}{2}\iMat)}$, for $n \in [N]_0 \setminus \{k_t\}$, 
    \myitem{1d} \label{alg:particleamala:weight_calculation} for $n \in [N]_0$, set
            $\smash{w_t^n \propto \PotentialAlt_t(\sta_{t-1:t}^{(n)}; \aux_t),}$
\end{enumerate}
and also replace $\qsemigroup_{t+1}(\ccdot)$ in the backward kernel in Step~\ref{alg:csmc:backward_sampling} by $\smash{\qsemigroupAlt_{t+1}(\ccdot; \aux_{t+1})}$.
\end{framedAlgorithm}

Step~\ref{alg:particleamala:mutation} marginally samples $\smash{\sta_t^n \sim \dN(\sta_t + \gradientIndicator \tfrac{\delta_t}{2} \nabla_{\sta_t} \log \pi_t(\sta_{1:t}), \delta_t \iMat)}$, for $n \neq k_t$. This follows from Lemma~\ref{lem:auxiliary_lemma_1} in Appendix~\ref{app:sec:integrating_out_the_auxiliary_variables}.

\begin{proposition}[validity of \gls{PARTICLEAMALA}] \label{prop:particleamala:validity}
  Sampling $\staAlt_{1:T}$ given $\sta_{1:T}$ via Algorithm~\ref{alg:particleamala} induces a Markov kernel $P_\particleamala(\staAlt_{1:T}|\sta_{1:T})$ which leaves $\pi_T$ invariant.
\end{proposition}

\subsection{Particle-MALA}
\label{subsec:particlemala}

In this section, we analytically integrate out the auxiliary variables $\aux_t$ appearing in the weights of the \gls{PARTICLEAMALA}. A single iteration of the resulting methodology -- which we term the \emph{\gls{PARTICLEMALA}} -- is as follows, where we write 
\begin{align}
  \log H_{t,\genericGrad}(\sta, \bar{\sta}) \coloneqq \frac{1}{\delta_t} \bigl[2 \genericGrad^\T (\bar{\sta} - \sta) - \tfrac{N}{N+1} \genericGrad^\T \genericGrad \bigr].\label{eq:particlemala:weight_factor}
\end{align}

\begin{framedAlgorithm}[\gls{PARTICLEMALA}] \label{alg:particlemala}
Implement Algorithm~\ref{alg:csmc} but replace the particle proposal (Step~\ref{alg:csmc:mutation}) and the weight calculation (Step~\ref{alg:csmc:weight_calculation}) by
\begin{enumerate}
    \myitem{1c}\label{alg:particlemala:mutation} sample $\smash{\aux_t \sim \dN(\sta_t + \gradientIndicator \tfrac{\delta_t}{2}
    \nabla_{\sta_t} \log \pi_t(\sta_{1:t})
    , \tfrac{\delta_t}{2} \iMat)}$, and $\smash{\sta_t^n \sim \dN(\aux_t, \tfrac{\delta_t}{2}\iMat)}$, for $n \in [N]_0 \setminus \{k_t\}$, 
    \myitem{1d} \label{alg:particlemala:weight_calculation} set $\smash{\bar{\sta}_t \coloneqq \tfrac{1}{N+1}\sum_{n=0}^{N} \sta_t^n}$ and, for $n \in [N]_0$, 
    \begin{align}
      w_t^n \propto \qsemigroup_t(\sta_{t-1:t}^{(n)}) H_{t,\gradientIndicator\frac{\delta_t}{2} \nabla_{\sta_t^n} \log \qsemigroup_t(\sta_{t-1:t}^{(n)})}(\sta_t^n, \bar{\sta}_t).
    \end{align}
\end{enumerate}
\end{framedAlgorithm}

Step~\ref{alg:particlemala:weight_calculation} pre-computes $\smash{\bar{\sta}_t}$ to ensure that the algorithm can still be implemented in $\bo(N)$ operations even though the weight of the $n$th particle now depends on the values of all $N+1$ particles. However, note that the auxiliary variables $\aux_t$ no longer appear in the weights.

\begin{remark}[\gls{PARTICLEAMALA} `exactly approximates' \gls{PARTICLEMALA}]\label{rem:particleamala_as_noisy_version_of_particlemala}
Note that the \gls{PARTICLEAMALA} differs from the \gls{PARTICLEMALA} only in the definition of the weights (and the backward-sampling weights). This allows us to interpret the former as a `noisy' version of the latter. 
Indeed,  write the unnormalised weight of the $n$th particle at time-$t$ in the \gls{PARTICLEAMALA} as $w_t^n(\aux_t)$, whilst $w_t^n$ denotes the corresponding weight under the \gls{PARTICLEMALA} (which does not depend on the auxiliary variable $\aux_t$). Then we have
    \begin{align}
    \frac{w_t^n(\aux_t)}{w_t^{k_t}(\aux_t)} 
    &= \frac{w_t^n}{w_t^{k_{\mathrlap{t}}}} \times 
    \frac{\proposalAlt_t^{-n}(\aux_t | \sta_t^{-n}, \sta_t^n; \calH_{t - 1})}{\proposalAlt_t^{-k_t}(\aux_t | \sta_t^{-k_t}, \sta_t^{k_t}; \calH_{t - 1})},
  \end{align}
  where $\smash{\proposalAlt_t^{-n}(\aux_t | \sta_t^{-n}, \sta_t^n; \calH_{t - 1}) = \dN(\aux_t; \bar{\sta}_t + \gradientIndicator \tfrac{\delta_t}{2(N+1)}\nabla_{\sta_t^n} \log \qsemigroup_t(\sta_{t-1:t}^{(n)}), \tfrac{\delta_t}{2(N+1)}\iMat)}$ is the conditional distribution of $\aux_t$ under the joint distribution of all random variables generated by Algorithm~\ref{alg:particleamala} up to (and including) time~$t$ assuming the reference particle at time $t$ is placed in position $n$ (and $\calH_{t-1}$ denotes the history of the particle system, i.e.\ all particles and ancestor indices up to time $t-1$). This conditional distribution follows from Lemma~\ref{lem:auxiliary_lemma_1} in Appendix~\ref{app:sec:integrating_out_the_auxiliary_variables}. 
  In particular, we therefore have 
  \begin{align}
    \E\biggl[ \frac{w_t^n(\aux_t)}{w_t^{k_t}(\aux_t)}\biggr] = \frac{w_t^n}{w_t^{k_{\mathrlap{t}}}},
  \end{align}
  where the expectation is taken w.r.t.\ $\smash{\proposalAlt_t^{-k_t}(\aux_t | \sta_t^{-k_t}, \sta_t^{k_t}; \calH_{t - 1})}$. Interestingly, for the \gls{PARTICLERWM} algorithm (recovered by setting $\gradientIndicator = 0$), the `auxiliary' and `marginal' variants are statistically equivalent.
\end{remark}

\begin{proposition}[validity of \gls{PARTICLEMALA}] \label{prop:particlemala:validity}
  Sampling $\staAlt_{1:T}$ given $\sta_{1:T}$ via Algorithm~\ref{alg:particlemala} induces a Markov kernel $P_{\text{\upshape\gls{PARTICLEMALA}}}(\staAlt_{1:T}|\sta_{1:T})$ which leaves $\pi_T$ invariant.
\end{proposition}

\subsection{Particle-aMALA+}
\label{subsec:particleamalaplus}

In this section, we extend the \gls{PARTICLEAMALA} in a different manner: we now modify the algorithm so that the proposal distributions incorporate gradients w.r.t.\ the joint smoothing distribution $\pi_T$ rather than w.r.t.\ the filters, $\pi_t$. This can be beneficial if there is a significant discrepancy between the marginal distribution of $\sta_t$ under the former and the latter as is typically the case if $D$ is large. Indeed, this discrepancy is likely the reason for the decay in performance of \gls{PARTICLEAMALA} and \gls{PARTICLEMALA} for very large $D$ visible in Figure~\ref{fig:lgssm_dimension}.

For $\MutationAlt_t(\sta_t| \sta_{t-1}; \aux_t)$ and $\PotentialAlt_t(\sta_{t-1:t}; \aux_t)$ still defined as in the \gls{PARTICLEAMALA} algorithm (i.e., as in \eqref{eq:particleamala:mutation} and \eqref{eq:particleamala:potential}),
we now write
\begin{samepage}
\begin{align*}
  \PotentialAlt_t(\sta_{t-2:t}; \aux_{t-1:t})%
  & \coloneqq \PotentialAlt_t(\sta_{t-1:t}; \aux_t)
  \frac{\dN(\aux_{t-1}; \sta_{t-1} + \gradientIndicator\tfrac{\delta_{t-1}}{2} 
  \nabla_{\sta_{t-1}} \log \pi_T(\sta_{1:T})
  , \tfrac{\delta_{t-1}}{2} \iMat) }{\dN(\aux_{t-1}; \sta_{t-1} + \gradientIndicator \tfrac{\delta_{t-1}}{2} 
  \nabla_{\sta_{t-1}} \log \pi_{t-1}(\sta_{1:t-1}) %
  , \tfrac{\delta_{t-1}}{2} \iMat)},\!\!\!\!\!\!\!\!\!\!
\end{align*}
\end{samepage}
as well as $\qsemigroupAlt_t(\sta_{t-2:t}; \aux_{t-1:t}) \coloneqq \MutationAlt_t(\sta_t|\sta_{t-1}; \aux_t) \PotentialAlt_t(\sta_{t-2:t}; \aux_{t-1:t})$, where we note that
\begin{align}
 \nabla_{\sta_t} \log \pi_T(\sta_{1:T}) = \nabla_{\sta_t} [\log \qsemigroup_t(\sta_{t-1:t}) + \log \qsemigroup_{t+1}(\sta_{t:t+1})].
\end{align}
A single iteration of the resulting `smoothing-gradient' methodology -- which we term the \emph{\gls{PARTICLEAMALAPLUS}} -- is then as follows.

\begin{framedAlgorithm}[\gls{PARTICLEAMALAPLUS}] \label{alg:particleamalaplus}
Implement Algorithm~\ref{alg:csmc} but replace the particle proposal (Step~\ref{alg:csmc:mutation}), the weight calculation (Step~\ref{alg:csmc:weight_calculation}), and backward sampling (Step~\ref{alg:csmc:backward_sampling}) by
\begin{enumerate}
    \myitem{1c}\label{alg:particleamalaplus:mutation} sample $\smash{\aux_t \sim \dN(\sta_t + \gradientIndicator \tfrac{\delta_t}{2}\nabla_{\sta_t}\log \pi_T(\sta_{1:T}), \tfrac{\delta_t}{2} \iMat)}$, and $\smash{\sta_t^n \sim \dN(\aux_t, \tfrac{\delta_t}{2}\iMat)}$, for $n \in [N]_0 \setminus \{k_t\}$, 
    \myitem{1d} \label{alg:particleamalaplus:weight_calculation} for $n \in [N]_0$, set
            $\smash{w_t^n \propto \PotentialAlt_t(\sta_{t-2:t}^{(n)}; \aux_{t-1:t}),}$
    \myitem{3} \label{alg:particleamalaplus:backward_sampling} for $t = T-1, \dotsc, 1$, sample $l_t = i \in [N]_0$ \gls{WP}
            \begin{align}
          \dfrac{W_t^{i} \qsemigroupAlt_{t+1}((\sta_{t-1:t}^{(i)}, \sta_{t+1}^{l_{t+1}}); \aux_{t:t+1}) \qsemigroupAlt_{t+2}((\sta_t^i, \sta_{t+1}^{l_{t+1}}, \sta_{t+2}^{l_{t+2}}); \aux_{t+1:t+2})}{\sum_{n=0}^N W_t^n \qsemigroupAlt_{t+1}((\sta_{t-1:t}^{(n)}, \sta_{t+1}^{l_{t+1}}); \aux_{t:t+1}) \qsemigroupAlt_{t+2}((\sta_t^n, \sta_{t+1}^{l_{t+1}}, \sta_{t+2}^{l_{t+2}}); \aux_{t+1:t+2})}.
          \end{align}
\end{enumerate}
\end{framedAlgorithm}

In Step~\ref{alg:particleamalaplus:backward_sampling}, we recall the convention that any quantity with `time' index $t > T$ should be ignored, so that $\qsemigroupAlt_{T+1} \equiv 1$.
Some comments about Algorithm~\ref{alg:particleamalaplus} are in order.
\begin{itemize}

    \item Step~\ref{alg:particleamalaplus:mutation} marginally samples $\smash{\sta_t^n \sim \dN(\sta_t + \gradientIndicator \tfrac{\delta_t}{2}\nabla_{\sta_t} \log \pi_T(\sta_{1:T}), \delta_t \iMat)}$, for $n \neq k_t$. This is in contrast to the \gls{PARTICLEAMALA} and \gls{PARTICLEMALA}, whose (marginal) proposal distribution is centred around $\sta_t + \gradientIndicator \tfrac{\delta_t}{2}\nabla_{\sta_t} \log \pi_t(\sta_{1:t})$.

    \item Steps~\ref{alg:particleamalaplus:weight_calculation} and \ref{alg:particleamalaplus:backward_sampling} are similar to the weight-calculation and backward-sampling steps in the previous algorithms. The only difference here is that the model is now no longer (first-order) Markov in the sense that the (incremental) weights at time $t$ now also depend on the state at time $t-2$.

\end{itemize}

\begin{proposition}[validity of \gls{PARTICLEAMALAPLUS}] \label{prop:particleamalaplus:validity}
  Sampling $\staAlt_{1:T}$ given $\sta_{1:T}$ via Algorithm~\ref{alg:particleamalaplus} induces a Markov kernel $P_\particlemala(\staAlt_{1:T}|\sta_{1:T})$ which leaves $\pi_T$ invariant.
\end{proposition}

\subsection{Relationship with other methods}
\label{subsec:particlemala:relationship_with_other_methods}

We end this section by relating the proposed algorithms to existing methodologies.

\begin{enumerate}

   \item \textbf{Generalisation of \gls{PARTICLERWM} and \gls{RWM}.} If $\gradientIndicator = 0$, then the algorithms introduced in this section (\gls{PARTICLEAMALA}, \gls{PARTICLEMALA} and \gls{PARTICLEAMALAPLUS}) do not make use of any gradient information and reduce to the \gls{PARTICLERWM} algorithm. In particular, if $T = N = 1$, they thus reduce to the \gls{RWM} algorithm.

   \item \glsunset{AMALA} \textbf{Generalisation of \gls{AMALA}.} \glsreset{AMALA} For $\gradientIndicator = 1$, the \gls{PARTICLEAMALA} (and similarly the \gls{PARTICLEAMALAPLUS}) algorithm generalise the \emph{\gls{AMALA}} from \citet{titsias2018auxiliary} in the sense that the former reduces to the latter if $T = N = 1$. This can be seen as follows, where we again suppress the `time' subscript $t = 1$ everywhere. Given that the current state of the Markov chain is $\sta = \sta^0$ (we can assume that $k = 0$ without loss of generality), Step~\ref{alg:particleamala:mutation} of Algorithm~\ref{alg:particleamala} first refreshes the auxiliary variable by sampling $\aux \sim \dN(\sta^0 + \tfrac{\delta}{2} \nabla \log \pi(\sta^0), \tfrac{\delta}{2}\iMat)$ and then proposes $\sta^1 \sim \dN(\aux, \tfrac{\delta}{2}\iMat)$. The remaining steps return $\staAlt \coloneqq \sta^1$ as the new state with acceptance probability $1 \wedge \alpha_{\text{\upshape\gls{AMALA}}}(\sta^0, \sta^1; \aux)$, where
    \begin{align}
      \alpha_{\text{\upshape\gls{AMALA}}}(\sta^0, \sta^1; \aux)  
      & \coloneqq \frac{1 - W^0}{1 - W^1}
      = \frac{\pi(\sta^1) \dN(\aux; \sta^1 + \tfrac{\delta}{2} \nabla \log \pi(\sta^1), \tfrac{\delta}{2}\iMat) \dN(\sta^0; \aux, \tfrac{\delta}{2}\iMat)}{\pi(\sta^0) \dN(\aux; \sta^0 + \tfrac{\delta}{2} \nabla \log \pi(\sta^0), \tfrac{\delta}{2}\iMat) \dN(\sta^1; \aux, \tfrac{\delta}{2}\iMat)}. \label{eq:amala_acceptance_ratio}
    \end{align}
    Otherwise, the old state $\staAlt \coloneqq \sta^0 = \sta$ is returned as the new state.  This induces the same Markov chain on $\spaceX$ as the \gls{AMALA} from \citet{titsias2018auxiliary} (the only difference relates to a re-centring of the auxiliary variables $\aux$ previously discussed in \citet{corenflos2023auxiliary} but this does not change the law of the Markov chain on the marginal space which does not include the auxiliary variable). 

    \item \glsunset{MALA} \textbf{Generalisation of \gls{MALA}.} \glsreset{MALA} Still taking $\gradientIndicator = 1$, the \gls{PARTICLEMALA} generalises the \emph{\gls{MALA}} \citep{besag1994representations} in the sense that the former reduces to the latter if $T = N = 1$. This can be seen as follows, where use the same notational conventions as in the case of \gls{AMALA} above.
    Step~\ref{alg:particlemala:mutation} of Algorithm~\ref{alg:particlemala} then marginally proposes $\sta^1 \sim  \dN(\sta^0 + \tfrac{\delta}{2} \nabla \log \pi(\sta^0), \delta \iMat )$. The remaining steps return $\staAlt \coloneqq \sta^1$ as the new state with acceptance probability $1 \wedge \alpha_{\text{\upshape\gls{MALA}}}(\sta^0, \sta^1)$, where
    \begin{align}
      \alpha_{\text{\upshape\gls{MALA}}}(\sta^0, \sta^1) 
      & \coloneqq \frac{1 - W^0}{1 - W^1}
      = \frac{\pi(\sta^1) \dN(\sta^0; \sta^1 + \tfrac{\delta}{2} \nabla \log \pi(\sta^1), \delta \iMat )}{\pi(\sta^0)\dN(\sta^1; \sta^0 + \tfrac{\delta}{2} \nabla \log \pi(\sta^0), \delta \iMat )}. \label{eq:mala_acceptance_ratio}
    \end{align}
    Otherwise, the old state $\staAlt \coloneqq \sta^0 = \sta$ is returned as the new state.
 
  In particular, Remark~\ref{rem:particleamala_as_noisy_version_of_particlemala} shows that we can view the \gls{AMALA} as a `noisy' version of \gls{MALA} \citep[as already mentioned in][]{titsias2018auxiliary} because, dropping the time subscript again, by Lemma~\ref{lem:auxiliary_lemma_1}:
  \begin{align}
     \alpha_{\text{\upshape\gls{AMALA}}}(\sta^0, \sta^1; \aux)  & = \alpha_{\text{\upshape\gls{MALA}}}(\sta^0, \sta^1) \frac{\dN(\aux; \bar{\sta} + \tfrac{\delta}{4}\nabla \log \pi(\sta^1), \tfrac{\delta}{4}\iMat)}{\dN(\aux; \bar{\sta} + \tfrac{\delta}{4}\nabla \log \pi(\sta^0), \tfrac{\delta}{4}\iMat)},
    \end{align}
    where $\bar{\sta} = (\sta^0 + \sta^1) / 2$, and hence
    \begin{align}
     \E[\alpha_{\text{\upshape\gls{AMALA}}}(\sta^0, \sta^1; \aux)] = \alpha_{\text{\upshape\gls{MALA}}}(\sta^0, \sta^1),
    \end{align}
    where the expectation is w.r.t.\ the conditional distribution of $\aux$ under the joint distribution of the random variables sampled in Step~\ref{alg:particleamala:mutation} of the \gls{PARTICLEAMALA}, i.e.\ w.r.t.\ $\dN(\bar{\sta} + \tfrac{\delta}{4}\nabla \log \pi(\sta^0), \tfrac{\delta}{4}\iMat)$. In other words, this algorithm is the same as \gls{MALA} except that the acceptance ratio is `randomised' in the sense that it is multiplied by a non-negative random variable whose expectation is $1$. Other examples of such algorithms can be found in \citet{ceperley1999penalty, nicholls2012coupled}; see also \citet[][Section~3.3.3]{finke2015extended} for a discussion as well as \citet[][page~2669]{andrieu2016establishing} for a simple argument showing that the asymptotic variance of \gls{AMALA} cannot be smaller than that of \gls{MALA}.

\end{enumerate}

%% file: 04_particlemgrad.tex
\subsection{Particle-aGRAD}
\label{subsec:particleagrad} 

The gradient-informed algorithms (\gls{PARTICLEMALA}, etc) developed in Section~\ref{sec:particlemala} can be expected to improve upon the \gls{PARTICLERWM} algorithm in the same way that \gls{AMALA}/\allowbreak\gls{MALA} improve upon the \gls{RWM} algorithm.
However, they may underperform compared to the \gls{CSMC} algorithm when the prior dynamics of the latent states are highly informative in the same way that \gls{MALA} can underperform relative to the \gls{IMH} algorithm (with prior as proposal) if the prior is highly informative. Additionally, note that the algorithms from Section~\ref{sec:particlemala} employ proposals that are \emph{separable} in the sense that, given the reference path, the marginal proposal distribution of $\sta_t^n$ does not depend on the ancestor particle $\smash{\sta_{t-1}^{a_{t-1}^n}}$ (that is, separability implies that the weight-calculation and resampling steps could be postponed until \emph{after} all particles have been proposed); such separable proposals can be expected to perform poorly if the latent states are highly correlated across time.

In this section, we further incorporate (conditionally) Gaussian prior dynamics into the particle proposals and thus interpolate between the \gls{CSMC} algorithm and the gradient-informed algorithms of Section~\ref{sec:particlemala}. 
Our construction generalises the \gls{AGRAD} and \gls{MGRAD} algorithms of \citet{titsias2018auxiliary}. %
In particular, the algorithms introduced in this section do not imply separable proposals, i.e., the proposal kernel for particle $\sta_t^n$ will generally depend on its ancestor particle $\smash{\sta_{t-1}^{a_{t-1}^n}}$.

Specifically, in this section, we consider the special case of the generic Feynman--Kac model from \eqref{eq:gamma_def} in which we can find a decomposition $\qsemigroup_t(\sta_{t-1:t}) = \Mutation_t(\sta_t|\sta_{t-1}) \Potential_t(\sta_{t-1:t})$, such that
\begin{align}
  \Mutation_t(\sta_t|\sta_{t-1}) = \dN(\sta_t; \priorMean_t(\sta_{t-1}), \priorVar_t(\sta_{t-1})), \label{eq:gaussian_dyn_gamma}
\end{align} 
is a Gaussian transition density whose mean $\priorMean_t(\sta_{t-1})$ and non-singular covariance matrix $\priorVar_t(\sta_{t-1})$ may depend on the previous state $\sta_{t-1}$, for $t > 1$; and that $\Mutation_1(\sta_1) = \dN(\sta_1; \priorMean_1, \priorVar_1)$. %
\begin{example}[state-space model, continued]\label{example:ssm:gaussian}
  The methods proposed in this section immediately apply with $\Mutation_t(\sta_t|\sta_{t-1}) \coloneqq f_t(\sta_t|\sta_{t-1})$ if the state-space model has conditionally Gaussian dynamics, i.e.\ if $f_t(\sta_t|\sta_{t-1}) = \dN(\sta_t; \priorMean_t(\sta_{t-1}), \priorVar_t(\sta_{t-1}))$, by taking $\Potential_t(\sta_{t-1:t}) = g_t(\obs_t|\sta_t)$. However, they may often still apply to state-space models with non-Gaussian dynamics via a change of measure, i.e., by taking $\Mutation_t(\sta_t|\sta_{t-1}) \coloneqq \dN(\sta_t; \priorMean_t(\sta_{t-1}), \priorVar_t(\sta_{t-1}))$ and $\Potential_t(\sta_{t-1:t}) = f_t(\sta_t|\sta_{t-1}) g_t(\obs_t|\sta_t) / \dN(\sta_t; \priorMean_t(\sta_{t-1}), \priorVar_t(\sta_{t-1}))$, or through a suitable transformation.
\end{example}

The first method proposed in this section is termed \emph{\gls{PARTICLEAGRAD}}. Conditional on the auxiliary variables $\aux_{1:T}$, it can be viewed as a \gls{CSMC} algorithm whose proposal kernels are those of the fully-adapted auxiliary particle filter for the state-space model defined by the Gaussian transitions $p(\sta_t|\sta_{t-1}) = \dN(\sta_t; \priorMean_t(\sta_{t-1}), \priorVar_t(\sta_{t-1}))$ from \eqref{eq:gaussian_dyn_gamma} and `pseudo observations' $\aux_t$ with $p(\aux_t|\sta_t) = \dN(\aux_t; \sta_t, \frac{\delta_t}{2}\iMat)$. We now write
\begin{align}
    \MutationAlt_t(\sta_t| \sta_{t-1}; \aux_t)
    & \coloneqq p(\sta_t|\sta_{t-1}, \aux_t)\\
    & \propto
    \dN(\sta_t; \priorMean_t(\sta_{t-1}), \priorVar_t(\sta_{t-1})) \dN(\aux_t; \sta_t, \tfrac{\delta_t}{2} \iMat)\\
    & \propto \dN(\sta_t; \priorMeanAlt_t(\sta_{t-1}, \aux_t), \priorVarAlt_t(\sta_{t-1})),\label{eq:particleagrad:mutation}\
   \shortintertext{with}
\priorMeanAlt_t(\sta, \aux)
   & \coloneqq \priorMean_t(\sta) +  \kalmanGain_t(\sta) [\aux -  \priorMean_t(\sta)], \label{eq:particleagrad_proposal_mean}\\
   \priorVarAlt_t(\sta) 
   &\coloneqq (\iMat - \kalmanGain_t(\sta)) \priorVar_t(\sta) = \tfrac{\delta_t}{2} \kalmanGain_t(\sta),\label{eq:particleagrad_proposal_variance}\\
    \kalmanGain_t(\sta) &\coloneqq (\priorVar_t(\sta) + \tfrac{\delta_t}{2} \iMat)^{-1} \priorVar_t(\sta),\\% = (\tfrac{\delta_t}{2} \priorVar_t(\sta)^{-1} + \iMat)^{-1},\\
\shortintertext{as well as}
    \PotentialAlt_t(\sta_{t-1:t}; \aux_t)
   &\coloneqq  \qsemigroup_t(\sta_{t-1:t}) \frac{\dN(\aux_t; \sta_t + \gradientIndicator \tfrac{\delta_t}{2}  \nabla_{\sta_t} \log \Potential_t(\sta_{t-1:t}), \frac{\delta_t}{2}\iMat)}{\MutationAlt_t(\sta_t| \sta_{t-1}; \aux_t)}, \label{eq:particleagrad:potential}
\end{align}
and $\qsemigroupAlt_t(\sta_{t-1:t}; \aux_t) \coloneqq \MutationAlt_t(\sta_t| \sta_{t-1}; \aux_t)\PotentialAlt_t(\sta_{t-1:t}; \aux_t)$. A single iteration of the \gls{PARTICLEAGRAD} algorithm is as follows.

\begin{framedAlgorithm}[\gls{PARTICLEAGRAD}] \label{alg:particleagrad}
Implement Algorithm~\ref{alg:csmc} but replace the particle proposal (Step~\ref{alg:csmc:mutation}) and the weight calculation (Step~\ref{alg:csmc:weight_calculation}) by
\begin{enumerate}
    \myitem{1c}\label{alg:particleagrad:mutation} sample $\smash{\aux_t \sim \dN(\sta_t + \gradientIndicator \tfrac{\delta_t}{2} \nabla_{\sta_t} \log \Potential_t(\sta_{t-1:t}), \tfrac{\delta_t}{2} \iMat)}$, and $\smash{\sta_t^n \sim \MutationAlt_t(\ccdot | \sta_{t-1}^{a_{t-1}^n}; \aux_t)}$, for $n \in [N]_0 \setminus \{k_t\}$, 
    \myitem{1d} \label{alg:particleagrad:weight_calculation} for $n \in [N]_0$, set $\smash{w_t^n \propto \PotentialAlt_t(\sta_{t-1:t}^{(n)}; \aux_t),}$
\end{enumerate}
and also replace $\qsemigroup_{t+1}(\ccdot)$ in the backward kernel in Step~\ref{alg:csmc:backward_sampling} by $\smash{\qsemigroupAlt_{t+1}(\ccdot; \aux_t)}$.
\end{framedAlgorithm}

\begin{proposition}[validity of \gls{PARTICLEAGRAD}] \label{prop:particleagrad:validity}
  Sampling $\staAlt_{1:T}$ given $\sta_{1:T}$ via Algorithm~\ref{alg:particleagrad} induces a Markov kernel $P_\particleagrad(\staAlt_{1:T}|\sta_{1:T})$ which leaves $\pi_T$ invariant.
\end{proposition}

\subsection{Particle-mGRAD}
\label{subsec:particlemgrad}

In this section, we analytically integrate out the auxiliary variables $\aux_t$ which appeared in the weights of the \gls{PARTICLEAGRAD} algorithm. Here we consider the case when the covariance matrices appearing in the conditionally Gaussian mutation kernel \eqref{eq:gaussian_dyn_gamma} do not depend on the previous state, i.e.,
\begin{equation}
  \priorVar_t(\sta_{t-1}) = \priorVar_t, \label{eq:constant_covariance_matrix}
\end{equation}
which then also implies that $\kalmanGain_t(\sta_{t-1}) = \kalmanGain_t$. %
A single iteration of the resulting methodology -- which we term the \emph{\gls{PARTICLEMGRAD}} algorithm -- is as follows, where we write
\begin{align*}\label{eq:particlemgrad:weight_factor}
   \log H_{t, \genericGrad}(\sta, \genericVec, \bar{\sta}, \bar{\genericVec}) 
   & = \tfrac{1}{2} (\sta - \genericVec)^\T ((\tfrac{\delta_t}{2}\kalmanGain_t)^{-1} + \gMat_t) (\sta - \genericVec)\\*
   & \qquad - [\tfrac{1}{2} N (\sta + \genericGrad)^\T \kalmanGain_t + (\sta - \genericVec)^\T] \gMat_t (\sta + \genericGrad)\\*
   & \qquad + (N+1) (\bar{\sta} - \bar{\genericVec})^\T \gMat_t (\genericVec + \genericGrad),%
\end{align*}
for $\gMat_t \coloneqq \tfrac{2}{\delta_t} (\iMat + N \kalmanGain_t)^{-1}$. 

\begin{framedAlgorithm}[\gls{PARTICLEMGRAD}] \label{alg:particlemgrad}
Implement Algorithm~\ref{alg:csmc} but replace the particle proposal (Step~\ref{alg:csmc:mutation}) and the weight calculation (Step~\ref{alg:csmc:weight_calculation}) by
\begin{enumerate}

    \myitem{1c}\label{alg:particlemgrad:mutation} sample $\smash{\aux_t \sim \dN(\sta_t + \gradientIndicator \tfrac{\delta_t}{2} \nabla_{\sta_t} \log \Potential_t(\sta_{t-1:t}), \tfrac{\delta_t}{2} \iMat)}$ and $\smash{\sta_t^n \sim \MutationAlt_t(\ccdot | \sta_{t-1}^{a_{t-1}^n}; \aux_t)}$, for $n \in [N]_0 \setminus \{k_t\}$, 
    
   \myitem{1d} \label{alg:particlemgrad:weight_calculation} set $\smash{\bar{\sta}_t \coloneqq \tfrac{1}{N+1} \sum_{n=0}^{N} \sta_t^n}$, 
    $\genericVec_t^n \coloneqq (\iMat - \kalmanGain_t) \priorMean_t(\sta_{t-1}^{a_{t-1}^n})$, $\smash{\bar{\genericVec}_t \coloneqq \tfrac{1}{N+1}\sum_{n = 0}^N \genericVec_t^n}$, and, for $n \in [N]_0$,
    \begin{align}
        w_t^n \propto \qsemigroup_t(\sta_{t-1:t}^{(n)}) H_{t,\gradientIndicator\frac{\delta_t}{2} 
         \nabla_{\sta_t^n} \log \Potential_t(\sta_{t-1:t}^{(n)})
        }(\sta_t^n, \genericVec_t^n, \bar{\sta}_t, \bar{\genericVec}_t).
    \end{align}
\end{enumerate}
\end{framedAlgorithm}

\begin{remark}[\gls{PARTICLEAGRAD} `exactly approximates' \gls{PARTICLEMGRAD}]
\label{rem:particleagrad_as_noisy_version_of_particlemgrad}
In analogue to the relationship between \gls{PARTICLEAMALA} and \gls{PARTICLEMALA} discussed in Remark~\ref{rem:particleamala_as_noisy_version_of_particlemala}, \gls{PARTICLEAGRAD} is a noisy version of \gls{PARTICLEMGRAD}. That is, letting $w_t^n(\aux_t)$ and $w_t^n$ be the unnormalised weights under \gls{PARTICLEAGRAD} and \gls{PARTICLEMGRAD}, respectively, we have 
  \begin{align}
    \E\biggl[ \frac{w_t^n(\aux_t)}{w_t^{k_t}(\aux_t)}\biggr] = \frac{w_t^n}{w_t^{k_{\mathrlap{t}}}},
  \end{align}
  where the expectation is taken with respect to the conditional distribution of $\aux_t$ under the joint distribution of all random variables generated by Algorithm~\ref{alg:particleagrad} up to (and including) time~$t$.
\end{remark}

\begin{proposition}[validity of \gls{PARTICLEMGRAD}] \label{prop:particlemgrad:validity}
  Sampling $\staAlt_{1:T}$ given $\sta_{1:T}$ via Algorithm~\ref{alg:particlemgrad} induces a Markov kernel $P_\particleagrad(\staAlt_{1:T}|\sta_{1:T})$ which leaves $\pi_T$ invariant.
\end{proposition}

\subsection{Particle-aGRAD+}
\label{subsec:particleagradplus}

While the algorithm of Section~\ref{alg:particleagrad} incorporates information from the smoothing distribution by merit of not modifying the latent dynamics, it may happen that the potential $\Potential_t(\sta_{t-1:t})$ strongly depends on $\sta_{t-1}$. In this case, considering the `myopic' gradient information $\nabla_{\sta_t} \log \Potential_t(\sta_{t-1:t})$ may not suffice to improve the mixing of the algorithm and information from $\sta_{t-1}$ may then be beneficial. Similarly to Section~\ref{subsec:particleamalaplus}, in this section, we extend the \gls{PARTICLEAGRAD} algorithm to incorporate gradients w.r.t. the `smoothing  potential' $\Potential_{1:T}(\sta_{1:T}) = \prod_{t=1}^T \Potential_t(\sta_{t-1:t})$ rather than w.r.t.\ the `filtering potential' $\prod_{s=1}^t \Potential_s(\sta_{s-1:s})$.

For $\MutationAlt_t(\sta_t| \sta_{t-1}; \aux_t)$ and $\PotentialAlt_t(\sta_{t-1:t}; \aux_t)$ still defined as in the \gls{PARTICLEAGRAD} algorithm (i.e., as in \eqref{eq:particleagrad:mutation} and \eqref{eq:particleagrad:potential}), 
we now write
\begin{align*}
   \PotentialAlt_t(\sta_{t-2:t}; \aux_{t-1:t})
   & \coloneqq \PotentialAlt_t(\sta_{t-1:t}; \aux_t)
  \frac{\dN(\aux_{t-1}; \sta_{t-1} + \gradientIndicator\tfrac{\delta_{t-1}}{2} 
   \nabla_{\sta_{t-1}} \log \Potential_{1:T}(\sta_{1:T}) %
   , \tfrac{\delta_{t-1}}{2} \iMat)}{\dN(\aux_{t-1}; \sta_{t-1} + \gradientIndicator\tfrac{\delta_{t-1}}{2} 
   \nabla_{\sta_{t-1}} \log \Potential_{t-1}(\sta_{t-2:t-1})
   , \tfrac{\delta_{t-1}}{2} \iMat)},\!\!\!\!\!\!\!\!\!\!
\end{align*}
as well as $\qsemigroupAlt_t(\sta_{t-2:t}; \aux_{t-1:t}) \coloneqq \MutationAlt_t(\sta_t| \sta_{t-1}; \aux_t) \PotentialAlt_t(\sta_{t-2:t}; \aux_{t-1:t})$, where we note that 
\begin{align}
\nabla_{\sta_t} \log \Potential_{1:T}(\sta_{1:T}) = \nabla_{\sta_t} \log [\Potential_t(\sta_{t-1:t}) + \log \Potential_{t+1}(\sta_{t:t+1})].
\end{align}
A single iteration of the resulting `smoothing-gradient' methodology -- which we term the \emph{\gls{PARTICLEAGRADPLUS}} algorithm -- is as follows.

\begin{framedAlgorithm}[\gls{PARTICLEAGRADPLUS}] \label{alg:particleagradplus}
Implement Algorithm~\ref{alg:csmc} but replace the particle proposal (Step~\ref{alg:csmc:mutation}), the weight calculation (Step~\ref{alg:csmc:weight_calculation}), and backward sampling (Step~\ref{alg:csmc:backward_sampling}) by
\begin{enumerate}
    \myitem{1c}\label{alg:particleagradplus:mutation} sample $\smash{\aux_t \sim \dN(\sta_t + \gradientIndicator \tfrac{\delta_t}{2}
    \nabla_{\sta_t} \log \Potential_{1:T}(\sta_{1:T}) %
    , \tfrac{\delta_t}{2} \iMat)}$, and $\smash{\sta_t^n \sim \MutationAlt_t(\sta_t| \sta_{t-1}; \aux_t)}$, for $n \in [N]_0 \setminus \{k_t\}$, 
    \myitem{1d} \label{alg:particleagradplus:weight_calculation} for $n \in [N]_0$, set $\smash{w_t^n \propto \PotentialAlt_t(\sta_{t-2:t}^{(n)}; \aux_{t-1:t}),}$
    \myitem{3} \label{alg:particleagradplus:backward_sampling} for $t = T-1, \dotsc, 1$, sample $l_t = i \in [N]_0$ \gls{WP}
    \begin{align}
  \dfrac{W_t^{i} \qsemigroupAlt_{t+1}((\sta_{t-1:t}^{(i)}, \sta_{t+1}^{l_{t+1}}); \aux_{t:t+1}) \qsemigroupAlt_{t+2}((\sta_t^i, \sta_{t+1}^{l_{t+1}}, \sta_{t+2}^{l_{t+2}}); \aux_{t+1:t+2})}{\sum_{n=0}^N W_t^n \qsemigroupAlt_{t+1}((\sta_{t-1:t}^{(n)}, \sta_{t+1}^{l_{t+1}}); \aux_{t:t+1}) \qsemigroupAlt_{t+2}((\sta_t^n, \sta_{t+1}^{l_{t+1}}, \sta_{t+2}^{l_{t+2}}); \aux_{t+1:t+2})}.
  \end{align}
\end{enumerate}
\end{framedAlgorithm}

Note that if $\Potential_t(\sta_{t-1:t}) = \Potential_t(\sta_t)$ does not depend on $\sta_{t-1}$, then the \gls{PARTICLEAGRADPLUS} algorithm coincides with the \gls{PARTICLEAGRAD} algorithm. However, when $\Potential_t(\sta_{t-1:t})$ varies highly in $\sta_{t-1}$, their behaviours may differ substantially.

\begin{proposition}[validity of \gls{PARTICLEAGRADPLUS}] \label{prop:particleagradplus:validity}
  Sampling $\staAlt_{1:T}$ given $\sta_{1:T}$ via Algorithm~\ref{alg:particleagradplus} induces a Markov kernel $P_\particleagradplus(\staAlt_{1:T}|\sta_{1:T})$ which leaves $\pi_T$ invariant.
\end{proposition}

\subsection{Twisted Particle-aGRAD(+)}
\label{subsec:twisted_particleagrad}

Recall that, conditionally on the auxiliary variables $\aux_{1:T}$, the \gls{PARTICLEAGRAD} algorithm could be viewed as a \gls{CSMC} algorithm whose proposal kernels $\MutationAlt_t(\sta_t| \sta_{t-1}; \aux_t) = p(\sta_t|\sta_{t-1}, \aux_t)$ are those of the fully-adapted auxiliary particle filter for the state-space model which is defined by the Gaussian transitions $p(\sta_t|\sta_{t-1}) = \dN(\sta_t; \priorMean_t(\sta_{t-1}), \priorVar_t(\sta_{t-1}))$ from \eqref{eq:gaussian_dyn_gamma} and observation densities $p(\aux_t|\sta_t) = \dN(\aux_t; \sta_t, \frac{\delta_t}{2}\iMat)$. 

In this section (and in this section only), we make the more restrictive assumption that the transition kernel from \eqref{eq:gaussian_dyn_gamma} is not only Gaussian but also affine, i.e., 
\begin{align}
  \priorMean_t(\sta_{t-1}) = \priorFactor_t \sta_{t-1} + \priorIntercept_t, \quad \text{and} \quad \priorVar_t(\sta_{t-1}) = \priorVar_t, \label{eq:affine_assumption}
\end{align}
for some $\priorFactor_t \in \reals^{D \times D}$, $\priorIntercept_t \in \reals^D$, and some covariance matrix $\priorVar_t \in \reals^{D \times D}$. Under~\eqref{eq:affine_assumption}, we can then go one step further and implement the fully \emph{twisted} particle filter \citep{Whiteley2014twisted, guarniero2017iterated, heng2020controlledSMC} proposal which conditions on \emph{all} future pseudo observations $\aux_{t:T}$. That is, we now write 
\begin{align}
    \MutationAlt_t(\sta_t| \sta_{t-1}; \aux_{t:T}) 
    & \coloneqq p(\sta_t|\sta_{t-1}, \aux_{t:T})\\
    & \propto
    \int_{\spaceX^{T-t}} \biggl[ \prod_{s=t}^T \dN(\sta_s; \priorFactor_s \sta_{s-1} + \priorIntercept_s,\priorVar_s) \dN(\aux_s; \sta_s, \tfrac{\delta_s}{2} \iMat) \biggr] \intDiff \sta_{t+1:T}\\
    & \propto \dN(\sta_t; \priorFactorAlt_t \sta_{t-1} + \priorInterceptAlt_t, \priorVarAlt_t),\label{eq:twistedparticleagrad:mutation}\\
    \PotentialAlt_t(\sta_{t-1:t}; \aux_{t:T})
    &\coloneqq  \qsemigroup_t(\sta_{t-1:t}) \frac{\dN(\aux_t; \sta_t + \gradientIndicator \tfrac{\delta_t}{2}  \nabla_{\sta_t} \log \Potential_t(\sta_{t-1:t}), \frac{\delta_t}{2}\iMat)}{\MutationAlt_t(\sta_t| \sta_{t-1}; \aux_{t:T})},\label{eq:twistedparticleagrad:potential}%
\end{align}
as well as $\qsemigroupAlt_t(\sta_{t-1:t}; \aux_{t:T}) \coloneqq \MutationAlt_t(\sta_t| \sta_{t-1}; \aux_{t:T})\PotentialAlt_t(\sta_{t-1:t}; \aux_{t:T})$. Here, $\priorInterceptAlt_t \in \reals^D$ and $\priorFactorAlt_t, \priorVarAlt_t \in \reals^{D \times D}$ can be obtained via Kalman-filtering recursions as explained in Appendix~\ref{app:sec:twisted_reference}.

A single iteration of the resulting methodology -- which we term the \emph{twisted \gls{PARTICLEAGRAD}} algorithm -- is then exactly as the \gls{PARTICLEAGRAD} (Algorithm~\ref{alg:particleagrad}), except that $\MutationAlt_t(\ccdot|\ccdot;\aux_t)$, $\PotentialAlt_t(\ccdot; \aux_t)$ and $\qsemigroupAlt_t(\ccdot;\aux_t)$ from Section~\ref{subsec:particleagrad} are replaced by $\MutationAlt_t(\ccdot|\ccdot;\aux_{t:T})$, $\PotentialAlt_t(\ccdot; \aux_{t:T})$ and $\qsemigroupAlt_t(\ccdot;\aux_{t:T})$ from this section. When the potential functions $\Potential_t(\sta_{t-1:t})$ vary in $\sta_{t-1}$, then we can further construct a \emph{twisted \gls{PARTICLEAGRADPLUS}} algorithm by replacing $\MutationAlt_t(\sta_t|\sta_{t-1}, \aux_t)$ in Algorithm~\ref{alg:particleagradplus} and in the denominator of $\PotentialAlt_{t}(\sta_{t-2:t}; \aux_{t-1:t})$ by $\MutationAlt_t(\sta_t|\sta_{t-1}, \aux_{t:T})$.

\begin{proposition}[validity of the twisted \gls{PARTICLEAGRAD}/\gls{PARTICLEAGRADPLUS}] \label{prop:twisted_particleagrad:validity}
  Sampling $\staAlt_{1:T}$ given $\sta_{1:T}$ via the twisted \gls{PARTICLEAGRAD} or twisted \gls{PARTICLEAGRADPLUS} algorithm induces a Markov kernel which leaves $\pi_T$ invariant.
\end{proposition}

\subsection{Relationship with other methods}
\label{subsec:particlemgrad:relationship_with_other_methods}

The algorithms proposed above relate to existing methods as follows. %

\glsreset{AGRAD}
\glsreset{MGRAD}

\begin{enumerate}

   \item \glsunset{AGRAD} \textbf{Generalisation of \gls{AGRAD}.} \glsreset{AGRAD} For $\gradientIndicator = 1$, the \gls{PARTICLEAGRAD} algorithm (and similarly the \gls{PARTICLEAGRADPLUS} algorithm as well as the twisted versions of either) generalises the \emph{\gls{AGRAD}} algorithm from \citet[called `aGrad-z' therein]{titsias2018auxiliary} in the sense that the former reduces to the latter if $T = N = 1$. This can be seen as follows, where we again suppress the `time' subscript $t = 1$ everywhere so that $\pi(\sta) \propto \Mutation(\sta) \Potential(\sta)$, where $\Mutation(\sta) = \dN(\sta; \priorMean, \priorVar)$. Given that the current state of the Markov chain is $\sta = \sta^0$ (we can assume that $k = 0$ without loss of generality), Step~\ref{alg:particleagrad:mutation} of Algorithm~\ref{alg:particleagrad} first refreshes the auxiliary variable by sampling $\aux \sim \dN(\sta^0 + \tfrac{\delta}{2} \nabla \log G(\sta^0), \tfrac{\delta}{2}\iMat)$ and then proposes $\sta^1 \sim \dN((\iMat - \kalmanGain)\priorMean + \kalmanGain \aux, \tfrac{\delta}{2}\kalmanGain)$, for $\kalmanGain = (\priorVar + \tfrac{\delta}{2}\iMat)^{-1} \priorVar$. The remaining steps return $\staAlt \coloneqq \sta^1$ as the new state with acceptance probability $1 \wedge \alpha_{\text{\upshape\gls{AGRAD}}}(\sta^0, \sta^1; \aux)$, where
    \begin{align}
      \!\!\!\!\alpha_{\text{\upshape\gls{AGRAD}}}(\sta^0, \sta^1; \aux)  
      & \coloneqq \frac{1 - W^0}{1 - W^1}\\
      & = \frac{\pi(\sta^1) \dN(\aux; \sta^1 + \tfrac{\delta}{2} \nabla \log G(\sta^1), \tfrac{\delta}{2}\iMat) \dN(\sta^0; (\iMat - \kalmanGain)\priorMean + \kalmanGain \aux, \tfrac{\delta}{2}\kalmanGain)}{\pi(\sta^0) \dN(\aux; \sta^0 + \tfrac{\delta}{2} \nabla \log G(\sta^0), \tfrac{\delta}{2}\iMat) \dN(\sta^1; (\iMat - \kalmanGain)\priorMean + \kalmanGain \aux, \tfrac{\delta}{2}\kalmanGain)}.\!\!\!\! \label{eq:agrad_acceptance_ratio}
    \end{align}
    Otherwise, the old state $\staAlt \coloneqq \sta^0 = \sta$ is returned as the new state. 

    \item \glsunset{MGRAD} \textbf{Generalisation of \gls{MGRAD}.} \glsreset{MGRAD} Still taking $\gradientIndicator = 1$, the \gls{PARTICLEMGRAD} algorithm generalises the \emph{\gls{MGRAD}} algorithm from \citet{titsias2018auxiliary} in the sense that the former reduces to the latter if $T = N = 1$. This can be seen as follows, where we use the same notational conventions as in the case of \gls{AGRAD} above.
    Step~\ref{alg:particlemgrad:mutation} of Algorithm~\ref{alg:particlemgrad} then marginally proposes $\sta^1 \sim  \dN((\iMat - \kalmanGain) \priorMean + \kalmanGain[\sta^0 + \tfrac{\delta}{2} \nabla \log \Potential(\sta^0)], \mgradProposalVar)$, where $\mgradProposalVar \coloneqq \tfrac{\delta}{2}\kalmanGain^2 + \kalmanGain$. The remaining steps return $\staAlt \coloneqq \sta^1$ as the new state with acceptance probability $1 \wedge \alpha_{\text{\upshape\gls{MGRAD}}}(\sta^0, \sta^1)$, where
    \begin{align}
      \alpha_{\text{\upshape\gls{MGRAD}}}(\sta^0, \sta^1) 
      & \coloneqq \frac{1 - W^0}{1 - W^1}
      = \frac{\pi(\sta^1)  \dN((\iMat - \kalmanGain) \priorMean + \kalmanGain[\sta^1 + \tfrac{\delta}{2} \nabla \log \Potential(\sta^1)], \mgradProposalVar)}{\pi(\sta^0) \dN((\iMat - \kalmanGain) \priorMean + \kalmanGain[\sta^0 + \tfrac{\delta}{2} \nabla \log \Potential(\sta^0)], \mgradProposalVar)}. \label{eq:mgrad_acceptance_ratio}
    \end{align}
    Otherwise, the old state $\staAlt \coloneqq \sta^0 = \sta$ is returned as the new state. In particular, by Remark~\ref{rem:particleagrad_as_noisy_version_of_particlemgrad}, in analogue to Section~\ref{subsec:particlemala:relationship_with_other_methods}, we can again interpret \gls{AGRAD} as a version of \gls{MGRAD} with `randomised' acceptance ratio.

   \item \textbf{Generalisation of a `preconditioned' \gls{PARTICLERWM} algorithm.} If $\gradientIndicator = 0$, then the \gls{PARTICLEAGRAD} and \gls{PARTICLEAGRADPLUS} algorithms reduce to a method recently proposed in \citet[Section~4.3]{corenflos2023auxiliary}, which can be seen as a `preconditioned' version of the \gls{PARTICLERWM} algorithm.

\end{enumerate}

\subsection{Interpolation between CSMC and Particle-MALA/Particle-aMALA}
\label{subsec:interpolation}

The \gls{PARTICLEMALA} (and related methods) proposed in Section~\ref{sec:particlemala} may be outperformed by the \gls{CSMC} algorithm in the case when the prior dynamics are highly informative -- in the same way that \gls{MALA} may be outperformed by the \gls{IMH} algorithm (with prior as proposal) if the prior dominates the posterior. For instance, in the extreme case that all the potential functions are constant, the \gls{CSMC} algorithm proposes $N$ trajectories (in addition to the reference path) that are \gls{IID} samples from $\pi_T$ (assuming an adaptive or low-variance conditional resampling scheme is used) while the $N$ trajectories proposed by \gls{PARTICLEMALA} are still highly correlated with the reference path. 

Put differently, the user is faced with the `tuning problem' of having to decide between the \gls{CSMC} algorithm on the one hand and the \gls{PARTICLEMALA} (and related methods) on the other hand. In this section, we show that the \gls{PARTICLEMGRAD} algorithm resolves this tuning problem in the sense that it can be viewed as interpolating between \gls{CSMC} and \gls{PARTICLEMALA}. Specifically, Proposition~\ref{prop:convergence_for_highly_informative_prior} shows that \gls{PARTICLEMGRAD} reduces to the \gls{CSMC} algorithm if the prior dynamics are highly informative. Conversely, Proposition~\ref{prop:convergence_for_weakly_informative_prior} shows that \gls{PARTICLEMGRAD} reduces to the \gls{PARTICLEMALA} if the prior dynamics are uninformative. The same results hold for the auxiliary-variable versions: \gls{PARTICLEAMALA} and \gls{PARTICLEAGRAD}.

We make the following assumptions (assumed to hold for all $t \in [T]$):
\begin{enumerate}
\renewcommand{\labelenumi}{\textbf{\theenumi}}
\renewcommand{\theenumi}{\textbf{A\arabic{enumi}}}
 \item \label{as:factorisation_over_time} For any $\sta_t \in \spaceX$, $\priorMean_t(\sta_{t-1}) = \priorMean_t$, $\priorVar_t(\sta_{t-1}) = \priorVar_t$ and $\Potential_t(\sta_{t-1:t}) = \Potential_t(\sta_t)$ are constant in $\sta_{t-1}$, with $\Potential_t$ uniformly bounded on $\spaceX$ and $\priorVar_t$ invertible.
  \item \label{as:log_potential_gradient_control} There exist $C_0, C_1 \geq 0$ such that $\gradientIndicator \lVert \nabla \log \Potential_t(\sta_t)\rVert_2 \leq C_0 + C_1 \lVert \sta_t \rVert_2$.
 \item \label{as:potential_as_distribution_with_finite_variance} $\max_{d \in [D]} \int_\spaceX x_{t,d}^2 \Potential_t(\sta_t) \intDiff \sta_t < \infty$, where $x_{t,d}$ is the $d$th component of $\sta_t$.
\end{enumerate}
Whenever $T > 1$, Assumption~\ref{as:factorisation_over_time} is strong because it requires the Feynman--Kac model to factorise over time. However, we expect that it could be relaxed at the cost of greatly complicating the arguments. Indeed, note that the model used in Figure~\ref{fig:lgssm_variance} does not satisfy this assumption. Assumption~\ref{as:log_potential_gradient_control} is rather mild, e.g.\ it holds in a state-space model with Gaussian measurement errors. %

In the following, for each $t \in [T]$, we will consider a sequence of prior covariance matrices $(\priorVar_{t,k})_{k \geq 1}$. We will therefore add the subscript $k$ to any quantity which depends on $\priorVar_{t,k}$. We also let $\lambda(\aMat)$ denote the set of eigenvalues of some matrix $\aMat$. The following propositions are proved in Appendix~\ref{app:sec:proof_of_the_interpolation_property}.

\begin{proposition}\label{prop:convergence_for_highly_informative_prior}
  For some $D, T, N \geq 1$, assume \ref{as:factorisation_over_time}--\ref{as:log_potential_gradient_control}, and assume that there exists a sequence $(\lambda_k)_{k \geq 1}$ in $(0, \infty)$ with $\max\{\lambda(\priorVar_{1,k}), \dotsc, \lambda(\priorVar_{T,k})\} \leq \lambda_k \to 0$ as $k \to \infty$. Then for any $\varepsilon > 0$, there exists a sequence $(F_{T, k})_{k \geq 1}$ of subsets of $\spaceX^T$ with $\lim_{k \to \infty} \pi_{T,k}(F_{T, k}) = 1$ such that
\begin{enumerate}
 \item \label{prop:convergence_for_highly_informative_prior:1} $\smash{\sup_{\sta_{1:T} \in F_{T, k}} \lVert  P_{\particlemgrad, k}(\ccdot|\sta_{1:T}) - P_{\csmc, k}(\ccdot|\sta_{1:T}) \rVert_\tv \in \bo(\lambda_k^{(1 - \varepsilon) / 4})}$;
  \item \label{prop:convergence_for_highly_informative_prior:2} $\smash{\sup_{\sta_{1:T} \in F_{T, k}} \lVert  P_{\particleagrad, k}(\ccdot|\sta_{1:T}) - P_{\csmc, k}(\ccdot|\sta_{1:T}) \rVert_\tv \in \bo(\lambda_k^{(1 - \varepsilon) / 4})}$.
\end{enumerate}
\end{proposition}

\begin{proposition}\label{prop:convergence_for_weakly_informative_prior}
   For some $D, T, N \geq 1$, assume \ref{as:factorisation_over_time}--\ref{as:potential_as_distribution_with_finite_variance}, and assume that there exists a sequence $(\lambda_k)_{k \geq 1}$ in $(0, \infty)$ with $\min\{\lambda(\priorVar_{1,k}), \dotsc, \lambda(\priorVar_{T,k})\} \geq \lambda_k \to \infty$ as $k \to \infty$. Then for any $\varepsilon > 0$, there exists a sequence $(F_{T, k})_{k \geq 1}$ of subsets of $\spaceX^T$ with $\lim_{k \to \infty} \pi_{T,k}(F_{T, k}) = 1$ such that
  \begin{enumerate}
   \item \label{prop:convergence_for_weakly_informative_prior:1} $\smash{\sup_{\sta_{1:T} \in F_{T, k}} \lVert P_{\particlemgrad, k}(\ccdot|\sta_{1:T}) - P_{\particlemala, k}(\ccdot|x)  \rVert_\tv \in \bo(\lambda_k^{-(1- \varepsilon) / 4})}$;
   \item \label{prop:convergence_for_weakly_informative_prior:2} 
   $\smash{\sup_{\sta_{1:T} \in F_{T, k}} \lVert P_{\particleagrad, k}(\ccdot|\sta_{1:T}) - P_{\particleamala, k}(\ccdot|x)  \rVert_\tv \in \bo(\lambda_k^{-(1- \varepsilon) / 4})}$.
  \end{enumerate}
\end{proposition}
As per Sections~\ref{subsubsec:csmc:relationship_with_classical_mcmc_algorithms}, \ref{subsec:particlemala:relationship_with_other_methods} and \ref{subsec:particlemgrad:relationship_with_other_methods}, taking $T= N = 1$ in Propositions~\ref{prop:convergence_for_highly_informative_prior} and \ref{prop:convergence_for_weakly_informative_prior} immediately imply that the \gls{AGRAD}/\gls{MGRAD} algorithm can be viewed as automatically interpolating between the \gls{IMH} algorithm with prior as proposal (if the prior is highly informative) and \gls{AMALA}/\gls{MALA} (if the prior is highly diffuse). To our knowledge, this interpretation has not been pointed out in the literature. It provides new intuition for the noteworthy performance of \gls{AGRAD}/\gls{MGRAD} in \citet{titsias2018auxiliary}.

\subsection{Complexity} 
\label{subsec:complexity}

An iteration of \gls{PARTICLEAGRAD} or \gls{PARTICLEAGRADPLUS} requires computing $T(N+1)$ gain matrices $\smash{\kalmanGain_t(\sta_{t-1}^{a_{t-1}^n}) \in \reals^{D \times D}}$; and all of these, in general, have a cubic cost in the latent-state dimension $D$. While this may be reasonable for small enough systems and will be helpful for informative likelihoods, the computational quickly outweighs the statistical benefits of the method. However, when the dynamics have additive noise \eqref{eq:constant_covariance_matrix}, $\kalmanGain_t$ does not depend on $\sta_{t-1}$. In this case, only $T$ gain matrices are needed and these can be pre-computed, only paying the cubic cost in the dimension upfront rather than at each iteration. 

The same applies for the \gls{PARTICLEMGRAD} algorithm for which we always require \eqref{eq:constant_covariance_matrix} to hold (the auxiliary variables could still be integrated out if \eqref{eq:constant_covariance_matrix} is relaxed, but only at the cost of a cubic computational complexity in the number of particles). %

However, as for the \gls{PARTICLERWM} algorithm and \gls{PARTICLEMALA}-type methods, we need to calibrate the step-size parameters $\delta_t$ which changes the gain matrices (so that pre-computation is not possible during the calibration stage). Thankfully, because $\kalmanGain_{t}$ and $\priorVar_t$ have the same eigenvectors no matter what $\delta_t$ is, it is possible to use similar spectral methods as in \citet{titsias2018auxiliary} to reduce the complexity of changing $\delta_t$ to quadratic. %

At first sight, the complexity of the twisted \gls{PARTICLEAGRAD} seems quadratic in $T$ as the proposal kernel $\MutationAlt_t(\sta_t|\sta_{t-1}, \aux_{t:T}) = \dN(\sta_t; \priorFactorAlt_t \sta_{t-1} + \priorInterceptAlt_t, \priorVarAlt_t)$ requires processing $T - t$ auxiliary variables for each time $t$. However, in Appendix~\ref{app:sec:twisted_reference}, we show how $\priorFactorAlt_t$, $\priorInterceptAlt_t$ and $\priorVarAlt_t$ can all be pre-computed based on standard Kalman filter recursions~\citep{kalman1960new}, preserving the linear cost in $T$ and $N$.

%% file: 05_simulations.tex
\subsection{Multivariate stochastic volatility model}

In this section, we illustrate the efficiency of our methods on a multivariate stochastic volatility model often used as a benchmark for high-dimensional sequential Monte Carlo methodology \citep[see, e.g.,][]{guarniero2017iterated}. This model is a state-space model with a non-linear observation equation:
\begin{align}
  g_t(\obs_t | \sta_t) = \dN(\obs_t; \zeroMat, \diag(\exp \sta_t)),
\end{align}
where $\exp$ is applied element-wise and where $\zeroMat$ is a $D$-dimensional vector of zeros. The prior on the latent variables is defined through auto-regressive Gaussian dynamics, i.e.\ for $t > 1$:
\begin{align}\label{eq:msv-dyn}
    f_t(\sta_t | \sta_{t-1}) = \dN(\sta_t; \priorMean_t(\sta_{t-1}), \priorVar_t)
\end{align}
where $\priorMean_t(\sta_{t-1}) \coloneqq \varphi \sta_{t-1}$ and  $\priorVar_t \in \smash{\reals^{D \times D}}$ has diagonal entries $\tau$ and off-diagonal entries $\tau \rho$. The initial distribution $f_1(\sta_1) = \dN(\sta_1; \priorMean_1, \priorVar_1)$ is the stationary distribution under the dynamics~\eqref{eq:msv-dyn}, i.e., $\priorMean_1 \coloneqq \zeroMat$ and $\priorVar_1 \coloneqq \priorVar_t / (1 - \varphi^2)$.
Here, $\varphi \in (-1, 1)$ is some autocorrelation coefficient, $\rho \in (-1, 1)$ is some intra-asset correlation coefficient and $\tau > 0$.

Throughout our experiments, we take $\varphi = 0.9$, $\rho = 0.25$, and $\tau \in \{0.1, 0.5, 1, 2\}$. The eigenvalues of $\priorVar_t$ are then proportional to $\tau$, i.e., a small value of $\tau$ corresponds to highly informative prior dynamics (as in Proposition~\ref{prop:convergence_for_highly_informative_prior}) while a large value of $\tau$ corresponds to weakly informative prior dynamics (as in Proposition~\ref{prop:convergence_for_weakly_informative_prior}). %
To make our observations robust to the choice of data set, for each $\tau$, we simulated $M = 5$ independent sets of $T = 128$ observations from the multivariate stochastic volatility model with $D = 30$, i.e., each state $\sta_t$ takes values in $\spaceX = \reals^{30}$. To make results more easily comparable,  experiments for different values of $\tau$ use the same random number generator seed.%

\subsection{Simulation study setup}

In addition to the methods proposed in Sections~\ref{sec:particlemala} and \ref{sec:particlemgrad} -- potentially without the use of gradient information by taking $\gradientIndicator = 0$ -- we consider the following benchmark methods:
\begin{enumerate}
    \item \textbf{\gls{CSMC}.} The \gls{CSMC} algorithm with bootstrap proposals (Algorithm~\ref{alg:csmc}).%
    
    \item \textbf{\gls{PARTICLERWM}.} The \gls{PARTICLERWM} algorithm (Algorithm~\ref{alg:particlerwm}) from \citet{finke2023conditional} (the special case of \gls{PARTICLEAMALA}/\gls{PARTICLEMALA}/\gls{PARTICLEAMALAPLUS} if $\gradientIndicator = 0$).
    
    \item \textbf{\textbf{\gls{MALA} and \gls{AMALA}.}} The $N$-proposal \gls{MALA} and \gls{AMALA} which correspond to the \gls{PARTICLEAMALA} and \gls{PARTICLEMALA} proposed in this work with a single time step (applied to the path-space representation of the Feynman--Kac model, i.e.\ with a single $(D \times T)$-dimensional state). %
    
    \item \textbf{\gls{AGRAD}.} The $N$-proposal \gls{AGRAD} algorithm, which corresponds to the \gls{PARTICLEAGRAD} proposed in this work with a single time step (again on the path space). %
    We note that we implemented \gls{AGRAD} using the auxiliary Kalman perspective of~\citet{corenflos2023auxiliary}, making the method complexity scale linearly with $T$ rather than quadratically with $T$ as in the original version of~\citet{titsias2018auxiliary}. %
    We do not compare to \gls{MGRAD} because computing its particle weights (and hence acceptance ratio) has quadratic complexity in $T$. %
\end{enumerate}

All algorithms use $N + 1 = 32$ particles, and those employing resampling use the conditional 'killing' resampling method~\citep{Karppinen2023bridge}, more stable than multinomial resampling, especially with highly informative priors. In each of $M = 5$ independent experiments, algorithms start from the same trajectory generated by a bootstrap particle filter using $32$ particles. The samplers run for \num{10000} steps to calibrate step-size parameters $\delta_t$, detailed below (note that calibration stabilises much faster). For \gls{CSMC}, which requires no calibration, the initial \num{10000} steps are discarded as warm-up. After calibration, $J=4$ independent chains start at the final calibration sample, running for $K=\num{50000}$ iterations, with the first \num{5000} discarded as burn-in to decorrelate the chains. Reported statistics are based on these $J$ independent chains.

The step-size parameters $\delta_t$ are calibrated for a \SI{75}{\percent} acceptance rate, as explained in Appendix~\ref{app:sec:step_size_adaptation}. This slightly exceeds recommendations by, e.g., \citet[][]{roberts2001optimal,titsias2018auxiliary}. This is because we use multiple proposals and the optimal acceptance rate is expected to increase accordingly. Here, 'acceptance rate at time $t$' refers to the relative frequency of with which the state $\sta_t$ is updated. Figure~\ref{fig:stoch_vol_acceptance_rate} in Appendix~\ref{app:subsec:calibrated_step_sizes_and_acceptance_rates} shows stable acceptance rates around \SI{75}{\percent} for all methods except \gls{CSMC} across all time steps. Figure~\ref{fig:stoch_vol_delta} in Appendix~\ref{app:subsec:calibrated_step_sizes_and_acceptance_rates} displays calibrated $\delta_t$ values.

Experiments ran on a shared computational cluster with identical configurations (32~GB RAM, four processor cores, on shared machines with $\num{2} \times \num{64}$-core AMD EPYC 7713 CPUs, clock speed 2.0 GHz). Nonetheless, cluster idiosyncrasies may be present, potentially impacting slower methods like \gls{PARTICLEAMALAPLUS} and \gls{PARTICLEMGRAD}.

\subsection{Breakdown of CSMC, aMALA and MALA}

Our results indicate that \gls{CSMC}, \gls{AMALA}, and \gls{MALA} failed to explore the right regions of the space for all of our chosen levels of informativeness of the latent dynamics ($\tau \in \{0.1, 0.5, 1, 2\}$). Specifically, Figures~\ref{fig:stoch_vol_marginals} and  \ref{fig:stoch_vol_energy_trace} in Appendix~\ref{app:subsec:breakdown} show that both the estimated marginal posterior means and also the energy traces of  \gls{CSMC}, \gls{AMALA} and \gls{MALA} differ substantially from those of all the other algorithms. Here, `energy trace' refers to $\log \pi_T(\sta_{1:T}) + \mathrm{const}$ computed on the sampled trajectories throughout the sampling procedure. Since \gls{CSMC}, \gls{AMALA} and \gls{MALA} thus do not produce reliable approximations of the distribution of interest, we omit these methods from our discussions in the sequel.

In the remainder of this section, we compare the remaining algorithms in terms of the \emph{\gls{ESS}} computed using the method of~\citet{Vehtari2021rank} with $J=4$ independent chains. We also compare the algorithms in terms of \gls{ESS} per second (\gls{ESS}/s). The latter corresponds to the time it would take to obtain a `perfect' sample using the Markov chain. In the main manuscript, we only show results for the median \gls{ESS} and averaged over all $T$ time steps. Appendix~\ref{app:subsec:ess} shows detailed results for the minimum and maximum \gls{ESS} and \gls{ESS}/s (which are qualitatively similar to the median case) separately for each time step $t = 1,\dotsc, T$.

\subsection{Benefits of exploiting gradient information}\label{subsec:exp_mala}

\begin{figure}
  \centering
  \includegraphics[]{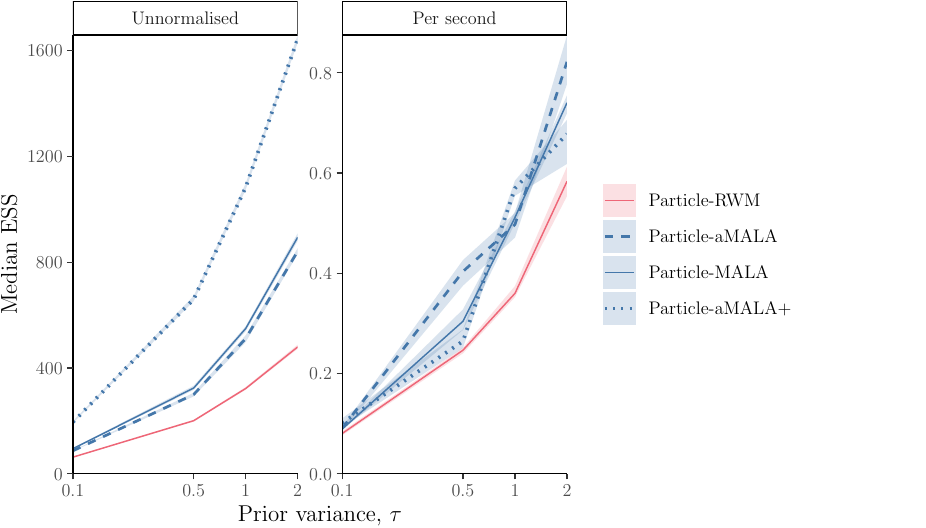}
  \caption{Performance of those proposed methods which do not require (conditionally or unconditionally) Gaussian prior dynamics compared with the existing \gls{PARTICLERWM} algorithm as a baseline. }
  \label{fig:stoch_vol_ess_med_1}
\end{figure}

Figure~\ref{fig:stoch_vol_ess_med_1} compares the median \gls{ESS} (`unnormalised') and median \gls{ESS}/s (`per second') of \gls{PARTICLEAMALA}, \gls{PARTICLEMALA} and \gls{PARTICLEAMALAPLUS}, i.e., for those methods which do not make any Gaussian assumption about the prior dynamics. Recall that these differ from the baseline: the \gls{PARTICLERWM} algorithm, only in the use of gradient information. Thus, the left panel in Figure~\ref{fig:stoch_vol_ess_med_1} illustrates the benefits (in terms of \gls{ESS}) of exploiting gradient information. Notably:
\begin{itemize}
    \item the improvement of \gls{PARTICLEMALA} over \gls{PARTICLEAMALA} is marginal at best. Possibly, the difference between both algorithms decreases with $N$ but this calls for further investigation;
    \item the `smoothing-gradient' variant \gls{PARTICLEAMALAPLUS} dominates all other alternatives for all values of $\tau$, with up to three times the performance of \gls{PARTICLERWM} and twice that of the `filter-gradient' variants \gls{PARTICLEAMALA} and \gls{PARTICLEMALA};
    \item the performance of all shown methods improves as $\tau$ increases: this is because the posterior distribution then decorrelates in time, and, therefore, the fact that they all use proposals which are separable (in the sense discussed in Section~\ref{sec:particlemgrad}) stops being penalising.
\end{itemize}
The right panel in Figure~\ref{fig:stoch_vol_ess_med_1} shows that the use of gradient information is still beneficial even when accounting for the cost of gradient calculation. However, the relative performance of the gradient-based methods is now less clear: whilst \gls{PARTICLEAMALAPLUS} has the highest sampling efficiency, it incurs additional overheads due to computing twice as many gradients as \gls{PARTICLEAMALA} and \gls{PARTICLEMALA} and due to dealing with non-Markovian potentials. 

\subsection{Benefits of exploiting Gaussian prior dynamics}\label{subsec:exp_tp}

In this section, we demonstrate that exploiting the latent (conditionally) Gaussian dynamics of the model (as done by \gls{PARTICLEAGRAD}, \gls{PARTICLEMGRAD} and twisted \gls{PARTICLEAGRAD}) can improve the sampling efficiency. 

\begin{figure}
  \centering
  \includegraphics[]{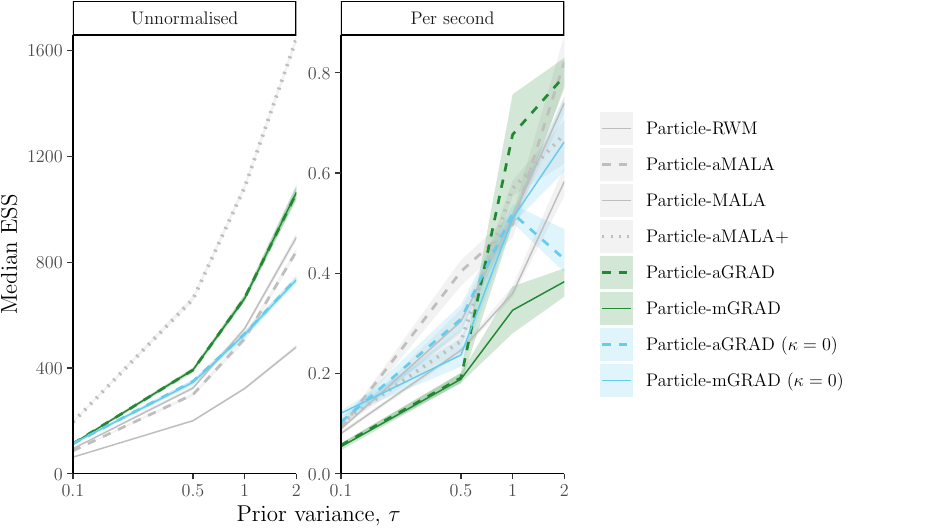}
  \caption{Performance of the proposed methods which require only \emph{conditionally} Gaussian prior dynamics \eqref{eq:gaussian_dyn_gamma}, i.e., $\Mutation_t(\sta_t|\sta_{t-1}) = \dN(\sta_t; \priorMean_t(\sta_{t-1}), \priorVar_t(\sta_{t-1}))$. The \gls{PARTICLEMGRAD} algorithm (with any $\gradientIndicator \in \{0,1\}$) also requires that $\priorVar_t(\sta_{t-1}) = \priorVar_t$ is constant to avoid superlinear computational complexity in $N$. Results that were already shown in the previous figure are greyed out.}
  \label{fig:stoch_vol_ess_med_2}
\end{figure}

First, in Figure~\ref{fig:stoch_vol_ess_med_2}, we illustrate the performance of those methods which require (at most) \emph{conditionally} Gaussian prior dynamics as in \eqref{eq:gaussian_dyn_gamma}, i.e., of \gls{PARTICLEAGRAD} and \gls{PARTICLEMGRAD} (note that the later also requires $\priorVar_t(\sta_{t-1}) = \priorVar_t$ \eqref{eq:constant_covariance_matrix} to retain linear computational complexity in $N$). In terms of \gls{ESS}, these methods improve upon the `filter-gradient' methods \gls{PARTICLEAMALA} and \gls{PARTICLEMALA} but they are still dominated by the `smoothing-gradient' method \gls{PARTICLEAMALAPLUS}. However, the picture is less clear when accounting for computation time.

\begin{figure}
  \centering
  \includegraphics[]{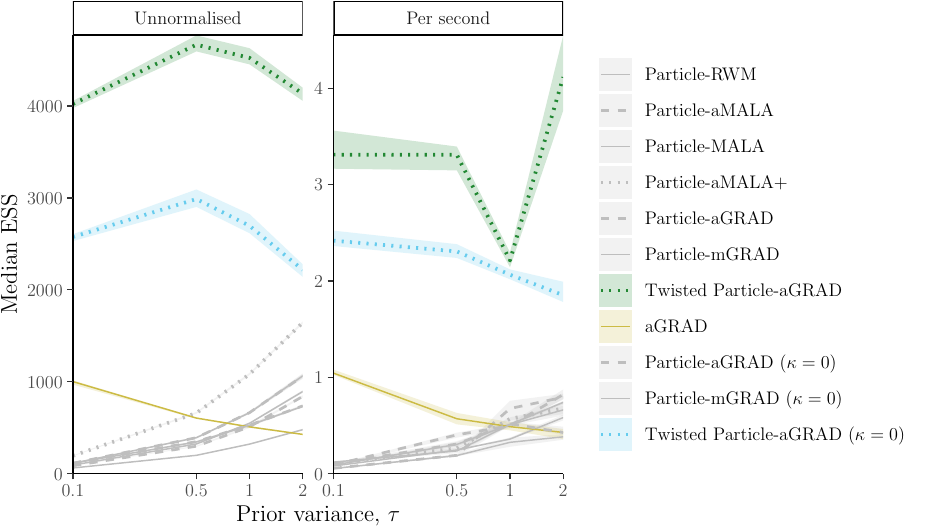}
   \caption{Performance of the proposed methods which require \emph{unconditionally} Gaussian prior dynamics \eqref{eq:affine_assumption}, i.e., $\Mutation_t(\sta_t|\sta_{t-1}) = \dN(\sta_t; \priorFactor_t \sta_{t-1} + \priorIntercept_t, \priorVar_t)$, compared with \gls{AGRAD} (which also requires \eqref{eq:affine_assumption}) as baseline. Results that were already shown in the previous two figures are greyed out. The abnormally large computation time of the twisted \gls{PARTICLEAGRAD} for $\kappa = \tau = 1$  was likely caused by some computational-cluster idiosyncrasies.}
  \label{fig:stoch_vol_ess_med_3}
\end{figure}

Second, in Figure~\ref{fig:stoch_vol_ess_med_3}, we illustrate the performance of the twisted \gls{PARTICLEAGRAD} which requires \emph{unconditionally} Gaussian prior dynamics as in \eqref{eq:affine_assumption}. As a baseline, we use the \gls{AGRAD} algorithm from \citet{titsias2018auxiliary} as it makes the same assumption. The twisted \gls{PARTICLEAGRAD} strongly outperforms this baseline and also all the other algorithms. Furthermore, the dominance of the twisted \gls{PARTICLEAGRAD} algorithm does not disappear when accounting for the computation time. This is because, in contrast to \gls{PARTICLEAMALAPLUS}, its modified model is still Markovian and because it only requires the computation of a single gradient per particle and time step. 

%% file: 06_conclusion.tex
\subsection{Summary}

We have proposed a methodology for Bayesian inference about the latent states in high-dimensional state-space models and beyond. Our methodology combines the \gls{CSMC} algorithm \citep{andrieu2010particle} with sophisticated `classical' \gls{MCMC} algorithms like \gls{MALA} \citep{besag1994representations}, \gls{AMALA} \citep{titsias2018auxiliary}, \gls{AGRAD}/\gls{MGRAD} \citep{titsias2011riemann, titsias2018auxiliary} or \gls{PCNL} \citep{cotter2013crank} to retain \emph{the best of both worlds:}
\begin{itemize}
    \item from the \gls{CSMC} algorithm, our methods retain the ability to exploit the model's `decorrelation-over-time' structure which permits favourable scaling with the number of time steps, $T$;
    \item from `classical' \gls{MCMC} algorithms, our methods retain the ability to use gradient-informed, local proposals which permits favourable scaling with the dimension of the states, $D$.
\end{itemize}
Most of our proposed algorithms (except the `marginal' ones) leverage an auxiliary-variable perspective recently proposed in \citet{corenflos2023auxiliary}. We name our algorithms  \gls{PARTICLEAMALA}, \gls{PARTICLEMALA},
\gls{PARTICLEAGRAD}, \gls{PARTICLEMGRAD} and \gls{PARTICLEPCNL}. This is motivated by the fact that if $T = N = 1$ (where $N \in \naturals$ is the number of particles), they reduce to the `classical' \gls{MCMC} algorithms: \gls{AMALA}, \gls{MALA}, \gls{PCNL}, \gls{AGRAD} and \gls{MGRAD}, respectively. Furthermore, if $T = 1$ but $N > 1$, our methods constitute novel multi-proposal versions of such `classical' \gls{MCMC} algorithms which may themselves be of interest with a view to exploiting parallelisation.

The generalisation of such `classical' \gls{MCMC} algorithms to $T > 1$ time steps is, however, not unique. And so we have presented additional variants named \gls{PARTICLEAMALAPLUS} and \gls{PARTICLEAGRADPLUS} and \emph{twisted} \gls{PARTICLEAGRAD}/\gls{PARTICLEAGRADPLUS}. These can be viewed as `lookahead' methods because their proposals employ `smoothing' rather than `filter' gradients or utilise information contained in future auxiliary variables. Notably, if $N = 1$ but $T > 1$, then the \gls{IMH}, \gls{RWM}, \gls{AMALA} and \gls{AGRAD} algorithm can still be recovered as a special case of slightly modified versions of the \gls{CSMC}, \gls{PARTICLERWM}, \gls{PARTICLEAMALAPLUS}, and twisted \gls{PARTICLEAGRADPLUS} algorithms (and also of the twisted \gls{PARTICLEAGRAD} algorithm if $\Potential_t(\sta_{t-1:t})$ is constant in $\sta_{t-1}$). Specifically, this modification would entail that the latter use no resampling (i.e., they instead set $\smash{a_t^n = n}$ for all $n \in [N]_0$ and all $t \in [T - 1]$), use \emph{ancestral tracing} instead of backward sampling (i.e., they instead set $\smash{l_t = a_t^{l_{t+1}}}$ for all $t \in [T - 1]$) and use $\delta_1 = \dotsc = \delta_T$.

We have further proved that the \gls{PARTICLEAGRAD}/\gls{PARTICLEMGRAD} algorithms have the desirable property that they naturally recover (a) the \gls{CSMC} algorithm if the prior dynamics are highly informative (i.e., if the target posterior distribution is dominated by the prior); (b) the \gls{PARTICLEAMALA}/\gls{PARTICLEMALA} if the prior dynamics are completely uninformative (i.e., if the target posterior distribution is dominated by the likelihood). This property independently helps explain the impressive performance of \gls{AGRAD} and \gls{MGRAD} reported in \citet{titsias2018auxiliary}.

Our methods have enabled Bayesian inference in a multivariate stochastic volatility model with $D = 30$ assets and $T = 128$ observations (\num{3840} unknowns in total) in which neither \gls{CSMC} nor \gls{AMALA}/\gls{MALA} gave reliable estimates. In particular, in this application, our twisted \gls{PARTICLEAGRAD} algorithm strongly outperformed the existing sophisticated \gls{AGRAD} algorithm -- even when accounting for computation time.

\subsection{Limitations}

The main limitations of our methods are the same as in all gradient-based `classical' \gls{MCMC} algorithms. First, they require continuously differentiable target densities (more precisely, the densities $\qsemigroup_t(\sta_{t-1:t})$ need to be computable and differentiable pointwise). This requirement is slightly softened for the methods of Section~\ref{sec:particlemgrad} where only the likelihood $\Potential_t(\sta_{t-1:t})$ is required to be differentiable, at the cost of needing (at least conditionally) Gaussian prior dynamics $\Mutation_t(\sta_t | \sta_{t-1})$. The favourable scaling with the dimension $D$ also typically requires target densities to be sufficiently smooth \citep[see, e.g.,][for counterexamples]{vogrinc2021counterexamples}. Second, while it improves mixing properties, locality in \gls{MCMC} is often detrimental when exploring multi-modal posteriors. This is inherited by our methods which, too, explore the space by local moves. %

\subsection{Extensions}

Our work opens up multiple avenues for further research.

\begin{itemize}
  \item The algorithms proposed in this work can be extended to more general graphical models, i.e., they can be combined with suitably `conditional' versions of the divide-\&-conquer sequential Monte Carlo algorithm from \citet{lindsten2017divide}. For instance, for a particular graphical model, such a `conditional' scheme was recently described in \citet[Section 3]{corenflos2022sequentialized}.%
  
  \item \gls{PARTICLEAMALA} can be incorporated straightforwardly into the methodology from \citet{corenflos2022sequentialized} to reduce the computation time per \gls{MCMC} update from $\bo(T)$ to $\bo(\log_2 T)$ (for some fixed dimension $D$) on parallel architectures. While less directly obvious (because of the non-Markovianity of the auxiliary target), the smoothing-gradient version \gls{PARTICLEAMALAPLUS} is likely parallelisable, too, by simply extending the framework to compute weight functions over three time steps rather than two. It is however less clear that \gls{PARTICLEMALA} is parallelisable, as the marginalisation has to be done across two time steps rather than one as presented in Section~\ref{subsec:particlemala}. %

  \item All our algorithms can be straightforwardly extended to use other resampling schemes than conditional multinomial resampling, e.g., conditional systematic resampling. 
  In fact, in our experiments, we used the conditional killing resampling which is stable under low-informative likelihoods \citep{Karppinen2023bridge}, a regime that may happen in our case when $\delta_t$ takes very small values at calibration time.

  \item In this work, we have left aside the question of choosing $\delta_t$ and have elected to take it to correspond to a \SI{75}{\percent} acceptance rate throughout. It is however clear that its optimal value (and the optimal value of the acceptance rate) depends on the number of proposals $N$ and on the dimension $D$. An optimal-scaling analysis \citep[see][and references therein]{roberts2001optimal} of the methods proposed in this work is therefore needed. An optimal-scaling analysis for a related algorithm without backward sampling and without gradient or prior-informed proposals can be found in \citet{malory2021bayesian}.
  
  \item In Section~\ref{sec:particlemgrad} (in which we propose \gls{PARTICLEAGRAD} and variations thereof), we have assumed that the covariance matrices $\priorVar_t(\sta_{t-1})$ (or $\priorVar_t)$ are non-singular. However, it is worth noting that proposal kernels used by the methods from Section~\ref{sec:particlemgrad} remain valid if the covariance matrices are singular, in the sense that they are \emph{still} absolutely continuous w.r.t.\ the true dynamics. However, the use of backward sampling is no longer possible for such degenerate dynamics. Instead, one must resort to ancestral tracing, i.e., taking $l_t \coloneqq a_t^{l_{t+1}}$, for $t = T-1,\dotsc, 1$. However, in this case, $N$ needs to grow with $T$ at a suitable rate which depends on the stability properties of the model, but at least linearly \citep{andrieu2018uniform,lindsten2015uniform}. An alternative is to fix $N$ but decrease the step sizes $\delta_t$ with $t$ (which would automatically occur when using adaptation based on acceptance rates as considered in work), as considered in \citet{malory2021bayesian} for a related method.

  \item Our proposed algorithms consider solely first-order gradient information.
  A natural extension would therefore be to incorporate second-order expansions or preconditioned and adaptive versions of the \gls{PARTICLEMALA} variants. Another obvious direction of study is to extend our methodology to other \gls{MCMC} kernels, such as
  the recently proposed Barker's robust proposal \citep{livingstone2022barker}, or non-reversible discrete-time kernels such as the discrete bouncy particle sampler~\citet{sherlock2022discrete}. Other natural extensions would consist of adapting the methodology to non-continuous spaces, e.g., using methods from \citet{zanella2020discrete, rhodes2022enhanced}, or constrained spaces.
\end{itemize}

%% file: appendix.tex
\section{Particle extensions of PCN(L)}
\label{app:sec:particlepcnl}
\glsreset{PCNL}

\subsection{Particle-aPCNL}
\label{app:subsec:particleapcnl} 

In this section, we extend the \emph{\gls{PCNL}} algorithm (and also the \emph{\gls{PCN}} algorithm recovered by setting $\gradientIndicator = 0$) \citep{cotter2013crank} to $T > 1$ time steps and $N > 1$. As a by-product, we derive an `auxiliary-variable' version of \gls{PCNL} which was mentioned, but not explicitly stated, in \citet{titsias2018auxiliary}. Throughout this section, we assume the prior dynamics are conditionally Gaussian, $\Mutation_t(\sta_t|\sta_{t-1}) = \dN(\sta_t; \priorMean_t(\sta_{t-1}), \priorVar_t(\sta_{t-1}))$ as in \eqref{eq:gaussian_dyn_gamma}. 

Throughout this section, we use the parametrisation of the \gls{PCNL} algorithm from \citet{titsias2018auxiliary}\footnote{In the parametrisation from \citet{cotter2013crank}, we would have $\delta_t \in [0, 2]$ $\beta_t = \frac{2 - \delta_t}{2 + \delta_t}$}, i.e., we set 
\begin{align}
  \beta_t \coloneqq \frac{2}{2 + \delta_t} \in (0, 1),
\end{align}
where $\delta_t > 0$ is again the step size at time $t$. Note that this implies that $\tfrac{1-\beta_t}{\beta_t} = \tfrac{\delta_t}{2}$. 

The first method proposed in this section is termed \emph{\gls{PARTICLEAPCNL}}. Conditional on the auxiliary variables $\aux_{1:T}$, it can be viewed as a \gls{CSMC} algorithm whose proposal kernels are those of the fully-adapted auxiliary particle filter for the state-space model defined by the Gaussian transitions $p(\sta_t|\sta_{t-1}) = \dN(\sta_t; \priorMean_t(\sta_{t-1}), \priorVar_t(\sta_{t-1}))$ from \eqref{eq:gaussian_dyn_gamma} and `pseudo observations' $\aux_t$ with $p(\aux_t|\sta_t) = \dN(\aux_t; \sta_t, \frac{\delta_t}{2}\priorVar_t(\sta_{t-1}))$. We now write
\begin{align}
    \MutationAlt_t(\sta_t|\sta_{t-1}, \aux_t)
    & \coloneqq p(\sta_t|\sta_{t-1}, \aux_t)\\
    & \propto \Mutation_t(\sta_t|\sta_{t-1}) \dN(\aux_t; \sta_t, \tfrac{\delta_t}{2} \priorVar_t(\sta_{t-1})) \label{eq:particleapcnl_faapf_proposal_interpretation}\\
    & \propto \dN(\sta_t; \priorMeanAlt_t(\sta_{t-1}, \aux_t), \priorVarAlt_t(\sta_{t-1})), \label{eq:particleapcnl:mutation}
  \shortintertext{with}
    \priorMeanAlt_t(\sta_{t-1}, \aux_t) 
    & \coloneqq \beta_t \aux_t + (1 - \beta_t) \priorMean_t(\sta_{t-1}), \label{eq:particleapcnl_proposal_mean}\\
    \priorVarAlt_t(\sta_{t-1}) 
    & \coloneqq (1-\beta_t) \priorVar_t(\sta_{t-1}), \label{eq:particleapcnl_proposal_variance}
 \end{align}
 as well as
 \begin{align}
    \PotentialAlt_t(\sta_{t-1:t}; \aux_t)
    & \coloneqq \qsemigroup_t(\sta_{t-1:t}) \frac{\dN(\aux_t; \sta_t + \gradientIndicator \tfrac{\delta_t}{2} \preconVar_t(\sta_{t-1}) \nabla_{\sta_t} \log \Potential_t(\sta_{t-1:t}), \tfrac{\delta_t}{2} \priorVar_t(\sta_{t-1}))}{\MutationAlt_t(\sta_t|\sta_{t-1}, \aux_t)}, \label{eq:particleapcnl:potential}
  \end{align}
and $\qsemigroupAlt_t(\sta_{t-1:t}; \aux_t) \coloneqq \MutationAlt_t(\sta_t| \sta_{t-1}; \aux_t)\PotentialAlt_t(\sta_{t-1:t}; \aux_t)$. Here, $\preconVar_t(\sta_{t-1}) \in \reals^{D \times D}$ is some preconditioning matrix whose choice is discussed in Section~\ref{app:subsec:choice_of_preconditioning_matrix} below.

A single iteration of the \gls{PARTICLEAPCNL} algorithm is then as follows.

\begin{framedAlgorithm}[\gls{PARTICLEAPCNL}] \label{alg:particleapcnl}
Implement Algorithm~\ref{alg:csmc} but replace the particle proposal (Step~\ref{alg:csmc:mutation}) and the weight calculation (Step~\ref{alg:csmc:weight_calculation}) by
\begin{enumerate}
    \myitem{1c}\label{alg:particleapcnl:mutation} sample $\smash{\aux_t \sim \dN(\sta_t + \gradientIndicator \tfrac{\delta_t}{2} \preconVar_t(\sta_{t-1}) \nabla_{\sta_t} \log \Potential_t(\sta_{t-1:t}), \tfrac{\delta_t}{2} \priorVar_t(\sta_{t-1}))}$, and $\smash{\sta_t^n \sim \MutationAlt_t(\ccdot | \sta_{t-1}^{a_{t-1}^n}; \aux_t)}$, for $n \in [N]_0 \setminus \{k_t\}$, 
    \myitem{1d} \label{alg:particleapcnl:weight_calculation} for $n \in [N]_0$, set $\smash{w_t^n \propto \PotentialAlt_t(\sta_{t-1:t}^{(n)}; \aux_t),}$
\end{enumerate}
and also replace $\qsemigroup_{t+1}(\ccdot)$ in the backward kernel in Step~\ref{alg:csmc:backward_sampling} by $\smash{\qsemigroupAlt_{t+1}(\ccdot; \aux_t)}$.
\end{framedAlgorithm}

\begin{proposition}[validity of \gls{PARTICLEAPCNL}] \label{prop:particleapcnl:validity}
  Sampling $\staAlt_{1:T}$ given $\sta_{1:T}$ via Algorithm~\ref{alg:particleapcnl} induces a Markov kernel $P_\particleapcnl(\staAlt_{1:T}|\sta_{1:T})$ which leaves $\pi_T$ invariant.
\end{proposition}

\subsection{Particle-PCNL}
\label{subsec:particlepcnl}

In this section, in analogy to the \gls{PARTICLEMALA} and \gls{PARTICLEMGRAD} algorithms from the main manuscript, we analytically integrate out the auxiliary variables $\aux_t$ which appeared in the weights of the \gls{PARTICLEAPCNL} algorithm. As in the case of the \gls{PARTICLEMGRAD} algorithm, we assume that the covariance matrices appearing in the conditionally Gaussian mutation kernel \eqref{eq:gaussian_dyn_gamma} do not depend on the previous state, i.e., $\priorVar_t(\sta_{t-1}) = \priorVar_t$ \eqref{eq:constant_covariance_matrix}.

A single iteration of the resulting methodology -- which we term the \emph{\gls{PARTICLEPCNL}} algorithm -- is as follows, where we write
\begin{align*}\label{eq:particlepcnl:weight_factor}
   \log H_{t, \genericGrad}(\sta, \genericVec, \bar{\sta}, \bar{\genericVec})
   & = \tfrac{1}{2} (\beta_t^{-1} + N + 1) (\sta - \genericVec)^\T \gMat_t  (\sta - \genericVec)\\*
  & \qquad - \tfrac{1}{2}N \beta_t (\sta + \genericGrad)^\T  \gMat_t (\sta + \genericGrad)\\*
  & \qquad +  (N+1) (\bar{\sta} - \bar{\genericVec})^\T \gMat_t (\genericVec + \genericGrad)\\*
  & \qquad - (\sta - \genericVec)^\T \gMat_t (\sta + \genericGrad), 
\end{align*}
for
\begin{align}
  \gMat_t \coloneqq \frac{\beta_t}{(1-\beta_t)(1+ N\beta_t)} \priorVar_t^{-1} = \frac{2 (\delta_t + 2)}{\delta_t(\delta_t + 2 + N)}\priorVar_t^{-1}.
\end{align}

\begin{framedAlgorithm}[\gls{PARTICLEPCNL}] \label{alg:particlepcnl}
Implement Algorithm~\ref{alg:csmc} but replace the particle proposal (Step~\ref{alg:csmc:mutation}) and the weight calculation (Step~\ref{alg:csmc:weight_calculation}) by
\begin{enumerate}

    \myitem{1c}\label{alg:particlepcnl:mutation} sample $\smash{\aux_t \sim \dN(\sta_t + \gradientIndicator \tfrac{\delta_t}{2} \preconVar_t(\sta_{t-1}) \nabla_{\sta_t} \log \Potential_t(\sta_{t-1:t}), \tfrac{\delta_t}{2} \priorVar_t)}$, and $\smash{\sta_t^n \sim \MutationAlt_t(\ccdot | \sta_{t-1}^{a_{t-1}^n}; \aux_t)}$, for $n \in [N]_0 \setminus \{k_t\}$, 
    
   \myitem{1d} \label{alg:particlepcnl:weight_calculation} set $\smash{\bar{\sta}_t \coloneqq \tfrac{1}{N+1} \sum_{n=0}^{N} \sta_t^n}$, 
    $\genericVec_t^n \coloneqq (1 - \beta_t) \priorMean_t(\sta_{t-1}^{a_{t-1}^n})$, $\smash{\bar{\genericVec}_t \coloneqq \tfrac{1}{N+1}\sum_{n = 0}^N \genericVec_t^n}$, and, for $n \in [N]_0$,
    \begin{align}
        w_t^n \propto \qsemigroup_t(\sta_{t-1:t}^{(n)}) H_{t,\gradientIndicator\frac{\delta_t}{2} 
         \preconVar_t(\sta_{t-1}^{a_{t-1}^n}) \nabla_{\sta_t^n} \log \Potential_t(\sta_{t-1:t}^{(n)})
        }(\sta_t^n, \genericVec_t^n, \bar{\sta}_t, \bar{\genericVec}_t).
    \end{align}
\end{enumerate}
\end{framedAlgorithm}

In the same way as outlined in Remarks~\ref{rem:particleamala_as_noisy_version_of_particlemala} and \ref{rem:particleagrad_as_noisy_version_of_particlemgrad}, the \gls{PARTICLEAPCNL} algorithm can be viewed as an `exact approximation' of the \gls{PARTICLEPCNL} algorithm. %

\begin{proposition}[validity of \gls{PARTICLEPCNL}] \label{prop:particlepcnl:validity}
  Sampling $\staAlt_{1:T}$ given $\sta_{1:T}$ via Algorithm~\ref{alg:particlepcnl} induces a Markov kernel $P_\particlepcnl(\staAlt_{1:T}|\sta_{1:T})$ which leaves $\pi_T$ invariant.
\end{proposition}

\subsection{Particle-aPCNL+}
\label{subsec:particleapcnlplus}

In this section, similar to the \gls{PARTICLEAMALAPLUS} and \gls{PARTICLEAGRADPLUS} algorithms from the main manuscript, we extend the \gls{PARTICLEAPCNL} algorithm to incorporate gradients w.r.t. the `smoothing' potential $\Potential_{1:T}(\sta_{1:T}) = \prod_{t=1}^T \Potential_t(\sta_{t-1:t})$ rather than w.r.t.\ the `filtering' potential $\prod_{s=1}^t \Potential_s(\sta_{s-1:s})$.

For $\MutationAlt_t(\sta_t| \sta_{t-1}; \aux_t)$ and $ \PotentialAlt_t(\sta_{t-1:t}, \aux_t)$ still defined as in the \gls{PARTICLEAPCNL} algorithm (i.e., as in \eqref{eq:particleapcnl:mutation} and \eqref{eq:particleapcnl:potential}), we now write
\begin{align*}
    \MoveEqLeft \PotentialAlt_t(\sta_{t-2:t}, \aux_{1:T})\\
    & \!\!\!\!\!\!\!\!\!\!\!\! \coloneqq \PotentialAlt_t(\sta_{t-1:t}, \aux_t) \frac{\dN(\aux_{t-1}; \sta_{t-1} + \gradientIndicator \tfrac{\delta_{t-1}}{2} \preconVar_{t-1}(\sta_{t-2}) \nabla_{\sta_{t-1}} \log \Potential_{1:T}(\sta_{1:T}), \tfrac{\delta_{t-1}}{2} \priorVar_{t-1}(\sta_{t-2}))}{\dN(\aux_{t-1}; \sta_{t-1} + \gradientIndicator \tfrac{\delta_{t-1}}{2} \preconVar_{t-1}(\sta_{t-2}) \nabla_{\sta_{t-1}} \log \Potential_{t-1}(\sta_{t-2:t-1}), \tfrac{\delta_{t-1}}{2} \priorVar_{t-1}(\sta_{t-2}))},
\end{align*}
as well as $\qsemigroupAlt_t(\sta_{t-2:t}; \aux_{t-1:t}) \coloneqq \MutationAlt_t(\sta_t| \sta_{t-1}; \aux_t) \PotentialAlt_t(\sta_{t-2:t}; \aux_{t-1:t})$, where we note that 
\begin{align}
\nabla_{\sta_t} \log \Potential_{1:T}(\sta_{1:T}) = \nabla_{\sta_t} [\log \Potential_t(\sta_{t-1:t}) + \log \Potential_{t+1}(\sta_{t:t+1})].
\end{align}
A single iteration of the resulting methodology -- which we term the \emph{\gls{PARTICLEAPCNLPLUS}} algorithm -- is as follows.

\begin{framedAlgorithm}[\gls{PARTICLEAPCNLPLUS}] \label{alg:particleapcnlplus}
Implement Algorithm~\ref{alg:csmc} but replace the particle proposal (Step~\ref{alg:csmc:mutation}), the weight calculation (Step~\ref{alg:csmc:weight_calculation}), and backward sampling (Step~\ref{alg:csmc:backward_sampling}) by
\begin{enumerate}
    \myitem{1c}\label{alg:particleapcnlplus:mutation} sample $\smash{\aux_t \sim \dN(\sta_t + \gradientIndicator \tfrac{\delta_t}{2} \preconVar_t(\sta_{t-1}) \nabla_{\sta_t} \log \Potential_T(\sta_{1:T}), \tfrac{\delta_t}{2} \priorVar_t(\sta_{t-1}))}$, and $\smash{\sta_t^n \sim \MutationAlt_t(\ccdot | \sta_{t-1}^{a_{t-1}^n}; \aux_t)}$, for $n \in [N]_0 \setminus \{k_t\}$, 
    \myitem{1d} \label{alg:particleapcnlplus:weight_calculation} for $n \in [N]_0$, set $\smash{w_t^n \propto \PotentialAlt_t(\sta_{t-2:t}^{(n)}; \aux_{t-1:t}),}$
    \myitem{3} \label{alg:particleapcnlplus:backward_sampling} for $t = T-1, \dotsc, 1$, sample $l_t = i \in [N]_0$ \gls{WP}
    \begin{align}
  \dfrac{W_t^{i} \qsemigroupAlt_{t+1}((\sta_{t-1:t}^{(i)}, \sta_{t+1}^{l_{t+1}}); \aux_{t:t+1}) \qsemigroupAlt_{t+2}((\sta_t^i, \sta_{t+1}^{l_{t+1}}, \sta_{t+2}^{l_{t+2}}); \aux_{t+1:t+2})}{\sum_{n=0}^N W_t^n \qsemigroupAlt_{t+1}((\sta_{t-1:t}^{(n)}, \sta_{t+1}^{l_{t+1}}); \aux_{t:t+1}) \qsemigroupAlt_{t+2}((\sta_t^n, \sta_{t+1}^{l_{t+1}}, \sta_{t+2}^{l_{t+2}}); \aux_{t+1:t+2})}.
  \end{align}
\end{enumerate}
\end{framedAlgorithm}

Note that if $\Potential_t(\sta_{t-1:t}) = \Potential_t(\sta_t)$ does not depend on $\sta_{t-1}$, then the \gls{PARTICLEAPCNLPLUS} algorithm coincides with the \gls{PARTICLEAPCNL} algorithm. However, when $\Potential_t(\sta_{t-1:t})$ varies highly in $\sta_{t-1}$, their behaviours may differ substantially. %

\begin{proposition}[validity of \gls{PARTICLEAPCNLPLUS}] \label{prop:particleapcnlplus:validity}
  Sampling $\staAlt_{1:T}$ given $\sta_{1:T}$ via Algorithm~\ref{alg:particleapcnlplus} induces a Markov kernel $P_\particleapcnlplus(\staAlt_{1:T}|\sta_{1:T})$ which leaves $\pi_T$ invariant.
\end{proposition}

\subsection{Twisted Particle-aPCNL(+)}
\label{app:subsec:twisted_particleapcnl}

In analogue to the twisted \gls{PARTICLEAGRAD} and twisted \gls{PARTICLEAGRADPLUS} algorithms, we can again construct `twisted' versions of the \gls{PARTICLEAPCNL} and \gls{PARTICLEAPCNLPLUS} algorithms, under the assumption that $\Mutation_t(\sta_t | \sta_{t-1}) = \dN(\sta_t; \priorFactor_t \sta_{t-1} + \priorIntercept_t, \priorVar_t)$, i.e., \eqref{eq:affine_assumption}. 

We start with the twisted \gls{PARTICLEAPCNL} algorithm. We now write 
\begin{align}
    \MutationAlt_t(\sta_t| \sta_{t-1}; \aux_{t:T}) 
    & \coloneqq p(\sta_t|\sta_{t-1}, \aux_{t:T})\\
    & \propto
    \int_{\spaceX^{T-t}} \biggl[ \prod_{s=t}^T \dN(\sta_s; \priorFactor_s \sta_{s-1} + \priorIntercept_s,\priorVar_s) \dN(\aux_s; \sta_s, \tfrac{\delta_s}{2} \priorVar_t) \biggr] \intDiff \sta_{t+1:T}\\
    & \propto \dN(\sta_t; \priorFactorAlt_t \sta_{t-1} + \priorInterceptAlt_t, \priorVarAlt_t),\label{eq:twistedparticleapcnl:mutation}\\
    \PotentialAlt_t(\sta_{t-1:t}; \aux_{t:T})
    & \coloneqq \qsemigroup_t(\sta_{t-1:t}) \frac{\dN(\aux_t; \sta_t + \gradientIndicator \tfrac{\delta_t}{2} \preconVar_t(\sta_{t-1}) \nabla_{\sta_t} \log \Potential_t(\sta_{t-1:t}), \tfrac{\delta_t}{2} \priorVar_t)}{\MutationAlt_t(\sta_t|\sta_{t-1}, \aux_{t:T})},\label{eq:twistedparticleapcnl:potential}
\end{align}
as well as $\qsemigroupAlt_t(\sta_{t-1:t}; \aux_{t:T}) \coloneqq \MutationAlt_t(\sta_t| \sta_{t-1}; \aux_{t:T})\PotentialAlt_t(\sta_{t-1:t}; \aux_{t:T})$. Here, $\priorInterceptAlt_t \in \reals^D$ and $\priorFactorAlt_t, \priorVarAlt_t \in \reals^{D \times D}$ can again be obtained via the Kalman-filtering recursions given in Appendix~\ref{app:sec:twisted_reference}.

A single iteration of the resulting methodology -- which we term the \emph{twisted \gls{PARTICLEAPCNL}} algorithm -- is then exactly as the \gls{PARTICLEAPCNL} algorithm (Algorithm~\ref{alg:particleapcnl}), except that $\MutationAlt_t(\ccdot|\ccdot;\aux_t)$, $\PotentialAlt_t(\ccdot; \aux_t)$ and $\qsemigroupAlt_t(\ccdot;\aux_t)$ from Section~\ref{app:subsec:particleapcnl} are replaced by $\MutationAlt_t(\ccdot|\ccdot;\aux_{t:T})$, $\PotentialAlt_t(\ccdot; \aux_{t:T})$ and $\qsemigroupAlt_t(\ccdot;\aux_{t:T})$ from this section. When the potential functions $\Potential_t(\sta_{t-1:t})$ vary in $\sta_{t-1}$, then we can further construct a \emph{twisted \gls{PARTICLEAGRADPLUS}} algorithm by replacing $\MutationAlt_t(\sta_t|\sta_{t-1}, \aux_t)$ in Algorithm~\ref{alg:particleapcnlplus} and in the denominator of $\PotentialAlt_{t}(\sta_{t-2:t}; \aux_{t-1:t})$ by $\MutationAlt_t(\sta_t|\sta_{t-1}, \aux_{t:T})$.

\begin{proposition}[validity of the twisted \gls{PARTICLEAPCNL}/\gls{PARTICLEAPCNLPLUS}] \label{prop:twisted_particleapcnl:validity}
  Sampling $\staAlt_{1:T}$ given $\sta_{1:T}$ via the twisted \gls{PARTICLEAPCNL} or twisted \gls{PARTICLEAPCNLPLUS} algorithm induces a Markov kernel which leaves $\pi_T$ invariant.
\end{proposition} 

\subsection{Choice of preconditioning matrix}
\label{app:subsec:choice_of_preconditioning_matrix}

There is some degree of freedom in choosing the preconditioning matrices $\preconVar_t(\sta_{t-1}) \in \reals^{D \times D}$ in the algorithms presented above. %
  \begin{enumerate}
   \item A simple option which does not require further assumptions is to take 
   \begin{align}
      \preconVar_t(\sta_{t-1}) \coloneqq \priorVar_t(\sta_{t-1}). \label{eq:choice_of_preconditioning_matrix:1}
   \end{align}
   
   \item If we make the stronger model assumption that $\priorMean_t(\sta_{t-1}) = \priorFactor_t \sta_{t-1} + \priorIntercept_t$ and $\priorVar_t(\sta_{t-1}) = \priorVar_t$ \eqref{eq:affine_assumption} (which is assumed to hold for the twisted versions of the \gls{PARTICLEAPCNL} and \gls{PARTICLEAPCNLPLUS} algorithms anyway), then we could alternatively set 
    \begin{align}
      \preconVar_t(\sta_{t-1}) = \preconVar_t \coloneqq \sum_{s=1}^T \marginalVar_{s, t}, \label{eq:choice_of_preconditioning_matrix:2}
   \end{align}
   where $\marginalVar_{s, t} \in \reals^{D \times D}$ is the block $(s, t)$ in the covariance matrix $\marginalVar \in \reals^{TD \times TD}$ of the prior dynamics $\Mutation_{1:T}(\sta_{1:T})$, i.e.,
   \begin{align}
     \marginalVar_{s, t} = 
     \begin{cases}
      \priorFactor_s \marginalVar_{s-1, t}, & \text{if $s > t$,}\\
      \marginalVar_{t, s}^\T, & \text{if $s < t$,}\\
      \marginalVar_t, & \text{if $s = t$,}
     \end{cases}
   \end{align}
   where $\marginalVar_t$ can be found via the recursion from Step~\ref{alg:twisted_proposal:general:1} of Algorithm~\ref{alg:twisted_proposal:general} from Appendix~\ref{app:sec:twisted_reference}. As discussed below, this specification has the potentially useful implication that the algorithm reduces to an `auxiliary-variable' version of the \gls{PCNL} algorithm on the path space in the absence of resampling and backward sampling.
   Unfortunately, evaluating the preconditioning matrices $\preconVar_t$ is likely to incur a quadratic computational complexity in $T$ which we prefer to avoid. 
   \item A compromise (which retains linear computational complexity in $T$) may be to truncate the above sum by setting
    \begin{align}
      \preconVar_t \coloneqq \sum_{s= (t-L) \vee 1}^{(t + L) \wedge T} \marginalVar_{s, t}, \label{eq:choice_of_preconditioning_matrix:3}
   \end{align}
   for some $L \in [T]_0$ (note that this still requires the model assumption \eqref{eq:affine_assumption}).
  \end{enumerate}

\subsection{Relationship with other methods}
\label{app:subsec:particlepcnl:relationship_with_other_methods}

The algorithms proposed above relate to existing methods as follows. 

\glsreset{APCNL}
\glsreset{PCNL}

\begin{enumerate}

   \item \glsunset{APCNL} \textbf{Generalisation of \gls{APCNL}.} \glsreset{APCNL} For $\gradientIndicator = 1$, the \gls{PARTICLEAPCNL} algorithm (and similarly the \gls{PARTICLEAPCNLPLUS} algorithm as well as the twisted versions of either) generalises an \emph{\gls{APCNL}} algorithm (which was mentioned but not explicitly derived in \citet{titsias2018auxiliary}) in the sense that the former reduces to the latter if $T = N = 1$. This can be seen as follows, where we again suppress the `time' subscript $t = 1$ everywhere so that $\pi(\sta) \propto \Mutation(\sta) \Potential(\sta)$, where $\Mutation(\sta) = \dN(\sta; \priorMean, \priorVar)$. We also take $\preconVar \coloneqq \priorVar$. Given that the current state of the Markov chain is $\sta = \sta^0$ (we can assume that $k = 0$ without loss of generality), Step~\ref{alg:particleapcnl:mutation} of Algorithm~\ref{alg:particleapcnl} first refreshes the auxiliary variable by sampling $\aux \sim \dN(\sta^0 + \tfrac{\delta}{2} \priorVar \nabla \log G(\sta^0), \tfrac{\delta}{2}\priorVar)$ and then proposes $\sta^1 \sim \dN((1-\beta)\priorMean + \beta \aux, (1-\beta)\priorVar)$. The remaining steps return $\staAlt \coloneqq \sta^1$ as the new state with acceptance probability $1 \wedge \alpha_{\text{\upshape\gls{APCNL}}}(\sta^0, \sta^1; \aux)$, where
    \begin{align}
      \MoveEqLeft\alpha_{\text{\upshape\gls{APCNL}}}(\sta^0, \sta^1; \aux)\\
      & \coloneqq \frac{1 - W^0}{1 - W^1}\\
      & = \frac{\pi(\sta^1) \dN(\aux; \sta^1 + \tfrac{\delta}{2} \priorVar \nabla \log G(\sta^1), \tfrac{\delta}{2}\priorVar) \dN(\sta^0; (1-\beta)\priorMean + \beta \aux, (1-\beta)\priorVar)}{\pi(\sta^0) \dN(\aux; \sta^0 + \tfrac{\delta}{2} \priorVar\nabla \log G(\sta^0), \tfrac{\delta}{2}\priorVar) \dN(\sta^1; (1-\beta)\priorMean + \beta \aux, (1-\beta)\priorVar)}. \label{eq:acpnl_acceptance_ratio}
    \end{align}
    Otherwise, the old state $\staAlt \coloneqq \sta^0 = \sta$ is returned as the new state. If $N = 1$ and $\gradientIndicator = 1$ but $T > 1$ then the \gls{APCNL} algorithm could still be recovered as a special case of (a slightly modified version of) the twisted \gls{PARTICLEAPCNLPLUS} algorithm (and also of the twisted \gls{PARTICLEAPCNL} algorithm if the potential functions $\Potential_t(\sta_{t-1:t})$ do not depend on $\sta_{t-1}$) if the latter uses no resampling and ancestral tracing instead of backward sampling, and if $\delta_1 = \dotsc = \delta_T$ and if the preconditioning matrices are specified via \eqref{eq:choice_of_preconditioning_matrix:2}. However, we do not recommend this choice of preconditioning matrix as it leads to squared computational complexity in $T$ (also incurred by \gls{APCNL}).

    \item \glsunset{PCNL} \textbf{Generalisation of \gls{PCNL}.} \glsreset{PCNL} Still taking $\gradientIndicator = 1$, the \gls{PARTICLEPCNL} algorithm generalises the \emph{\gls{PCNL}} algorithm \citep{cotter2013crank} in the sense that the former reduces to the latter if $T = N = 1$. This can be seen as follows, where we use the same notational conventions as in the case of \gls{APCNL} above.
    Step~\ref{alg:particlepcnl:mutation} of Algorithm~\ref{alg:particlepcnl} then marginally proposes $\sta^1 \sim  \dN((1-\beta) \priorMean + \beta[\sta^0 + \tfrac{\delta}{2} \priorVar \nabla \log \Potential(\sta^0)], (1-\beta^2)\priorVar)$. The remaining steps return $\staAlt \coloneqq \sta^1$ as the new state with acceptance probability $1 \wedge \alpha_{\text{\upshape\gls{PCNL}}}(\sta^0, \sta^1)$, where
    \begin{align}
      \alpha_{\text{\upshape\gls{PCNL}}}(\sta^0, \sta^1) 
      & \coloneqq \frac{1 - W^0}{1 - W^1}\\
      & = \frac{\pi(\sta^1)  \dN((1-\beta) \priorMean + \beta[\sta^1 + \tfrac{\delta}{2} \nabla \log \Potential(\sta^1)], (1-\beta^2)\priorVar)}{\pi(\sta^0) \dN((1 - \beta) \priorMean + \beta[\sta^0 + \tfrac{\delta}{2} \nabla \log \Potential(\sta^0)], (1-\beta^2)\priorVar)}. \label{eq:pcnl_acceptance_ratio}
    \end{align}
    Otherwise, the old state $\staAlt \coloneqq \sta^0 = \sta$ is returned as the new state. In particular, %
    in analogue to Section~\ref{subsec:particlemala:relationship_with_other_methods}, we can again interpret \gls{APCNL} as a version of \gls{PCNL} with `randomised' acceptance ratio.
  
\end{enumerate}

\section{Twisted proposals}
\label{app:sec:twisted_reference}

In this section, we detail the mutation kernel $\MutationAlt_t(\sta_t|\sta_{t-1}; \aux_{t:T})$ used by the twisted \gls{PARTICLEAGRAD}/\gls{PARTICLEAGRADPLUS} \eqref{eq:twistedparticleagrad:mutation} and twisted \gls{PARTICLEAPCNL}/\gls{PARTICLEAPCNLPLUS} \eqref{eq:twistedparticleapcnl:mutation} algorithms. This mutation kernel can be thought of as the fully-twisted particle-filter proposal for the state-space model which is defined by the Gaussian transitions $p(\sta_t|\sta_{t-1}) = \Mutation_t(\sta_t|\sta_{t-1}) = \dN(\sta_t; \priorFactor_t \sta_{t-1} + \priorIntercept_t, \priorVar_t)$ from \eqref{eq:affine_assumption} and observation densities $p(\aux_t|\sta_t) = \dN(\aux_t; \sta_t, \frac{\delta_t}{2}\vMat_t)$, where we take $\vMat_t \coloneqq \iMat$ in the case of the twisted \gls{PARTICLEAGRAD} or twisted \gls{PARTICLEAGRADPLUS} algorithm and $\vMat_t \coloneqq \priorVar_t$ in the case of the twisted \gls{PARTICLEAPCNL} or twisted \gls{PARTICLEAPCNLPLUS} algorithm:
\begin{align}
  \MutationAlt_t(\sta_t| \sta_{t-1}; \aux_{t:T})  
  & = p(\sta_t|\sta_{t-1}, \aux_{1:T}) = \dN(\sta_t; \priorFactorAlt_t \sta_{t-1} + \priorInterceptAlt_t, \priorVarAlt_t).
\end{align}

\paragraph{General algorithm.}
Algorithm~\ref{alg:twisted_proposal:general} explains how the twisted-proposal parameters $\priorInterceptAlt_t \in \reals^D$ and $\priorFactorAlt_t, \priorVarAlt_t \in \reals^{D \times D}$ can be calculated at linear complexity in $T$, independently of the total number of particles, $N$. Notably, Algorithm~\ref{alg:twisted_proposal:general} does not require $\priorVar_t$ to be invertible.

\begin{framedAlgorithm}[twisted-proposal parameters] 
\label{alg:twisted_proposal:general}
At the start of an iteration of the twisted \gls{PARTICLEAGRAD}, \gls{PARTICLEAGRADPLUS}, \gls{PARTICLEAPCNL} or \gls{PARTICLEAPCNLPLUS} algorithm (after having sampled all the auxiliary variables $\aux_{1:T}$ upfront -- e.g., as in Algorithm~\ref{alg:generic_auxiliary_csmc} from Appendix~\ref{app:subsec:generic_auxiliary_algorithm} -- which is possible because these only depend on the reference path),
\begin{enumerate}
  \item \label{alg:twisted_proposal:general:1} recursively compute the moments of $p(\sta_t) = \dN(\sta_t; \marginalMean_t, \marginalVar_t)$, for $t = 1, \dotsc, T$, as
  \begin{align}
    \marginalMean_t & \coloneqq \priorFactor_t \marginalMean_{t-1} + \priorIntercept_t,\\
    \marginalVar_t & \coloneqq \priorFactor_t \marginalVar_{t-1} \priorFactor_t^\T + \priorVar_t,
  \end{align}
  if $t > 1$, and with initial condition $\marginalMean_1 = \priorIntercept_1$ and $\marginalVar_1 = \priorVar_1$, 
  
  \item \label{alg:twisted_proposal:general:2} recursively compute the moments of the time-reversed state transition kernels $p(\sta_t|\sta_{t+1}) = \dN(\sta_t; \priorFactor_t^{\leftarrow} \sta_{t+1} + \priorIntercept_t^{\leftarrow}, \priorVar_t^{\leftarrow})$, for $t = T -1, \dotsc, 1$, as
  \begin{align}
    \priorFactor_t^{\leftarrow} 
    & \coloneqq \priorFactor_{t+1} \marginalVar_t \marginalVar_{t+1}^{-1},\\
    \priorIntercept_t^{\leftarrow} 
    & \coloneqq \marginalMean_t - \priorFactor_t^{\leftarrow} \marginalMean_{t+1},\\
    \priorVar_t^{\leftarrow} & \coloneqq \marginalVar_t - \priorFactor_t^{\leftarrow} \marginalVar_t \priorFactor_{t+1}^\T.
 \end{align}

 \item \label{alg:twisted_proposal:general:3} run the Kalman filtering recursion for the time-reversed state-space model (i.e., with observation densities $p(\aux_t|\sta_t) = \dN(\aux_t; \sta_t, \frac{\delta_t}{2}\vMat_t)$, initial distribution $p(\sta_t)$ and time-reversed state transitions $p(\sta_t|\sta_{t+1})$ found in Steps~\ref{alg:twisted_proposal:general:1} and \ref{alg:twisted_proposal:general:2}) to compute the moments of $p(\sta_t|\aux_{t:T}) = \dN(\sta_t; \kalmanMean_{t|t}^{\leftarrow}, \kalmanVar_{t|t}^{\leftarrow})$, for $t = T, \dotsc, 1$,
 
 \item \label{alg:twisted_proposal:general:4} set $\priorFactorAlt_1 \coloneqq \zeroMat_{D \times D}$, $\priorInterceptAlt_1 \coloneqq \kalmanMean_{1|1}^{\leftarrow}$ as well as $\priorVarAlt_1 \coloneqq \kalmanVar_{1|1}^{\leftarrow}$, and, for $t = 2, \dotsc, T$, 
 \begin{align}
   \priorFactorAlt_t & \coloneqq \kalmanVar_{t|t}^{\leftarrow} \priorFactor_{t-1}^{\leftarrow} (\priorVar_{t-1}^{\leftarrow} + \priorFactor_{t-1}^{\leftarrow} \kalmanVar_{t|t}^{\leftarrow} \{\priorFactor_{t-1}^{\leftarrow}\}^\T)^{-1},\\
   \priorInterceptAlt_t & \coloneqq \kalmanMean_{t|t}^{\leftarrow} - \priorFactorAlt_t (\priorFactor_{t-1}^{\leftarrow} \kalmanMean_{t|t}^{\leftarrow} + \priorIntercept_{t-1}^{\leftarrow}), \\
  \priorVarAlt_t & \coloneqq (\iMat - \priorFactorAlt_t \priorFactor_{t-1}^{\leftarrow}) \kalmanVar_{t|t}^{\leftarrow}.
\end{align}
\end{enumerate}
\end{framedAlgorithm}

Algorithm~\ref{alg:twisted_proposal:general} is justified by the decomposition
\begin{align}
  \MutationAlt_t(\sta_t| \sta_{t-1}; \aux_{t:T})  
  & \propto p(\sta_{t-1:t}|\aux_{t:T})\\
  & = p(\sta_{t-1}|\sta_t) p(\sta_t|\aux_{t:T})\\
  & = \dN(\sta_t; \priorFactor_t^{\leftarrow} \sta_{t+1} + \priorIntercept_t^{\leftarrow}, \priorVar_t^{\leftarrow}) \dN(\sta_t; \kalmanMean_{t|t}^{\leftarrow}, \kalmanVar_{t|t}^{\leftarrow}) \label{eq:twisted_proposal:general_penultimate_line}\\
  & \propto \dN(\sta_t; \priorFactorAlt_t \sta_{t-1} + \priorInterceptAlt_t, \priorVarAlt_t), \label{eq:twisted_proposal:general_ultimate_line}
\end{align}
where, as described in Algorithm~\ref{alg:twisted_proposal:general}, $p(\sta_t|\aux_{t:T}) = \dN(\sta_t; \kalmanMean_{t|t}^{\leftarrow}, \kalmanVar_{t|t}^{\leftarrow})$ is the time-$t$ filter for the time-reversed state-space model with the same observation densities $p(\aux_t|\sta_t) = \dN(\aux_t; \sta_t, \frac{\delta_t}{2}\vMat_t)$ as before but with initial distribution $p(\sta_t) = \dN(\sta_t; \marginalMean_T, \marginalVar_T)$ and time-reversed state transitions $p(\sta_t|\sta_{t+1}) = \dN(\sta_t; \priorFactor_t^{\leftarrow} \sta_{t+1} + \priorIntercept_t^{\leftarrow}, \priorVar_t^{\leftarrow})$. Thus, in Algorithm~\ref{alg:twisted_proposal:general}:
\begin{itemize}
  \item Step~\ref{alg:twisted_proposal:general:1} calculates the marginal prior distributions of the states. To see this, note that this step is effectively the Kalman-filter recursion without observations. Note that if the prior dynamics are stationary with stationary distribution $\dN(\marginalMean, \marginalVar)$, then Step~\ref{alg:twisted_proposal:general:1} can be skipped (because then $\marginalMean_t = \marginalMean$ and $\marginalVar_t = \marginalVar$, for any $t \in [T]$).

  \item Step~\ref{alg:twisted_proposal:general:2} computes the parameters of the time-reversed transition kernels and follows from standard Gaussian algebra~\citep[see, e.g.,][Section~A.1]{sarkka2023bayesian} by noting that 
  \begin{equation}
    \begin{bmatrix}
        \sta_t \\ \sta_{t+1}
    \end{bmatrix} \sim \dN\biggl(
    \begin{bmatrix}
        \marginalMean_t \\ \marginalMean_{t+1} %
    \end{bmatrix}, 
    \begin{bmatrix}
        \marginalVar_t & \priorFactor_{t+1} \marginalVar_t \\
        \marginalVar_t \priorFactor_{t+1}^\T & \marginalVar_{t+1}
    \end{bmatrix}
    \biggr).
\end{equation}

\item Step~\ref{alg:twisted_proposal:general:4} derives \eqref{eq:twisted_proposal:general_ultimate_line} from \eqref{eq:twisted_proposal:general_penultimate_line} and corresponds to a single update step of a Kalman filter \citep[Chapter 6, Equation~6.21]{sarkka2023bayesian}, where $\sta_{t-1}$ plays the r\^ole of an observation.
\end{itemize}

\paragraph{Alternative algorithm for invertible covariance matrices.} If $\priorVar_t$ is invertible for all $t \in [T]$, then the twisted-proposal parameters can be alternatively computed via Algorithm~\ref{alg:twisted_proposal:invertible} which may be slightly simpler to implement for some users and which may provide additional numerical advantages in the case of explosive prior dynamics.

\begin{framedAlgorithm}[twisted-proposal parameters: alternative] 
\label{alg:twisted_proposal:invertible} At the start of an iteration of the twisted \gls{PARTICLEAGRAD}, \gls{PARTICLEAGRADPLUS}, \gls{PARTICLEAPCNL} or \gls{PARTICLEAPCNLPLUS} algorithm (after having sampled all the auxiliary variables $\aux_{1:T}$ upfront -- e.g., as in Algorithm~\ref{alg:generic_auxiliary_csmc} from Appendix~\ref{app:subsec:generic_auxiliary_algorithm} -- which is possible because these only depend on the reference path),
\begin{enumerate}
  \item run the Kalman filtering recursion to compute the moments of $p(\sta_t| \aux_{1:t-1}) = \dN(\sta_t; \kalmanMean_{t|t-1}, \kalmanVar_{t|t-1})$, for $t = 1, \dotsc, T$,
  \item run the Kalman smoothing (a.k.a.\ Rauch--Tung--Striebel smoothing) recursion to compute the moments of $p(\sta_t|\aux_{1:T}) = \dN(\sta_t; \kalmanMean_{t|T}, \kalmanVar_{t|T})$, for $t = T, T -1, \dotsc, 1$,
  \item for $t \in [T]$, set
      \begin{align}
      \priorVarAlt_t & \coloneqq [\priorVar_t^{-1} + \kalmanVar_{t|T}^{-1} - \kalmanVar_{t|t-1}^{-1}]^{-1},\\
       \priorFactorAlt_t & \coloneqq \priorVarAlt_t [\priorVar_t^{-1} \priorIntercept_t + \kalmanVar_{t|T}^{-1}  \kalmanMean_{t|T} - \kalmanVar_{t|t-1}^{-1} \kalmanMean_{t|t-1}],\\
      \priorInterceptAlt_t & \coloneqq \priorVarAlt_t \priorVar_t^{-1} \priorFactor_t.
    \end{align}
\end{enumerate}
\end{framedAlgorithm}

Algorithm~\ref{alg:twisted_proposal:invertible} is justified by the decomposition
\begin{align}
  p(\sta_t|\sta_{t-1}, \aux_{1:T})
  & \propto p(\sta_{t-1:t}, \aux_{1:T})\\
  & = p(\sta_{t-1}|\sta_t, \aux_{1:T}) p(\sta_t|\aux_{1:T})\\
  &  \propto \frac{p(\sta_t|\sta_{t-1})}{p(\sta_t|\aux_{1:t-1})} p(\sta_t|\aux_{1:T})\\
  & = \frac{\dN(\sta_t; \priorFactor_t \sta_{t-1} + \priorIntercept_t, \priorVar_t)}{\dN(\sta_t; \kalmanMean_{t|t-1}, \kalmanVar_{t|t-1})} \dN(\sta_t; \kalmanMean_{t|T}, \kalmanVar_{t|T})\\
  & \propto \dN(\sta_t; \priorFactorAlt_t \sta_{t-1} + \priorInterceptAlt_t, \priorVarAlt_t).
\end{align}

\section{Integrating out the auxiliary variables}
\label{app:sec:integrating_out_the_auxiliary_variables}

In this section, we prove a few lemmata which are used in subsequent sections. 
\begin{itemize}
    \item \textbf{Lemmata~\ref{lem:determinant_of_special_block_matrix} and \ref{lem:inverse_of_special_block_matrix}.}  Lemmata~\ref{lem:determinant_of_special_block_matrix} and \ref{lem:inverse_of_special_block_matrix} derive the determinant and inverse of a certain simple block matrix which appears repeatedly in the remainder of this section and also in Appendix~\ref{app:sec:proof_of_the_interpolation_property}.
    \item \textbf{Lemma~\ref{lem:auxiliary_lemma_1}.} Lemma~\ref{lem:auxiliary_lemma_1} will allow us to derive marginal proposal distributions of the various algorithms, i.e., the distribution of $\smash{\sta_t^{-k_t} = (\sta_t^0, \dotsc, \sta_t^{k_t - 1}, \sta_t^{k_t + 1}, \dotsc, \sta_t^{N})}$ conditional on $k_t$, $\sta_t^{k_t} = \sta_t$ and all the particles and ancestor indices with time indices $s < t$, but with the auxiliary variables $\aux_{1:T}$ integrated out. 
    \item \textbf{Lemma~\ref{lem:auxiliary_lemma_2}.} Lemma~\ref{lem:auxiliary_lemma_2} will allow us to evaluate the particle weights used in the `marginal' algorithms (\gls{PARTICLEMALA}, \gls{PARTICLEMGRAD} and \gls{PARTICLEPCNL}) at linear complexity in $N$ although the weight of the $n$th particle depends on the values of all other particles.
\end{itemize} 

\subsection{Properties of a particular block matrix}

Let $\iMat_M$ denote the $(M \times M)$ identity matrix. When $M = D$ we continue to leave out the subscript. Furthermore, let $\unitMat_{M \times N} \in \{1\}^{M \times N}$ and denote a matrix in which every element is $1$. For matrices $\aMat, \bMat \in \reals^{D \times D}$, define the block matrix
\begin{align}
  \myBlockMat{N}(\aMat, \bMat) 
  & \coloneqq \iMat_N \otimes \aMat + \unitMat_{N \times N} \otimes \bMat
  = 
  \begin{bmatrix}
   \aMat+\bMat      & \bMat        & \dotsc & \bMat\\
   \bMat      & \aMat+\bMat        & \ddots & \vdots\\
   \vdots & \ddots   & \ddots & \bMat\\
   \bMat      & \dotsc   & \bMat      & \aMat+\bMat
  \end{bmatrix} \in \reals^{(DN) \times (DN)}. \label{eq:definition_of_special_block_matrix}
\end{align}

\begin{lemma}\label{lem:determinant_of_special_block_matrix}
For $N, D \in \naturals$, let $\aMat, \bMat \in \reals^{D \times D}$. Then,
\begin{align}
  \det(\myBlockMat{N}(\aMat, \bMat)) = \det(\aMat)^{N-1} \det(\aMat + N\bMat).
\end{align}
\end{lemma}
\begin{namedProof}
  Subtracting the last row of $\myBlockMat{N}(\aMat, \bMat)$ from all other rows and then adding the sum of the first $N-1$ columns to the last column gives the upper-triangular block matrix
  \begin{align}
  \begin{bmatrix}
   \aMat      & \zeroMat        & \dotsc & \zeroMat\\
   \bMat      & \ddots   & \ddots & \vdots\\
   \vdots & \ddots   & \aMat      & \zeroMat\\
   \bMat      & \dotsc   & \bMat      & \aMat+N\bMat
  \end{bmatrix}.
\end{align}
 This proves the result. \hfill \qedwhite
\end{namedProof}

\begin{lemma}\label{lem:inverse_of_special_block_matrix}
  For $N, D \in \naturals$, let $\aMat, \bMat \in \reals^{D \times D}$, such that $\aMat$ and $(\aMat + N\bMat)$ are invertible. Then
  \begin{align}
    \myBlockMat{N}(\aMat, \bMat)^{-1} = \myBlockMat{N}(\aMat^{-1}, - (\aMat + N\bMat)^{-1} \bMat \aMat^{-1}).
  \end{align}
\end{lemma}
\begin{namedProof}
  We must have $\myBlockMat{N}(\aMat, \bMat) \myBlockMat{N}(\fMat, \gMat) = \iMat_{DN}$ and hence
  \begin{align}
    (\aMat + \bMat)(\fMat+\gMat) + (N-1)\bMat \gMat &= \iMat,\\
    \bMat(\fMat+\gMat) + (\aMat+\bMat)\gMat + (N-2)\bMat \gMat & = \zeroMat.
  \end{align}
  This implies $\fMat = \aMat^{-1}$ and $\gMat = - (\aMat + N\bMat)^{-1} \bMat \aMat^{-1}$. \hfill \qedwhite
\end{namedProof}

\subsection{Conditional and marginal proposal distributions}

In this section, for any tuple $(\mathbf{z}_0, \dotsc, \mathbf{z}_N)$ of values in $\spaceX \coloneqq \reals^D$ and any $n \in [N]_0$, we write $\mathbf{z}_{-n} \coloneqq (\mathbf{z}_0, \dotsc, \mathbf{z}_{n-1}, \mathbf{z}_{n+1}, \dotsc, \mathbf{z}_N)$. Given some $N \in \naturals$, $n \in [N]_0$, $\sta_n, \genericGrad_n \in \spaceX$, we consider the following joint distribution on $\spaceX^{N+1}$:
\begin{align}
  q_{-n}(\aux, \sta_{-n}| \sta_n)%
  & \coloneqq \dN(\aux; \sta_n + \genericGrad_n, \eMat) \prodSubstackAligned{m}{=}{0}{m}{\neq}{n}^N \dN(\sta_m; \genericVec_m + \hMat_m \aux, \dMat_m), \label{eq:generic_joint_proposal}
\end{align}
where, for any $m \in [N]_0$, $\genericVec_m \in \spaceX$ and $\hMat_m \in \reals^{D \times D}$, and $\dMat_m, \eMat \in \reals^{D \times D}$ are positive definite and symmetric. 

To simplify the presentation -- and with some abuse of notation since we use the same symbols for tuples and their vectorised versions -- we write
\begin{align}
 \sta_{-n} & \coloneqq  
 \begin{bmatrix}
   \sta_0 \\
   \vdots\\
   \sta_{n-1}\\
   \sta_{n+1}\\
   \vdots\\
   \sta_n 
 \end{bmatrix}, \quad
 \genericVec_{-n} \coloneqq  
 \begin{bmatrix}
   \genericVec_0 \\
   \vdots\\
   \genericVec_{n-1}\\
   \genericVec_{n+1}\\
   \vdots\\
   \genericVec_n 
 \end{bmatrix}, \quad 
 \hMat_{-n} \coloneqq 
  \begin{bmatrix}
   \hMat_0\\
   \vdots\\
   \hMat_{n-1}\\
   \hMat_{n+1}\\
   \vdots\\
   \hMat_n
  \end{bmatrix}, \quad 
  \dMat_{-n} \coloneqq \diag\left(
  \begin{bmatrix}
   \dMat_0\\
   \vdots\\
   \dMat_{n-1}\\
   \dMat_{n+1}\\
   \vdots\\
   \dMat_n
  \end{bmatrix}\right),
\end{align}
where, in the last expression, $\diag$ induces a block-diagonal matrix. With this notation, we can formulate \eqref{eq:generic_joint_proposal} equivalently as
\begin{align}
  q_{-n}(\aux, \sta_{-n}| \sta_n) = \dN(\aux; \sta_n + \genericGrad_n, \eMat) \dN(\sta_{-n}; \genericVec_{-n} + \hMat_{-n} \aux, \dMat_{-n}). \label{eq:generic_joint_proposal_alt}
\end{align}

\begin{lemma}\label{lem:auxiliary_lemma_1}
For any $n \in [N]_0$ and $\sta_n \in \spaceX$,
\begin{enumerate}
    \item \label{lem:auxiliary_lemma_1:marginal} the marginal distribution of $\sta_{-n}$ given $\sta_n$ under \eqref{eq:generic_joint_proposal_alt} is
\begin{align}
  q_{-n}(\sta_{-n}| \sta_n) 
  & = \dN(\sta_{-n}; \boldsymbol{\mu}_{-n}, \boldsymbol{\Sigma}_{-n}),
\end{align}
where
\begin{align}
  \boldsymbol{\mu}_{-n} 
  & \coloneqq \genericVec_{-n} + \hMat_{-n} (\sta_n + \genericGrad_n),\\
  \boldsymbol{\Sigma}_{-n} 
  & \coloneqq
  \dMat_{-n} + \hMat_{-n} \eMat \hMat_{-n}^\T;
\end{align}
    \item  \label{lem:auxiliary_lemma_1:conditional} the conditional distribution of $\aux$ given $\sta_{-n}$ and $\sta_n$ under \eqref{eq:generic_joint_proposal_alt} is
    \begin{align}
        \!\!\!q_{-n}(\aux | \sta_{-n}, \sta_n)
        &= \dN(\aux; \sta_n + \genericGrad_n + \mathbf{K}[\sta_{-n} - \genericVec_{-n} - \hMat_{-n}(\sta_n + \genericGrad_n)], (\iMat - \mathbf{K}\hMat_{-n})\eMat),\!\!\!
    \end{align}
    where $\mathbf{K} \coloneqq \eMat \hMat_{-n}^\T (\dMat_{-n} + \hMat_{-n} \eMat \hMat_{-n}^\T)^{-1}$.
    \end{enumerate}
\end{lemma}
\begin{namedProof}
  This follows by simple algebra~\citep[see, e.g.,][Appendix~A.1]{sarkka2023bayesian}. \hfill \qedwhite
\end{namedProof}

\begin{lemma}\label{lem:auxiliary_lemma_2}
 Assume now that $\hMat_m = \hMat$ and $\dMat_m = \dMat$, for any $m \in [N]_0$. Then, with the notation from Lemma~\ref{lem:auxiliary_lemma_1}, 
 \begin{align}
       q_{-n}(\sta_{-n} | \sta_n)
       & \propto H_{\genericGrad_n}(\sta_n, \genericVec_n, \bar{\sta}, \bar{\genericVec}) I((\sta_m - \genericVec_m)_{m=0}^N),
  \end{align}
  where 
  \begin{enumerate}
      \item $\mathbf{z}_{0:N} \mapsto I(\mathbf{z}_{0:N})$ is invariant under any permutation of its arguments;
      \item $\bar{\sta} \coloneqq \frac{1}{N+1} \sum_{m=0}^N \sta_n$, and $\bar{\genericVec} \coloneqq \frac{1}{N+1} \sum_{m=0}^N \genericVec_n$, and 
  \begin{align}
  \MoveEqLeft \log H_{\genericGrad}(\sta, \genericVec, \bar{\sta}, \bar{\genericVec})\\
  & = \tfrac{1}{2} (\sta - \genericVec)^\T (\dMat^{-1} + \gMat) (\sta - \genericVec)\\
  & \qquad - \tfrac{1}{2}N (\sta + \genericGrad)^\T \hMat^\T (\dMat^{-1} - N \gMat) \hMat (\sta + \genericGrad)\\
  & \qquad -  (N+1) (\bar{\sta} - \bar{\genericVec})^\T [\gMat (\sta - \genericVec) - (\dMat^{-1} - N \gMat) \hMat(\sta + \genericGrad)]\\
  & \qquad - (\sta - \genericVec)^\T (\dMat^{-1} - N \gMat)\hMat(\sta + \genericGrad),
\end{align}
 whose evaluation complexity does not depend on $N$. Here,
 \begin{align}
   \gMat 
   & \coloneqq
   (\dMat + N \hMat \eMat\hMat^{\T})^{-1} \hMat \eMat \hMat^\T \dMat^{-1} \label{eq:gmatrix:1}\\
   & = \dMat^{-1} \hMat\eMat (\eMat + N \eMat \hMat^\T \dMat^{-1} \hMat\eMat)^{-1} \eMat \hMat^\T \dMat^{-1}. \label{eq:gmatrix:2}
 \end{align}
  \end{enumerate}
\end{lemma}

\begin{namedProof}
The equivalence of \eqref{eq:gmatrix:1} and \eqref{eq:gmatrix:2} follows from the \emph{push-through identity} \citep{henderson1981deriving}.
By assumption, $\boldsymbol{\Sigma}_{-n} = \myBlockMat{N}(\dMat, \hMat \eMat \hMat^\T)$. Thus,
Lemma~\ref{lem:inverse_of_special_block_matrix} gives
\begin{align}
 \boldsymbol{\Sigma}_{-n}^{-1} = \myBlockMat{N}(\dMat^{-1}, - \gMat).
\end{align}
In particular, letting ${\otimes}$ be the Kronecker product, this implies that
\begin{align}
  \boldsymbol{\Sigma}_{-n}^{-1} \hMat_{-n}
  & = \unitMat_{N \times 1} \otimes [(\dMat^{-1} - N\gMat)\hMat],\\
  \hMat_{-n}^\T \boldsymbol{\Sigma}_{-n}^{-1} \hMat_{-n}
  & = N \hMat^\T (\dMat^{-1} - N\gMat) \hMat.
\end{align}
Therefore, defining
\begin{gather}
 \sta \coloneqq  
 \begin{bmatrix}
   \sta_0 \\
   \vdots\\
   \sta_N 
 \end{bmatrix}, \quad
 \genericVec \coloneqq  
 \begin{bmatrix}
   \genericVec_0 \\
   \vdots\\
   \genericVec_N 
 \end{bmatrix}, \quad \boldsymbol{\Sigma}
 \coloneqq \myBlockMat{N+1}(\dMat^{-1}, -\gMat)^{-1},
\end{gather}
we have
\begin{align*}
  \MoveEqLeft q_{-n}(\sta_{-n}|\sta_n)\\
  & \propto \exp\bigl(-\tfrac{1}{2}\bigl[ (\sta_{-n} - \genericVec_{-n} - \hMat_{-n} (\sta_n + \genericGrad_n))^\T \boldsymbol{\Sigma}_{-n}^{-1} (\sta_{-n} - \genericVec_{-n} - \hMat_{-n} (\sta_n + \genericGrad_n))\bigr]\bigr)\\
  & = \exp\bigl(-\tfrac{1}{2}\bigl[(\sta_{-n} - \genericVec_{-n})^\T \boldsymbol{\Sigma}_{-n}^{-1} (\sta_{-n} - \genericVec_{-n})\\*
  & \qquad\qquad\quad + (\sta_n + \genericGrad_n)^\T \hMat_{-n}^\T \boldsymbol{\Sigma}_{-n}^{-1} \hMat_{-n} (\sta_n + \genericGrad_n)\\*
  & \qquad\qquad\quad - 2 (\sta_{-n} - \genericVec_{-n})^\T \boldsymbol{\Sigma}_{-n}^{-1} \hMat_{-n} (\sta_n + \genericGrad_n)\bigr]\bigr)\\
  & = \exp\bigl(-\tfrac{1}{2}\bigl[(\sta - \genericVec)^\T \boldsymbol{\Sigma}^{-1} (\sta - \genericVec)\\*
  & \qquad\qquad\quad - (\sta_n - \genericVec_n)^\T (\dMat^{-1} - \gMat) (\sta_n - \genericVec_n)\\*
  & \qquad\qquad\quad + 2 (\sta_{-n} - \genericVec_{-n})^\T  [\unitMat_{N \times 1} \otimes \gMat] (\sta_n - \genericVec_n)\\*
  & \qquad\qquad\quad + N (\sta_n + \genericGrad_n)^\T \hMat^\T (\dMat^{-1} - N\gMat)\hMat (\sta_n + \genericGrad_n)\\*
  & \qquad\qquad\quad - 2 (\sta_{-n} - \genericVec_{-n})^\T \boldsymbol{\Sigma}_{-n}^{-1} \hMat_{-n} (\sta_n + \genericGrad_n)\bigr]\bigr)\\ 
  & = \exp\bigl(-\tfrac{1}{2}\bigl[(\sta - \genericVec)^\T \boldsymbol{\Sigma}^{-1} (\sta - \genericVec)\\
  & \qquad\qquad\quad - (\sta_n - \genericVec_n)^\T (\dMat^{-1} - \gMat) (\sta_n - \genericVec_n)\\*
  & \qquad\qquad\quad + N (\sta_n + \genericGrad_n)^\T \hMat^\T (\dMat^{-1} - N\gMat) \hMat (\sta_n + \genericGrad_n)\\*
  & \qquad\qquad\quad + 2 (\sta - \genericVec)^\T \left\{\unitMat_{(N+1) \times 1} \otimes  [\gMat (\sta_n - \genericVec_n) - (\dMat^{-1} - N\gMat) \hMat(\sta_n + \genericGrad_n)]\right\}\\*
  & \qquad\qquad\quad - 2(\sta_n - \genericVec_n)^\T [\gMat (\sta_n - \genericVec_n) - (\dMat^{-1} - N\gMat)\hMat(\sta_n + \genericGrad_n)]\bigr]\bigr)\\
  & = H_{\genericGrad_n}(\sta_n, \genericVec_n, \bar{\sta}, \bar{\genericVec}) I((\sta_m - \genericVec_m)_{m=0}^N),
\end{align*}
with 
\begin{align}
  I((\sta_m - \genericVec_m)_{m=0}^N) \propto \exp(-\tfrac{1}{2} (\sta - \genericVec)^\T \boldsymbol{\Sigma}^{-1} (\sta - \genericVec)). 
\end{align}
This completes the proof. \hfill \qedwhite
\end{namedProof}

\section{Generic algorithms and proof of Propositions~\ref{prop:particleamala:validity}--\ref{prop:twisted_particleapcnl:validity}}
\label{app:sec:invariance_proofs}

In this section, we prove that the algorithms proposed in this work leave $\target_T$ invariant. To this end, we first prove the validity of two generic algorithms. 
\begin{itemize}
    \item \textbf{Generic auxiliary algorithm.} The first generic algorithm includes auxiliary variables $\aux_t$ in the space and admits the `auxiliary-variable' based algorithms: \gls{PARTICLEAMALA}, \gls{PARTICLEAGRAD}, \gls{PARTICLEAPCNL}  as well as their smoothing-gradient (`+') and twisted versions, as special cases. Its proof extends the auxiliary-variable interpretation of the \gls{PARTICLERWM} algorithm which was given in \citet{corenflos2023auxiliary}. 
     \item \textbf{Generic marginal algorithm.} The second generic algorithm integrates out the auxiliary variables and admits the `marginal' algorithms from the main manuscript (\gls{PARTICLEMALA}, \gls{PARTICLEMGRAD}, \gls{PARTICLEPCNL}). Its proof relies on an argument previously given in \citet{finke2016embedded}. 
\end{itemize}

\subsection{Generic auxiliary algorithm}
\label{app:subsec:generic_auxiliary_algorithm}

Define an extended target distribution
\begin{align}
  \targetAlt_T(\sta_{1:T}, \aux_{1:T}) 
  & \coloneqq \target_T(\sta_{1:T}) \prod_{t=1}^T \dN(\aux_t; \sta_t + \Phi_t(\sta_{1:T}), \eMat_t(\sta_{t-1:t})), \label{eq:generic_extended_distribution_with_auxiliary_variables}
\end{align}
where, for any $t \in [T]$, $\Phi_t \colon \spaceX^T \to \spaceX$ is a function satisfying
\begin{align}
 \Phi_t(\sta_{1:T}) = 
 \begin{cases}
   \genericGrad_T(\sta_{t-T:T}), & \text{if $t = T$,}\\
   \genericGrad_t(\sta_{t-1:t}) + \genericGradAlt_t(\sta_{t:t+1}), & \text{otherwise,}
 \end{cases}
\end{align}
and $\eMat_t(\sta_{t-1:t}) \in \reals^{D \times D}$ is some positive-definite symmetric matrix.
Additionally, let 
\begin{align}
  \qsemigroupAlt_t(\sta_{t-2:t}; \aux_{1:T})
  & = \MutationAlt_t(\sta_t|\sta_{t-1}; \aux_{1:T}) \PotentialAlt_t(\sta_{t-2:t}; \aux_{1:T}),
\end{align}
for some mutation kernel 
$\MutationAlt_t(\sta_t|\sta_{t-1}; \aux_{1:T})$ and some potential function $\PotentialAlt_t(\sta_{t-2:t}; \aux_{1:T})$ (both of which may depend on some or all of $\aux_1, \dotsc, \aux_t$) such that
\begin{align}
  \targetAlt_T(\sta_{1:T}|\aux_{1:T}) \propto \prod_{t=1}^T \qsemigroupAlt_t(\sta_{t-2:t}; \aux_{1:T}).
\end{align}

\begin{framedAlgorithm}[generic auxiliary algorithm] \label{alg:generic_auxiliary_csmc}
Given $\sta_{1:T} \in \spaceX^T$, sample 
\begin{align}
 \aux_t \sim \dN(\sta_t + \Phi_t(\sta_{1:T}), \eMat_t(\sta_{t-1:t})),
\end{align}
for any $t = 1, \dotsc,T$ and then
\begin{enumerate}
    \item \label{alg:generic_auxiliary_csmc:1} for $t = 1, \dotsc, T$, 
    \begin{enumerate}
        \item sample $k_t$ from a uniform distribution on $[N]_0$ and set $\sta_t^{k_t} \coloneqq \sta_t$,
        \item if $t > 1$, set $\smash{a_{t-1}^{k_t} \coloneqq k_{t-1}}$ and sample $a_{t-1}^n = i$ \gls{WP} $\smash{W_{t-1}^i}$, for $n \in [N]_0 \setminus \{k_t\}$,
        
        \item \label{alg:generic_auxiliary_csmc:mutation} sample %
        $\smash{\sta_t^n \sim \MutationAlt_t(\ccdot|\sta_{t-1}^{a_{t-1}^n}; \aux_{1:T})}$ for $n \in [N]_0 \setminus \{k_t\}$, 
        
        \item \label{alg:generic_auxiliary_csmc:weight_calculation} for $n \in [N]_0$, set $w_t^n \propto \PotentialAlt_t(\sta_{t-2:t}^{(n)}; \aux_{1:T})$,
        
        \item \label{alg:generic_auxiliary_csmc:weight_normalisation} for $n \in [N]_0$, set $W_t^n \coloneqq  w_t^n / \sum_{m = 0}^N w_t^m$;%
    \end{enumerate}
    
    \item \label{alg:generic_auxiliary_csmc:2} sample $i \in [N]_0 \setminus \{k_T\}$ \gls{WP} $\dfrac{W_T^i}{1 - W_T^{k_T}}$; set $l_T \coloneqq i$ \gls{WP} $1 \wedge \dfrac{1- W_T^{k_T}}{1 - W_T^i}$; otherwise, set $l_T \coloneqq k_T$;

    \item \label{alg:generic_auxiliary_csmc:backward_sampling} for $t = T-1, \dotsc, 1$, sample $l_t = i \in [N]_0$ \gls{WP}
    \begin{align}
          \dfrac{W_t^{i} \qsemigroupAlt_{t+1}((\sta_{t-1:t}^{(i)}, \sta_{t+1}^{l_{t+1}}); \aux_{1:T})  \qsemigroupAlt_{t+2}((\sta_t^{i}, \sta_{t+1}^{l_{t+1}}, \sta_{t+2}^{l_{t+2}}); \aux_{1:T})}{\sum_{n=0}^N W_t^n  \qsemigroupAlt_{t+1}((\sta_{t-1:t}^{(n)}, \sta_{t+1}^{l_{t+1}}); \aux_{1:T}) \qsemigroupAlt_{t+2}((\sta_t^{n}, \sta_{t+1}^{l_{t+1}}, \sta_{t+2}^{l_{t+2}}); \aux_{1:T})};
    \end{align}
    \item \label{alg:generic_auxiliary_csmc:4} return $\staAlt_{1:T} \coloneqq (\sta_1^{l_1}, \dotsc, \sta_t^{l_T})$.
\end{enumerate}
\end{framedAlgorithm}

\begin{proposition}[validity of the generic auxiliary algorithm] \label{prop:validity_of_auxiliary_algorithms}
 Sampling $\staAlt_{1:T}$ given $\sta_{1:T}$ via Algorithm~\ref{alg:generic_auxiliary_csmc} induces a Markov kernel $P(\staAlt_{1:T}|\sta_{1:T})$ which leaves $\target_T$ invariant.
\end{proposition}
\begin{namedProof}[of Proposition~\ref{prop:validity_of_auxiliary_algorithms}]
The extended distribution from \eqref{eq:generic_extended_distribution_with_auxiliary_variables} admits $\target_T(\sta_{1:T})$ as a marginal. Therefore, a valid \gls{MCMC} update for sampling from this extended distribution is given by alternating the following two steps. Given $\sta_{1:T} \in \spaceX^T$,
    \begin{enumerate}
        \item sample $\aux_t \sim \dN(\sta_t + \Phi_t(\sta_{1:T}), \eMat_t(\sta_{t-1:t}))$, for $t = 1,\dotsc, T$;
        \item run a standard \gls{CSMC} algorithm with backward sampling (as in Algorithm~\ref{alg:csmc}) targeting $\targetAlt_T(\sta_{1:T}| \aux_{1:T})$
        but with $\Mutation_t(\sta_t|\sta_{t-1})$, $\Potential_t(\sta_{t-1:t})$, and $\qsemigroup_t(\sta_{t-1:t})$ replaced by $\MutationAlt_t(\sta_t|\sta_{t-1}; \aux_{1:T})$, $\PotentialAlt_t(\sta_{t-2:t}; \aux_{1:T})$ and $\qsemigroupAlt_t(\sta_{t-2:t}; \aux_{1:T})$, and with appropriate adjustments (e.g., of the backward kernels) to account for the possibility that the model may only be second-order Markov.
    \end{enumerate}
    These to steps are equivalent to Algorithm~\ref{alg:generic_auxiliary_csmc}. \hfill \qedwhite
\end{namedProof}

\subsection{Generic marginal algorithm}

Consider the same setting as above but now assume that for any $t \in [T]$, $\genericGradAlt_t \equiv \zeroMat$, so that $\Phi_t(\sta_{1:T}) = \genericGrad_t(\sta_{t-1:t})$ as well as that $\eMat_t(\sta_{t-1:t}) = \eMat_t$ is independent of $\sta_{t-1:t}$. %

Furthermore, assume that $\MutationAlt_t(\sta_t|\sta_{t-1}; \aux_{1:T}) = \MutationAlt_t(\sta_t|\sta_{t-1}; \aux_t)$ only depends on the $t$th auxiliary variable $\aux_t$ and, specifically, is a Gaussian distribution of the following form:
\begin{align}
  \MutationAlt_t(\sta_t|\sta_{t-1}; \aux_t) 
  & \coloneqq \dN(\sta_t; \genericVec_t(\sta_{t-1}) + \hMat_t \aux_t,  \dMat_t),
\end{align}
where $\genericVec_t(\sta) \in \spaceX$ whilst $\hMat_t, \dMat_t \in \reals^{D \times D}$ do not depend on $\sta_{t-1}$ and define
\begin{align}
  \proposalAlt_t^{-n}(\sta_t^{-n}, \aux_t|\sta_t^n; \mathcal{H}_{t-1}) \coloneqq \dN(\aux_t; \sta_t^n + \genericGrad_t(\sta_{t-1:t}^{(n)}), \eMat_t) \prodSubstackAligned{m}{=}{0}{m}{\neq}{n}^N \MutationAlt_t(\sta_t^m|\sta_{t-1}^{a_{t-1}^m}; \aux_t),
\end{align}
where $\mathcal{H}_{t-1}$ is the history of the particle system up to time $t-1$, i.e., all particles and ancestor indices with `time' subscript $s \leq t-1$. By Lemma~\ref{lem:auxiliary_lemma_1} from Appendix~\ref{app:sec:integrating_out_the_auxiliary_variables}, we obtain a closed-form expression for
\begin{align}
  \proposalAlt_t^{-n}(\sta_t^{-n} |\sta_t^n; \mathcal{H}_{t-1}) \coloneqq \int_{\spaceX}\proposalAlt_t^{-n}(\sta_t^{-n}, \aux_t|\sta_t^n; \mathcal{H}_{t-1}) \intDiff \aux_t.
\end{align}

\begin{framedAlgorithm}[generic marginal algorithm] \label{alg:generic_marginal_csmc}
Given $\sta_{1:T} \in \spaceX^T$:
\begin{enumerate}
    \item \label{alg:generic_marginal_csmc:1} for $t = 1, \dotsc, T$, 
    \begin{enumerate}
        \item sample $k_t$ from a uniform distribution on $[N]_0$ and set $\sta_t^{k_t} \coloneqq \sta_t$,
        \item if $t > 1$, set $\smash{a_{t-1}^{k_t} \coloneqq k_{t-1}}$ and sample $a_{t-1}^n = i$ \gls{WP} $\smash{W_{t-1}^i}$, for $n \in [N]_0 \setminus \{k_t\}$,
        
        \item \label{alg:generic_marginal_csmc:mutation} sample $\smash{\sta_t^{-k_t} \sim \proposalAlt_t^{-k_t}(\sta_t^{-k_t} |\sta_t^{k_t}; \mathcal{H}_{t-1})}$\\(e.g.\ by sampling $\smash{\aux_t \sim \dN(\sta_t + \genericGrad_t(\sta_{t-1:t}), \eMat_t)}$ and then $\smash{\sta_t^n \sim \MutationAlt_t(\ccdot|\sta_{t-1}^{a_{t-1}^n}; \aux_t)}$ for $n \in [N]_0 \setminus \{k_t\}$), 
        
        \item \label{alg:generic_marginal_csmc:weight_calculation} for $n \in [N]_0$, set $\smash{w_t^n \propto \qsemigroup_t(\sta_{t-1:t}^{(n)})\proposalAlt_t^{-n}(\sta_t^{-n} |\sta_t^n; \mathcal{H}_{t-1}) }$,%
        
        \item \label{alg:generic_marginal_csmc:weight_normalisation} for $n \in [N]_0$, set $W_t^n \coloneqq  w_t^n / \sum_{m = 0}^N w_t^m$;%
    \end{enumerate}
    
    \item \label{alg:generic_marginal_csmc:2} sample $i \in [N]_0 \setminus \{k_T\}$ \gls{WP} $\dfrac{W_T^i}{1 - W_T^{k_T}}$; set $l_T \coloneqq i$ \gls{WP} $1 \wedge \dfrac{1- W_T^{k_T}}{1 - W_T^i}$; otherwise, set $l_T \coloneqq k_T$;

    \item \label{alg:generic_marginal_csmc:backward_sampling} for $t = T-1, \dotsc, 1$, sample $l_t = i \in [N]_0$ \gls{WP}
          $
          \dfrac{W_t^i \qsemigroup_{t+1}(\sta_t^i, \sta_{t+1}^{l_{t+1}}) }{\sum_{n=0}^N W_t^n \qsemigroup_{t+1}(\sta_t^n, \sta_{t+1}^{l_{t+1}})};
          $
    \item \label{alg:generic_marginal_csmc:4} return $\staAlt_{1:T} \coloneqq (\sta_1^{l_1}, \dotsc, \sta_t^{l_T})$.
\end{enumerate}
\end{framedAlgorithm}

Algorithm~\ref{alg:generic_marginal_csmc} can be implemented in $\bo(N)$ operations because Lemma~\ref{lem:auxiliary_lemma_1} from Appendix~\ref{app:sec:integrating_out_the_auxiliary_variables}  allows us to write the weight in Step~\ref{alg:generic_marginal_csmc:weight_calculation} as
\begin{align}
  w_t^n & \propto 
  \qsemigroup_t(\sta_{t-1:t}^{(n)})\proposalAlt_t^{-n}(\sta_t^{-n} |\sta_t^n; \mathcal{H}_{t-1})\\
  & \propto \qsemigroup_t(\sta_{t-1:t}^{(n)}) \smash{H_{t, \genericGrad_t(\sta_{t-1:t}^{(n)})}(\sta_t^n, \genericVec_t^n, \bar{\sta}_t, \bar{\genericVec}_t)}, \label{eq:generic_marginal_algorithm:weight_1}
  \end{align}
where $\smash{\genericVec_t^n \coloneqq \genericVec_t(\sta_{t-1}^{a_{t-1}^n})}$, $\smash{\bar{\sta}_t \coloneqq \tfrac{1}{N+1}\sum_{m=0}^N \sta_t^m}$, $\smash{\bar{\genericVec}_t \coloneqq \tfrac{1}{N+1}\sum_{m=0}^N \genericVec_t^m}$ and
  \begin{align*}
   \log H_{t, \genericGrad}(\sta, \genericVec, \bar{\sta}, \bar{\genericVec})
   & = \tfrac{1}{2} (\sta - \genericVec)^\T (\dMat_t^{-1} + \gMat_t) (\sta - \genericVec)\\*
  & \qquad - \tfrac{1}{2}N (\sta + \genericGrad)^\T \hMat_t^\T (\dMat_t^{-1} - N \gMat_t) \hMat_t (\sta + \genericGrad)\\*
  & \qquad -  (N+1) (\bar{\sta} - \bar{\genericVec})^\T [\gMat_t (\sta - \genericVec) - (\dMat_t^{-1} - N \gMat_t) \hMat_t(\sta + \genericGrad)]\\*
  & \qquad - (\sta - \genericVec)^\T (\dMat_t^{-1} - N \gMat_t)\hMat_t(\sta + \genericGrad), \label{eq:generic_marginal_algorithm:weight_2}
\end{align*}
with $\smash{\gMat_t \coloneqq (\dMat + N \hMat_t \eMat_t\hMat_t^\T)^{-1} \hMat_t \eMat_t \hMat_t^\T \dMat_t^{-1}
}$ (see \eqref{eq:gmatrix:2} for an alternative expression).

\begin{proposition}[validity of the generic marginal algorithm] \label{prop:validity_of_marginal_algorithms}
 Sampling $\staAlt_{1:T}$ given $\sta_{1:T}$ via Algorithm~\ref{alg:generic_marginal_csmc} induces a Markov kernel $P(\staAlt_{1:T}|\sta_{1:T})$ which leaves $\target_T$ invariant.
\end{proposition}

\begin{namedProof}[of Proposition~\ref{prop:validity_of_marginal_algorithms}]

   We begin with a few observations.
   \begin{enumerate}

    \item %
    Since the unnormalised weights satisfy
   \begin{align}
        w_t^n \propto \qsemigroup_t(\sta_{t-1:t}^{(n)}) \proposalAlt_t^{-n}(\sta_t^{-n} | \sta_t^n; \mathcal{H}_{t-1}),
    \end{align}
    we have that
    \begin{align}
        \proposalAlt_t^{-k_t}(\sta_t^{-k_t} | \sta_t^{k_t}; \mathcal{H}_{t-1}) = \frac{w_t^{k_t}}{w_t^{l_t}} \frac{\qsemigroup_t(\sta_{t-1:t}^{(l_t)})}{\qsemigroup_t(\sta_{t-1:t}^{(k_t)})} \proposalAlt_t^{-l_t}(\sta_t^{-l_t} | \sta_t^{l_t}; \mathcal{H}_{t-1}).
    \end{align}

    \item For a given set of final-time weights $\{W_T^n\}_{n \in [N]_0}$, let $R_T(\ccdot|\ccdot; \mathcal{H}_{T})$ be the $\sum_{n=0}^N W_T^n \delta_n$-invariant Markov kernel used in Step~\ref{alg:generic_marginal_csmc:2} of Algorithm~\ref{alg:generic_marginal_csmc}. That is, sampling $l_T \sim R_T(\ccdot |k_T; \mathcal{H}_{T})$ could be the forced-move update; or, in the more common specification of \gls{CSMC} algorithms \citep{andrieu2010particle}, i.e.\ without the forced-move update, we would simply have $R_T(l_T |k_T; \mathcal{H}_{T}) = W_T^{l_T}$. In either case, it can then be verified that
   \begin{align}
    W_T^{k_T} R_T(l_T|k_T; \mathcal{H}_{T}) = W_T^{l_T} R_T(k_T|l_T; \mathcal{H}_{T}),
   \end{align}
   for any $k_T, l_T \in [N]_0$.

   \item Under Algorithm~\ref{alg:generic_marginal_csmc}, we have the following identities (with probability $1$): $\sta_t = \sta_t^{k_t}$ and $\sta_t' = \sta_t^{l_t}$, for $1 \leq t \leq T$, as well as $a_{t-1}^{k_t} = k_{t-1}$, for any $1 < t \leq T$.

   \end{enumerate}
   Putting these observations together then shows that the Algorithm~\ref{alg:generic_marginal_csmc} targets the following extended distribution (i.e., this is the distribution of all random variables obtained if we first sampled $\sta_{1:T} \sim \target_T$ and then ran Algorithm~\ref{alg:generic_marginal_csmc}):
    \begin{align*}
       &\quad\frac{\target_T(\sta_{1:T})}{(N+1)^T} \delta_{\sta_{1:T}}(\sta_{1:T}^{k_{1:T}})\biggl[\prod_{t=1}^T \proposalAlt_t^{-k_t}(\sta_t^{-k_t} |\sta_t^{k_t}; \mathcal{H}_{t-1})\biggr] \biggl[\prod_{t=2}^T \delta_{k_{t-1}}(a_{t-1}^{k_t}) \prodSubstackAligned{n}{=}{0}{n}{\neq}{k_t}^N W_{t-1}^{a_{t-1}^n}\biggr]\\*
      & \quad\times R_T(l_T|k_T; \mathcal{H}_T) \biggl[\prod_{t=1}^{T-1} \frac{w_t^{l_t} \qsemigroup_{t+1}(\sta_t^{l_t}, \sta_{t+1}^{l_{t+1}})}{\sum_{m=0}^N w_t^m \qsemigroup_{t+1}(\sta_t^m, \sta_{t+1}^{l_{t+1}})}\biggr] \delta_{\sta_{1:T}^{l_{1:T}}}(\staAlt_{1:T})\\
      & = \frac{\target_T(\staAlt_{1:T})}{(N+1)^T} \delta_{\staAlt_{1:T}}(\sta_{1:T}^{l_{1:T}})\biggl[\prod_{t=1}^T \proposalAlt_t^{-l_t}(\sta_t^{-l_t} |\sta_t^{l_t}; \mathcal{H}_{t-1})\biggr]\\*
      & \qquad \times \biggl[\prod_{t=2}^T \frac{w_{t-1}^{a_{t-1}^{l_t}}\qsemigroup_t(\sta_{t-1}^{a_{t-1}^{l_t}}, \sta_t^{l_{t+1}})}{\sum_{m=0}^N w_{t-1}^m \qsemigroup_t(\sta_{t-1}^m, \sta_t^{l_t})} \prodSubstackAligned{n}{=}{0}{n}{\neq}{l_t}^N W_{t-1}^{a_{t-1}^n}\biggr]\\*
      & \qquad \times R_T(k_T|l_T; \mathcal{H}_T) \biggl[\prod_{t=1}^{T-1} \delta_{a_{t-1}^{k_t}}(k_{t-1})\biggr] \delta_{\sta_{1:T}^{k_{1:T}}}(\sta_{1:T}),
    \end{align*}
    where the r.h.s.\ is the distribution obtained if we first sampled $\staAlt_{1:T}\sim \target_T$ and then ran Algorithm~\ref{alg:generic_marginal_csmc} algorithm but with ancestor sampling \citep{lindsten2012ancestor} instead of backward sampling. This is a modified version of the proof technique from \citet{finke2016embedded}.  In other words, if $\sta_{1:T} \sim \target_T$ and if $\staAlt_{1:T}$ is sampled via Algorithm~\ref{alg:generic_marginal_csmc}, then $\staAlt_{1:T} \sim \target_T$. This completes the proof. \hfill \qedwhite
\end{namedProof}

\subsection{Invariance of the algorithms}
We can now easily verify the validity of the `auxiliary' algorithms (\gls{PARTICLEAMALA}, \gls{PARTICLEAMALAPLUS}, \gls{PARTICLEAGRAD}, \gls{PARTICLEAGRADPLUS}, \gls{PARTICLEAPCNL}, \gls{PARTICLEAPCNLPLUS}, and twisted \gls{PARTICLEAGRAD}/\allowbreak\gls{PARTICLEAGRADPLUS}/\allowbreak\gls{PARTICLEAPCNL}/\gls{PARTICLEAPCNLPLUS}) by noting that these are special cases of Algorithm~\ref{alg:generic_auxiliary_csmc}, and the validity of the `marginal' algorithms (\gls{PARTICLEMALA}, \gls{PARTICLEMGRAD}, \gls{PARTICLEPCNL}) by noting that these are special cases of Algorithm~\ref{alg:generic_marginal_csmc}. %

\begin{namedProof}[of Proposition~\ref{prop:particleamala:validity}]
  This follows by taking $\genericGrad_t(\sta_{t-1:t}) \coloneqq \gradientIndicator \tfrac{\delta_t}{2}\nabla_{\sta_t} \log \qsemigroup_t(\sta_{t-1:t})$, $\genericGradAlt_t \equiv \zeroMat$, $\MutationAlt_t(\sta_t|\sta_{t-1}; \aux_{1:T}) = \dN(\sta_t; \aux_t, \tfrac{\delta_t}{2} \iMat)$ and $\eMat_t \equiv \tfrac{\delta_t}{2} \iMat$ %
  in Proposition~\ref{prop:validity_of_auxiliary_algorithms}. \hfill \qedwhite
\end{namedProof}

\begin{namedProof}[of Proposition~\ref{prop:particlemala:validity}]
 This follows from Proposition~\ref{prop:validity_of_marginal_algorithms} with the same setting as in Proposition~\ref{prop:particleamala:validity}. In particular, in this case, $\genericVec_t \equiv \zeroMat$, $\hMat_t = \iMat$, $\dMat_t = \eMat_t = \tfrac{\delta_t}{2} \iMat$. Consequently, \eqref{eq:generic_marginal_algorithm:weight_2} then simplifies to  \eqref{eq:particlemala:weight_factor},
where we have used that $\gMat_t = [\tfrac{\delta_t}{2}(N+1)]^{-1} \iMat = \dMat_t^{-1} / (N+1)$ and $\dMat_t^{-1} - N\gMat_t = \gMat_t$. \hfill \qedwhite
\end{namedProof}

\begin{namedProof}[of Proposition~\ref{prop:particleamalaplus:validity}]
  This follows in the same way as the proof of Proposition~\ref{prop:particleamala:validity} except that now $\genericGradAlt_t(\sta_{t:t+1}) = \gradientIndicator \frac{\delta_t}{2} \nabla_{\sta_t}  \log \qsemigroup_{t+1}(\sta_{t:t+1})$. \hfill \qedwhite
\end{namedProof}

\begin{namedProof}[of Proposition~\ref{prop:particleagrad:validity}]
  This follows in the same way as the proof of Proposition~\ref{prop:particleamala:validity} except that now $\genericGrad_t(\sta_{t-1:t}) \coloneqq \gradientIndicator \tfrac{\delta_t}{2}\nabla_{\sta_t} \log \Potential_t(\sta_{t-1:t})$, and $\MutationAlt_t(\sta_t|\sta_{t-1}; \aux_{1:T}) = \dN(\sta_t; \priorMeanAlt_t(\sta_{t-1}, \aux_t), \priorVarAlt_t(\sta_{t-1}))$, where $\priorMeanAlt_t(\sta_{t-1}, \aux_t)$ and $\priorVarAlt_t(\sta_{t-1})$ are defined in \eqref{eq:particleagrad_proposal_mean} and \eqref{eq:particleagrad_proposal_variance}. \hfill \qedwhite
\end{namedProof}

\begin{namedProof}[of Proposition~\ref{prop:particlemgrad:validity}]
  This follows from Proposition~\ref{prop:validity_of_marginal_algorithms} with the same setting as in Proposition~\ref{prop:particleagrad:validity}. In particular, in this case, $\hMat_t = \kalmanGain_t \coloneqq (\priorVar_t + \tfrac{2}{\delta_t}\iMat)^{-1}  \priorVar_t$, $\dMat_t = \tfrac{\delta_t}{2}\kalmanGain_t$ and $\eMat_t = \tfrac{\delta_t}{2} \iMat$. Consequently,
 \eqref{eq:generic_marginal_algorithm:weight_2} then simplifies to \eqref{eq:particlemgrad:weight_factor}, 
where we have used that $\kalmanGain_t$ is symmetric, that $\hMat_t^\T \dMat_t^{-1} = \dMat_t^{-1} \hMat_t = \tfrac{2}{\delta_t} \iMat$ and hence
\begin{align}
  \dMat_t^{-1} - N\gMat_t
  & = 
  \kalmanGain_t^{-1} \gMat_t = \gMat_t \kalmanGain_t^{-1}.
\end{align}
This completes the proof. \hfill \qedwhite
\end{namedProof}

\begin{namedProof}[of Proposition~\ref{prop:particleagradplus:validity}]
  This follows in the same way as the proof of Proposition~\ref{prop:particleagrad:validity} except that now $\genericGradAlt_t(\sta_{t:t+1}) = \gradientIndicator \frac{\delta_t}{2} \nabla_{\sta_t} \log G_{t+1}(\sta_{t:t+1})$. \hfill \qedwhite
\end{namedProof}

\begin{namedProof}[of Proposition~\ref{prop:twisted_particleagrad:validity}]
  This follows in the same way as the proof of Propositions~\ref{prop:particleagrad:validity} and \ref{prop:particleagradplus:validity}, respectively, but with $\MutationAlt_t(\sta_t|\sta_{t-1}; \aux_{1:T}) = \dN(\sta_t; \priorFactorAlt_t \sta_{t-1} + \priorInterceptAlt_t, \priorVarAlt_t)$. \hfill \qedwhite
\end{namedProof}

\begin{namedProof}[of Proposition~\ref{prop:particleapcnl:validity}]
  This follows in the same way as the proof of Proposition~\ref{prop:particleamala:validity} except that now $\genericGrad_t(\sta_{t-1:t}) \coloneqq \gradientIndicator \tfrac{\delta_t}{2} \preconVar_t(\sta_{t-1}) \nabla_{\sta_t} \log \Potential_t(\sta_{t-1:t})$,  $\eMat_t(\sta_{t-1:t}) \coloneqq \tfrac{\delta_t}{2} \priorVar_t(\sta_{t-1})$ and $\MutationAlt_t(\sta_t|\sta_{t-1}; \aux_{1:T}) = \dN(\sta_t; \priorMeanAlt_t(\sta_{t-1}, \aux_t), \priorVarAlt_t(\sta_{t-1}))$, where $\priorMeanAlt_t(\sta_{t-1}, \aux_t)$ and $\priorVarAlt_t(\sta_{t-1})$ are defined in \eqref{eq:particleapcnl_proposal_mean} and \eqref{eq:particleapcnl_proposal_variance}. \hfill \qedwhite
\end{namedProof}

\begin{namedProof}[of Proposition~\ref{prop:particlepcnl:validity}]
  This follows from Proposition~\ref{prop:validity_of_marginal_algorithms} with the same setting as in Proposition~\ref{prop:particleapcnl:validity}. In particular, in this case, $\hMat_t = \beta_t \iMat$, $\dMat_t = (1-\beta_t) \priorVar_t$ and $\eMat_t = \tfrac{\delta_t}{2} \priorVar_t$. Consequently, \eqref{eq:generic_marginal_algorithm:weight_2} then simplifies to \eqref{eq:particlepcnl:weight_factor}, 
  where we have used that $\eMat \hMat_t^\T \dMat_t^{-1} = \dMat_t^{-1} \hMat_t \eMat = \iMat$ and hence
  \begin{align}
      \dMat_t^{-1} + \gMat_t & = (\beta_t^{-1} + N + 1) \gMat_t,\\
      \dMat_t^{-1} - N\gMat_t
      & = 
     \beta_t^{-1} \gMat_t.
  \end{align}
This completes the proof. \hfill \qedwhite
\end{namedProof}

\begin{namedProof}[of Proposition~\ref{prop:particleapcnlplus:validity}]
  This follows in the same way as the proof of Proposition~\ref{prop:particleapcnl:validity} except that now $\genericGradAlt_t(\sta_{t:t+1}) = \gradientIndicator \frac{\delta_t}{2} \preconVar_t(\sta_t) \nabla_{\sta_t} \log G_{t+1}(\sta_{t:t+1})$. \hfill \qedwhite
\end{namedProof}

\begin{namedProof}[of Proposition~\ref{prop:twisted_particleapcnl:validity}]
  This follows in the same way as the proof of Propositions~\ref{prop:particleapcnl:validity} and \ref{prop:particleapcnlplus:validity}, respectively, but with $\MutationAlt_t(\sta_t|\sta_{t-1}; \aux_{1:T}) = \dN(\sta_t; \priorFactorAlt_t \sta_{t-1} + \priorInterceptAlt_t, \priorVarAlt_t)$. \hfill \qedwhite
\end{namedProof}

\section{Proof of Propositions~\ref{prop:convergence_for_highly_informative_prior} and \ref{prop:convergence_for_weakly_informative_prior}}
\label{app:sec:proof_of_the_interpolation_property}

\subsection{Preliminaries}
For some given $N \in \naturals$, let $\Psi^n$ denote either the \emph{Boltzmann selection function} (with the convention $h^0 \coloneqq 0$):
\begin{align}
 \Psi^n(h^{1:N})
 & \coloneqq 
  \frac{\exp(h^n)}{1 + \sum_{m=0}^N \exp(h^m)},
\end{align}
or the \emph{Rosenbluth--Teller selection function}:
\begin{align}
 \Psi^n(h^{1:N})
 & \coloneqq 
 \begin{dcases}
  \frac{\exp(h^n)}{1 + \sum_{m=1}^N \exp(h^m) - 1 \wedge \exp(h^n)}, & \text{if $n > 0$,}\\
  1- \sum_{l=1}^N \Psi^l(h^{1:N}), & \text{if $n = 0$.}
  \end{dcases}
\end{align}
In either case, $\Psi^n$ is Lipschitz continuous with constant denoted $[\Psi^n]_\lip$.

\subsection{Marginal MCMC kernels in the special case: \texorpdfstring{$T = 1$}{T = 1}}

For the moment, we assume that $T = 1$. To simplify the notation, we drop the `time' subscripts $t = 1$. With this convention, for some bounded and differentiable $\Potential: \reals^D \to (0, \infty)$, define
\begin{align}
  \pi(\sta) & \propto \dN(\sta; \priorMean, \priorVar) \Potential(\sta).
\end{align}
The $\pi$-invariant Markov kernels induced by the (non-auxiliary variable based) algorithms discussed in this work can then be written as
\begin{align}
  P_\placeholder(\diff \staAlt|\sta)
  & = \sum_{l = 0}^N  \int_{\spaceX^{N+2}} \delta_{\sta}(\intDiff \sta^0) q_\placeholder^{-0}(\diff \sta^{-0}|\sta^0) \Psi^l(\{h_\placeholder^n(\sta^{0:N})\}_{n=1}^N) \delta_{\sta^l}(\diff \staAlt),
\end{align}
where have appealed to symmetry to always place the reference `path' in position $0$, and with
\begin{align}
  h_\placeholder^n(\sta^{0:N}) 
  & \coloneqq \log q_\placeholder^{-n}(\sta^{-n}|\sta^n) - \log q_\placeholder^{-0}(\sta^{-0}|\sta^0),\\
  q_\placeholder^{-n}(\sta^{-n}|\sta^n)
  & = \dN(\sta^{-n}; \priorMean_{\placeholder}(\sta^n), \priorVar_{\placeholder}),
\end{align}
where $\priorMean_{\placeholder}(\sta^n) \in \reals^{ND}$ is a suitable mean vector (which may depend on $\sta^n \in \reals^D$), $\priorVar_{\placeholder} \in \reals^{(ND) \times (ND)}$ a suitable variance, and where we again slightly abuse notation to let $\sta^{-n}$ represent both the tuple $(\sta^0, \dotsc, \sta^{n-1}, \sta^{n+1}, \dotsc, \sta^N)$ and its vectorised form
\begin{align}
  \sta^{-n} \coloneqq \myVec(\sta^{-n}) =  
  \begin{bmatrix}
    \sta^0\\
    \vdots\\
    \sta^{n-1}\\
    \sta^{n+1}\\
    \vdots\\
    \sta^N
  \end{bmatrix} \in \reals^{ND}.
\end{align}
Additionally, `$\placeholder$' is a placeholder for `$\csmc$', `$\particlemala$', or `$\particlemgrad$'. Specifically, by the developments from Section~\ref{app:sec:integrating_out_the_auxiliary_variables} (Lemma~\ref{lem:auxiliary_lemma_1} and its proof), and recalling that the block matrix operator $\myBlockMat{N}$ was defined in \eqref{eq:definition_of_special_block_matrix},
\begin{align}
  \priorMean_{\particlemgrad}(\sta^n) & = \unitMat_{N \times 1} \otimes [\priorMean + \kalmanGain(\sta^n + \genericGrad(\sta^n) - \priorMean)],\\
  \priorVar_{\particlemgrad} & = \tfrac{\delta}{2} \myBlockMat{N}(\kalmanGain, \kalmanGain^2) = \tfrac{\delta}{2}[\iMat_{N} \otimes \kalmanGain + \unitMat_{N \times N} \otimes \kalmanGain^2],\\
  \priorMean_{\csmc}(\sta^n) & = \unitMat_{N \times 1} \otimes \priorMean,\\
  \priorVar_{\csmc} & = \myBlockMat{N}(\priorVar, \zeroMat_{D \times D}) = \iMat_N \otimes \priorVar,\\
  \priorMean_{\particlemala}(\sta^n) & = \unitMat_{N \times 1} \otimes [\sta^n + \genericGrad(\sta^n) + \genericGradAltAlt(\sta^n)],\\
  \priorVar_{\particlemala} & = \tfrac{\delta}{2}\myBlockMat{N}(\iMat, \iMat) = \tfrac{\delta}{2} [\iMat_{ND} + \unitMat_{N \times N} \otimes \iMat],
\end{align}
where $\genericGrad(\sta) \coloneqq \gradientIndicator \tfrac{\delta}{2} \nabla \log \Potential(\sta)$ and $\genericGradAltAlt(\sta) \coloneqq \gradientIndicator \tfrac{\delta}{2} \nabla \log \Mutation(\sta)$ and with $\kalmanGain \coloneqq (\tfrac{\delta}{2} \priorVar^{-1} + \iMat)^{-1} = \priorVar (\priorVar + \tfrac{\delta}{2}\iMat)^{-1} = (\priorVar + \tfrac{\delta}{2}\iMat)^{-1} \priorVar$. %

Key to our proofs will be the following bound which follows from the triangle inequality and a telescoping-sum decomposition (here: $\placeholderAlt$ and $\placeholderAltAlt$ are again placeholders which take values in $\{\csmc, \particlemala, \particlemgrad\}$): 
\begin{align}
\MoveEqLeft \lVert P_{\placeholderAlt}(\ccdot|\sta) - P_{\placeholderAltAlt}(\ccdot|\sta) \rVert_\tv\\
& \leq \lVert q_{\placeholderAlt}^{-0}(\ccdot|\sta) - q_{\placeholderAltAlt}^{-0}(\ccdot|\sta) \rVert_\tv\\
& \quad + \sum_{l = 0}^N \int_{\spaceX^{N+1}} \delta_{\sta}(\diff \sta^0) q_{\placeholderAlt}^{-0}(\diff \sta^{-0}|\sta^0) \lvert \Psi^l(\{h_{\placeholderAlt}^n(\sta^{0:N})\}_{n=1}^N) - \Psi^l(\{h_{\placeholderAltAlt}^n(\sta^{0:N})\}_{n=1}^N) \rvert\\
& \leq \sqrt{\kl(q_{\placeholderAlt}^{-0}(\ccdot |\sta) \| q_{\placeholderAltAlt}^{-0}(\ccdot|\sta))}\\
& \quad + \sum_{l = 0}^N [\Psi^l]_\lip \int_{\spaceX^{N+1}} \delta_{\sta}(\diff \sta^0) q_{\placeholderAlt}^{-0}(\diff \sta^{-0}|\sta^0) \sum_{n=1}^N \lvert h_{\placeholderAlt}^n(\sta^{0:N}) - h_{\placeholderAltAlt}^n(\sta^{0:N}) \rvert\\
& \leq C \Bigl[\sqrt{D_{\placeholderAlt, \placeholderAltAlt}^0(\sta)} + \sum_{n=0}^N D_{\placeholderAlt, \placeholderAltAlt}^n(\sta) \Bigr]. \label{eq:generic_tv_bound}
\end{align}
Here, the penultimate line follows from Pinsker's inequality and the Lipschitz continuity of the selection function; $C \geq 0$ is some constant which may depend on these Lipschitz constants and $N$ and $D$; for the last inequality, we have defined
\begin{align}
D_{\placeholder, \placeholderAltAlt}^n(\sta)
& \coloneqq \int_{\spaceX^{N+1}} \delta_{\sta}(\diff \sta^0) q_{\placeholderAlt}^{-0}(\diff \sta^{-0}|\sta^0)  \lvert \log q_{\placeholderAlt}^{-n}(\sta^{-n} | \sta^n) - \log q_{\placeholderAltAlt}^{-n}(\sta^{-n}|\sta^n) \rvert. \label{eq:d_quantity}
\end{align}

\subsection{Proofs of Part~1}

\begin{namedProof}[of Part~\ref{prop:convergence_for_highly_informative_prior:1} of Proposition~\ref{prop:convergence_for_highly_informative_prior}]
  By Assumption~\ref{as:factorisation_over_time}, the model factorises over time and so do the \gls{CSMC} and \gls{PARTICLEMGRAD} algorithms. Hence, without loss of generality, we prove the result in the case that $T = 1$ (and we drop the `time' subscript $t = 1$ hereafter). Throughout the proof, we will also make repeated use of the fact that the eigenvalues of $\kalmanGain_k$ are given by $(2 \lambda_{k, d}) / (2 \lambda_{k, d} + \delta)$, for $d \in [D]$.
  
  For $\varepsilon \geq 1$ the result is trivially true but meaningless. Fix $\varepsilon \in (0, 1)$. 
  \begin{align}
    F_k \coloneqq \bigl\{\sta \in \reals^D \,\big|\, \lVert \sta -\priorMean \rVert_2 \leq \lambda_k^{(1 - \varepsilon) / 2}\bigr\},
  \end{align}
  denote a ball of radius $\smash{\lambda_k^{(1 - \varepsilon) / 2}}$ around $\priorMean$, for any $k \geq 1$. We then have $\pi_k(F_k) = (1 + H_k)^{-1}$, where, letting $F_k^\compl \coloneqq \spaceX \setminus F_k$:
  \begin{align}
    H_k
    & \coloneqq \frac{\int_{F_k^\compl} \Potential(\sta) \dN(\diff \sta; \priorMean, \priorVar_k)}{\int_{F_k} \Potential(\sta) \dN(\diff \sta; \priorMean, \priorVar_k)}\\
    & \leq \frac{\sup_{x \in \spaceX} \Potential(\sta)}{\inf_{\sta \in F_k} \Potential(\sta)} \frac{\int_{F_k^\compl} \dN(\diff \sta; \priorMean, \priorVar_k)}{\int_{F_k} \dN(\diff \sta; \priorMean, \priorVar_k)}\\ 
    & \leq \frac{\sup_{x \in \spaceX} \Potential(\sta)}{\inf_{\sta \in F_k} \Potential(\sta)} \frac{\int_{\spaceX} \dN(\diff \sta; \zeroMat, \iMat) \ind\{\lVert \sta\rVert_2 > \lambda_k^{-\varepsilon/2}\}}{\int_{\spaceX} \dN(\diff \sta; \zeroMat, \iMat) \ind\{\lVert \sta\rVert_2 \leq \lambda_k^{-\varepsilon/2}\}}\\ 
    & \to 0,
  \end{align}
  as $k \to \infty$, where we have used that $G$ is bounded and that $(\inf_{\sta \in F_k} \Potential(\sta))_{k \geq 1}$ is an increasing sequence in $(0, \infty)$ (since $(F_k)_{k \geq 1}$ is decreasing and $F_1$ is compact).

  By the decomposition from \eqref{eq:generic_tv_bound}, all that remains is to control the terms 
  \begin{align}
      \sup_{\sta^0 \in F_k} D_{\csmc, \particlemgrad, k}^n(\sta^0),
  \end{align}
  for arbitrary $n \in [N]_0$. 

  Firstly, by Lemma~\ref{lem:determinant_of_special_block_matrix} from Appendix~\ref{app:sec:integrating_out_the_auxiliary_variables}, letting $\lambda(\priorVar_k) = \{\lambda_{k,1}, \dotsc, \lambda_{k,D}\}$ denote the eigenvalues of $\priorVar_k$ and noting that $\kalmanGain_k$ is simultaneously diagonalisable with $\kalmanGain_k^2$:
  \begin{align}
  \MoveEqLeft \lvert \log(\det(\priorVar_{\csmc, k}))- \log(\det(\priorVar_{\particlemgrad, k}))\rvert \\
  & = \Bigl\lvert \sum_{d=1}^D N \log \lambda_{k, d} - (N-1) \log \Bigl(\frac{\delta \lambda_{k,d}}{2 \lambda_{k,d} + \delta}\Bigr) - \log\left(\frac{\delta \lambda_{k,d}}{2 \lambda_{k,d} + \delta} + \frac{2 \delta N \lambda_{k,d}^2}{(2 \lambda_{k,d} + \delta)^2}\right)\Bigr\rvert\\
  & = \Bigl\lvert \sum_{d=1}^D N \log \Bigl( \frac{2 \lambda_{k, d} + \delta}{\delta}\Bigr) + \log \Bigl(\frac{\delta \lambda_{k,d}}{2 \lambda_{k,d} + \delta}\Bigr) - \log\left(\frac{\delta \lambda_{k,d}}{2 \lambda_{k,d} + \delta} + \frac{2 \delta N \lambda_{k,d}^2}{(2 \lambda_{k,d} + \delta)^2}\right)\Bigr\rvert\\
  & = \Bigl\lvert \sum_{d=1}^D N \log \Bigl( \frac{2 \lambda_{k, d} + \delta}{\delta}\Bigr) + \log \Bigl( \frac{2 \lambda_{k, d} + \delta}{2 \lambda_{k, d} (N+1) + \delta}\Bigr) \Bigr\rvert
  \in \bo(\lambda_k). \label{eq:prop:convergence_for_highly_informative_prior:proof_result:1}
 \end{align}
 
 Secondly, by Lemma~\ref{lem:inverse_of_special_block_matrix} from Appendix~\ref{app:sec:integrating_out_the_auxiliary_variables}, 
 \begin{align}
     \priorVar_{\particlemgrad, k}^{-1} = \frac{2}{\delta} \mathcal{M}_N(\kalmanGain_k^{-1}, -(\iMat + N \kalmanGain_k)^{-1}),
 \end{align}
 and with the conventions that the sum symbol $\sum_i$ is shorthand for $\sum_{i \in [N]_0 \setminus \{n\}}$, that $\sum_j$ is shorthand for $\sum_{j \in [N]_0 \setminus \{n\}}$, that $\sum_{i \neq j}$ is shorthand for $\sum_{j\in [N]_0 \setminus \{n, i\}}$, and again writing $\genericGrad(\sta) = \gradientIndicator \tfrac{\delta}{2}  \nabla \log \Potential(\sta)$ we obtain:
   \begin{align}
    \MoveEqLeft \bigl\lvert (\sta^{-n} - \priorMean_\csmc(\sta^n))^\T \priorVar_{\csmc, k}^{-1} (\sta^{-n} - \priorMean_\csmc(\sta^n))\\
    & \!\!\!\!\!\!\!\! -  (\sta^{-n} - \priorMean_{\particlemgrad, k}(\sta^n))^\T \priorVar_{\particlemgrad, k}^{-1}(\sta^{-n} - \priorMean_{\particlemgrad, k}(\sta^n)) \bigr\rvert\\
    & = \Bigl\lvert \sum_i (\sta^i - \priorMean)^\T \priorVar_k^{-1}(\sta^i - \priorMean) \\
    & \quad- \frac{2}{\delta}\sum_i (\sta^i - \genericGrad(\sta^n) - \priorMean)^\T \kalmanGain_k^{-1}(\sta^i - \genericGrad(\sta^n) - \priorMean) \\
    & \quad+ \frac{2}{\delta}\sum_i\sum_j (\sta^i - \genericGrad(\sta^n) - \priorMean)^\T (\iMat + N \kalmanGain_k)^{-1}(\sta^j - \genericGrad(\sta^n) - \priorMean)\Bigr\lvert\!\!\!\!\!\!\!\!\!\!\!\!\!\!\!\!\\
    & = \Bigl\lvert \frac{2}{\delta} \sum_i \sum_j (\sta^i - \priorMean)^\T (\iMat + N \kalmanGain_k)^{-1} (\sta^j - \priorMean)\\
    & \quad + \frac{2}{\delta} \sum_i (\sta^i - \priorMean)^\T \Bigl(\frac{\delta}{2}\priorVar_k^{-1} - \kalmanGain_k^{-1}\Bigr) (\sta^i - \priorMean)\\
    & \quad + \frac{2}{\delta} \sum_i (\sta^i - \priorMean)^\T \Bigl[\iMat + \Bigl(\frac{4(N-1)}{\delta}-1\Bigr) (\iMat + N \kalmanGain_k)^{-1}\kalmanGain_k\Bigr] (\sta^n - \genericGrad(\sta^n) - \priorMean)\\
    & \quad + N (\sta^n - \genericGrad(\sta^n) - \priorMean)^\T \Bigl[ \kalmanGain_k + \Bigl(\frac{2(N-1)}{\delta} - 1\Bigr) \kalmanGain_k (\iMat + N \kalmanGain_k)^{-1} \kalmanGain_k \Bigr](\sta^n - \genericGrad(\sta^n) - \priorMean) \Bigr\lvert\!\!\!\!\!\!\!\!\!\!\!\!\!\!\!\!\\
    & \leq \frac{2}{\delta} \sum_i \sum_j \lVert \sta^i - \priorMean\rVert_2 \lVert \sta^j - \priorMean\rVert_2 \lVert (\iMat + N \kalmanGain_k)^{-1} \rVert_{2,2} \\
    & \quad + \frac{2}{\delta} \sum_i \lVert \sta^i - \priorMean \rVert_2 \lVert \sta^i - \priorMean \rVert_2  \Bigl\lVert\frac{\delta}{2}\priorVar_k^{-1} - \kalmanGain_k^{-1}\Bigr\rVert_{2,2} \\
    & \quad + \frac{2}{\delta} \sum_i \lVert \sta^i - \priorMean \rVert_2 \lVert \sta^n - \genericGrad(\sta^n) - \priorMean \rVert_2 \Bigl\lVert \iMat + \Bigl(\frac{4 (N - 1)}{\delta}-1\Bigr) (\iMat + N \kalmanGain_k)^{-1} \kalmanGain_k\Bigr\rVert_{2,2}\!\!\!\!\!\!\!\!\\
    & \quad + N \lVert \sta^n - \genericGrad(\sta^n) - \priorMean \rVert_2^2 \Bigl\lVert \kalmanGain_k + \Bigl(\frac{2 (N - 1)}{\delta} - 1\Bigr) \kalmanGain_k (\iMat + N \kalmanGain_k)^{-1} \kalmanGain_k \Bigr\rVert_{2,2}\\
    & \leq C \Bigl[ \sum_i \sum_j \lVert \sta^i - \priorMean\rVert_2 \lVert \sta^j - \priorMean\rVert_2  \\
    & \qquad\quad + (1 +  \lVert \sta^n - \priorMean \rVert_2) \sum_i \lVert \sta^i - \priorMean \rVert_2 \\
    & \qquad\quad + \lambda_k (1 + \lVert \sta^n - \priorMean \rVert_2)^2 \Bigr],\label{eq:prop:convergence_for_highly_informative_prior:proof_result:2}
  \end{align}
  for some constant $C \geq 0$ which only depends on $N$, $\delta$ and $\priorMean$. Here, we have used that all the matrices inside the operator norms are simultaneously diagonalisable with $\priorVar_k$ (so that the operator norms can be bounded above by some function of $\lambda_k$):
  \begin{align}
      \lVert (\iMat + N \kalmanGain_k)^{-1} \rVert_{2,2} &\leq 1, \\
      \Bigl\lVert\frac{\delta}{2}\priorVar_k^{-1} - \kalmanGain_k^{-1}\Bigr\rVert_{2,2} &= 1,\\
      \Bigl\lVert \kalmanGain_k + \Bigl(\frac{2 (N - 1)}{\delta} - 1\Bigr) \kalmanGain_k (\iMat + N \kalmanGain_k)^{-1} \kalmanGain_k \Bigr\rVert_{2,2} &\leq C' \frac{2\lambda_k}{2\lambda_k + \delta} \leq C' \frac{2}{\delta}\lambda_k,\\
      \Bigl\lVert \iMat + \Bigl(\frac{4 (N - 1)}{\delta}-1\Bigr) (\iMat + N \kalmanGain_k)^{-1} \kalmanGain_k\Bigr\rVert_{2,2} & \leq C'',
  \end{align}
  for other constants $C', C'' \geq 0$.

  Furthermore, by definition of $(F_k)_{k \geq 1}$,
  \begin{align}
     \sup_{\sta \in F_k} \lVert \sta -\priorMean\rVert_2 \in \bo(\lambda_k^{(1 - \varepsilon) / 2}).
     \label{eq:prop:convergence_for_highly_informative_prior:bound_on_compact_set1}
  \end{align}
  Consequently, for $i, j \in [N]_0$:
  \begin{align}
    \MoveEqLeft\sup_{\sta^0 \in F_k} \int_{\spaceX^N} \dN(\diff \sta^{-0}; \priorMean_{\csmc}, \priorVar_{\csmc,k}) \lVert \sta^i - \priorMean \rVert_2 \lVert \sta^j - \priorMean \rVert_2\\
    & \in 
    \begin{cases}
      \bo(\lambda_k^{(1-\varepsilon)}), & \text{if $i = j = 0$,}\\
      \bo(\lambda_k^{(2-\varepsilon)/2}), & \text{if either $i = 0$ or $j = 0$,}\\
      \bo(\lambda_k), & \text{if neither $i = 0$ nor $j = 0$,}\\
    \end{cases}
    \label{eq:prop:convergence_for_highly_informative_prior:proof_result:3}
  \end{align}
  as $\lambda_k \to 0$, and where the last two cases follow from the Cauchy--Schwarz inequality. Similarly, for $i \in [N]_0$,
  \begin{align}
    \sup_{\sta^0 \in F_k} \int_{\spaceX^N} \dN(\diff \sta^{-0}; \priorMean_{\csmc}, \priorVar_{\csmc,k}) \lVert \sta^i - \priorMean \rVert_2 %
    & \in 
    \begin{cases}
      \bo(\lambda_k^{(1-\varepsilon)/2}), & \text{if $i = 0$,}\\
      \bo(\lambda_k^{1/2}), & \text{if $i \neq 0$,}
    \end{cases}
    \label{eq:prop:convergence_for_highly_informative_prior:proof_result:4}
  \end{align}
  as $\lambda_k \to 0$. Combining the bounds from \noeqref{eq:prop:convergence_for_highly_informative_prior:proof_result:2}
  \noeqref{eq:prop:convergence_for_highly_informative_prior:proof_result:3} \eqref{eq:prop:convergence_for_highly_informative_prior:proof_result:1}--\eqref{eq:prop:convergence_for_highly_informative_prior:proof_result:4} then shows that 
   \begin{align}
    \MoveEqLeft \sup_{\sta^0 \in F_k} D_{\csmc, \particlemgrad, k}^n(\sta^0) 
    \in \bo(\lambda_k^{(1 - \varepsilon) / 2}),
  \end{align}
  for any $n \in [N]_0$. Plugging these bounds into \eqref{eq:generic_tv_bound} completes the proof. \hfill \qedwhite
\end{namedProof}

\begin{namedProof}[of Part~\ref{prop:convergence_for_weakly_informative_prior:1} of Proposition~\ref{prop:convergence_for_weakly_informative_prior}]
  By Assumption~\ref{as:factorisation_over_time}, the model factorises over time and so do the \gls{PARTICLEMALA} and \gls{PARTICLEMGRAD} algorithms. Hence, without loss of generality, we again only prove the result in the case that $T = 1$ (and we again drop the `time' subscript $t = 1$ hereafter).
    
  For $\varepsilon \geq 1$ the result is trivially true but meaningless. Fix $\varepsilon \in (0, 1)$. Since $\Potential$ is integrable (by Assumption~\ref{as:potential_as_distribution_with_finite_variance}) and since $\pi_k$ is invariant to scaling of $\Potential$ by a positive constant factor, we assume that $\int_\spaceX \Potential(\sta) \intDiff \sta$ = 1, without loss of generality, so that $\Potential$ can be viewed as a density (and we will also use the symbol $\Potential$ to denote the corresponding distribution). Let $\priorMean_\Potential$ and $\priorVar_\Potential$ be mean and variance of $G$ (which exist by Assumption~\ref{as:potential_as_distribution_with_finite_variance}) and define
  \begin{align}
   F_k & \coloneqq  \bigl\{\sta \in \spaceX \,\big|\, \lVert \sta-\priorMean\rVert_2 \vee \sqrt{(\sta-\priorMean_\Potential)^\T \priorVar_\Potential^{-1}(\sta-\priorMean_\Potential)} < \lambda_k^{\varepsilon/2} \bigr\}.
  \end{align}
  We then have $\pi_k(F_k) = (1 + H_k)^{-1}$, where, letting $\mathbf{Y} \sim \Potential$ and letting $F_k^\compl \coloneqq \spaceX \setminus F_k$:
  \begin{align}
    H_k 
    & \coloneqq \frac{\int_{F_k^\compl} \Potential(\sta) \dN(\diff \sta; \priorMean, \priorVar_k)}{\int_{F_k} \Potential(\sta) \dN(\diff \sta; \priorMean, \priorVar_k)}\\
    & \leq \frac{\int_{F_k^\compl} \Potential(\sta)\intDiff \sta}{\inf_{\sta \in F_k}\exp(-\tfrac{1}{2} \lVert \sta - \priorMean\rVert_2^2 \lambda_k^{-1}) \int_{F_k} \Potential(\sta)  \intDiff \sta}\\
    & \leq  \frac{\int_{F_k^\compl} \Potential(\sta)\intDiff \sta}{\int_{F_k} \Potential(\sta)  \intDiff \sta} \exp(\tfrac{1}{2}\lambda_k^{\varepsilon-1})\\
    & = \Prob(\mathbf{Y} \in F_k^\compl) \frac{\exp(\tfrac{1}{2}\lambda_k^{\varepsilon-1})}{\int_{F_k} \Potential(\sta)  \intDiff \sta}\\
    & \leq \Prob\bigl(\sqrt{(\mathbf{Y}-\priorMean_\Potential)^\T \priorVar_\Potential^{-1}(\mathbf{Y}-\priorMean_\Potential)} \geq \lambda_k^{\varepsilon/2}\bigr) \frac{\exp(\tfrac{1}{2}\lambda_k^{\varepsilon-1})}{\int_{F_k} \Potential(\sta)  \intDiff \sta}\\
    & \leq \frac{D}{\lambda_k^{\varepsilon}} \frac{\exp(\tfrac{1}{2}\lambda_k^{\varepsilon-1})}{\int_{F_k} \Potential(\sta)  \intDiff \sta}\\
    & \to 0.
  \end{align}
  The penultimate line follows from the (multidimensional) Chebyshev's inequality and the last line uses that $F_k \to \spaceX$ as $k \to \infty$.

  By the decomposition from \eqref{eq:generic_tv_bound}, all that remains is to control the terms
  \begin{align}
    \sup_{\sta^0 \in F_k} D_{\particlemala, \particlemgrad, k}^n(\sta^0), 
   \end{align}
    for arbitrary $n \in [N]_0$

  Firstly, by Lemma~\ref{lem:determinant_of_special_block_matrix} from Appendix~\ref{app:sec:integrating_out_the_auxiliary_variables}, letting $\lambda(\priorVar_k) = \{\lambda_{k,1}, \dotsc, \lambda_{k,D}\}$ denote the eigenvalues of $\priorVar_k$ and noting that $\kalmanGain_k$ is simultaneously diagonalisable with $\kalmanGain_k^2$:
  \begin{align}
  \MoveEqLeft \lvert \log(\det(\priorVar_{\particlemala}))- \log(\det(\priorVar_{\particlemgrad, k}))\rvert \\
  & = \Bigl\lvert \sum_{d=1}^D N \log\Bigl(\frac{2 \lambda_{k, d} + \delta}{2 \lambda_{k, d}}\Bigr) + \log\Bigl(\frac{2\lambda_{k, d} + \delta}{2 \lambda_{k, d} + \delta / (N+1)}\Bigr)\Bigr\rvert
  \in \bo(\lambda_k^{-1}). \label{eq:prop:convergence_for_weakly_informative_prior:proof_result:1}
 \end{align}
 Secondly, by Lemma~\ref{lem:inverse_of_special_block_matrix} from Appendix~\ref{app:sec:integrating_out_the_auxiliary_variables}, and again with the conventions that $\sum_i$ is shorthand for $\sum_{i \in [N]_0 \setminus \{n\}}$, that $\sum_j$ is shorthand for $\sum_{j \in [N]_0 \setminus \{n\}}$, that $\sum_{i \neq j}$ is shorthand for $\sum_{j\in [N]_0 \setminus \{n, i\}}$, and writing $\genericGrad(\sta) \coloneqq \gradientIndicator \tfrac{\delta}{2}  \nabla \log \Potential(\sta)$ as well as $\genericGradAltAlt_k(\sta) \coloneqq \gradientIndicator \tfrac{\delta}{2} \nabla \log \Mutation_k(\sta) = \gradientIndicator \tfrac{\delta}{2} \priorVar_k^{-1}(\priorMean - \sta)$, so that $\genericGrad(\sta) + \genericGradAltAlt_k(\sta) = \gradientIndicator \tfrac{\delta}{2} \nabla \log \pi_k(\sta)$:
   \begin{align}
    \MoveEqLeft \bigl\lvert (\sta^{-n} - \priorMean_{\particlemala, k}(\sta^n))^\T \priorVar_{\particlemala}^{-1} (\sta^{-n} - \priorMean_{\particlemala, k}(\sta^n)\\
    & \!\!\!\!\!\!\!\! -  (\sta^{-n} - \priorMean_{\particlemgrad, k}(\sta^n))^\T \priorVar_{\particlemgrad, k}^{-1}(\sta^{-n} - \priorMean_{\particlemgrad, k}(\sta^n)) \bigr\rvert\\
    & \leq C \lambda_k^{-1} \Bigl[ \sum_i \sum_j \lVert \sta^i - \priorMean\rVert_2 \lVert \sta^j - \priorMean\rVert_2  + (1 + \lVert \sta^n - \priorMean \rVert_2) \sum_{i=0}^N \lVert \sta^i - \priorMean \rVert_2 \Bigr],\label{eq:prop:convergence_for_weakly_informative_prior:proof_result:2}
  \end{align}
  for some constant $C \geq 0$ which only depends on $N$, $\delta$ and $\priorMean$. %
  Here, we have followed the same steps as for \eqref{eq:prop:convergence_for_highly_informative_prior:proof_result:2} and used that all the matrices inside the operator norms are simultaneously diagonalisable with $\priorVar_k$ (so that the operator norms can be bounded above by some function of $\lambda_k^{-1}$).

 Furthermore, by definition of $F_k$, we have
  \begin{align}
     \sup_{\sta \in F_k} \lVert \sta -\priorMean\rVert_2 \in \bo(\lambda_k^{\varepsilon / 2}),\label{eq:prop:convergence_for_weakly_informative_prior:bound_on_compact_set1}
  \end{align}
  as $\lambda_k \to \infty$. Consequently, for $i, j \in [N]_0$, by the Cauchy--Schwarz inequality:
  \begin{align}
    \MoveEqLeft\sup_{\sta^0 \in F_k} \int_{\spaceX^N} \dN(\diff \sta^{-0}; \priorMean_{\particlemala, k}, \priorVar_{\particlemala}) \lVert \sta^i - \priorMean \rVert_2 \lVert \sta^j - \priorMean \rVert_2\\
    & \in 
    \begin{cases}
      \bo(\lambda_k^\varepsilon), & \text{if $i = j = 0$,}\\
      \bo(\lambda_k^{\varepsilon/2}), & \text{if either $i = 0$ or $j = 0$,}\\
      \bo(1), & \text{if neither $i = 0$ nor $j = 0$,}\\
    \end{cases}
    \label{eq:prop:convergence_for_weakly_informative_prior:proof_result:3}
  \end{align}
  as $\lambda_k \to \infty$. Similarly, for $i \in [N]_0$,
  \begin{align}
    \sup_{\sta^0 \in F_k} \int_{\spaceX^N} \dN(\diff \sta^{-0}; \priorMean_{\particlemala, k}, \priorVar_{\particlemala}) \lVert \sta^i - \priorMean \rVert_2 %
    & \in 
    \begin{cases}
      \bo(\lambda_k^{\varepsilon/2}), & \text{if $i = 0$,}\\
      \bo(1), & \text{if $i \neq 0$,}
    \end{cases}
    \label{eq:prop:convergence_for_weakly_informative_prior:proof_result:4}
  \end{align}
  as $\lambda_k \to \infty$. Combining the bounds from \noeqref{eq:prop:convergence_for_weakly_informative_prior:proof_result:2}
  \noeqref{eq:prop:convergence_for_weakly_informative_prior:proof_result:3} \eqref{eq:prop:convergence_for_weakly_informative_prior:proof_result:1}--\eqref{eq:prop:convergence_for_weakly_informative_prior:proof_result:4} then shows that 
   \begin{align}
    \MoveEqLeft \sup_{\sta^0 \in F_k} D_{\particlemala, \particlemgrad, k}^n(\sta^0) 
    \in \bo(\lambda_k^{- (1 - \varepsilon) / 2}),
  \end{align}
  for any $n \in [N]_0$. Plugging these bounds into \eqref{eq:generic_tv_bound} completes the proof. \hfill \qedwhite
\end{namedProof}

\subsection{Auxiliary MCMC kernels in the special case: \texorpdfstring{$T=1$}{T = 1}}

The $\pi$-invariant Markov kernels induced by the auxiliary-variable based algorithms discussed in this work can then be written as
\begin{align}
  P_\placeholder(\diff \staAlt|\sta)
  & = \sum_{l = 0}^N  \int_{\spaceX^{N+3}} \delta_{\sta}(\intDiff \sta^0) q_\placeholder^{-0}(\diff \sta^{-0} \times \diff \aux| \sta^{-0}, \sta^0) \Psi^l(\{h_\placeholder^n(\sta^{0:N}, \aux)\}_{n=1}^N) \delta_{\sta^l}(\diff \staAlt),
\end{align}
where have appealed to symmetry to always place the reference `path' in position $0$, where `$\placeholder$' is now a placeholder for `$\particleagrad$', `$\particleamala$', or `$\csmc$' and with
\begin{align}
  q_\placeholder^{-n}(\sta^{-n}, \aux|\sta^n) 
  & = q_\placeholder^{-n}(\sta^{-n}|\sta^n) q_\placeholder^{-n}(\aux| \sta^{-n},\sta^n),\\
  h_\placeholder^n(\sta^{0:N}, \aux)
  & \coloneqq \log q_\placeholder^{-n}(\sta^{-n}|\sta^n) - \log q_\placeholder^{-0}(\sta^{-0}|\sta^0)\\
  & \quad + \ind\{\placeholder \neq \csmc\}[\log q_\placeholder^{-n}(\aux| \sta^{-n}, \sta^n)  - \log q_\placeholder^{-0}(\aux| \sta^{-0}, \sta^0)],\\
  q_\placeholder^{-n}(\sta^{-n}|\sta^n)
  & = \dN(\sta^{-n}; \priorMean_{\placeholder}(\sta^n), \priorVar_{\placeholder}),\\
  q_\placeholder^{-n}(\aux| \sta^{-n},\sta^n)
  & = \dN(\aux; \priorMeanAltAlt_{\placeholder}(\sta^{-n}, \sta^n), \priorVarAltAlt_{\placeholder}),
\end{align}
where again $\priorMean_{\placeholder}(\sta^n) \in \reals^{ND}$ and $\priorMeanAltAlt_{\placeholder}(\sta^{-n}, \sta^n) \in \reals^{D}$ are suitable mean vector, and $\priorVar_{\placeholder} \in \reals^{(ND) \times (ND)}$, $\priorVarAltAlt_{\placeholder} \in \reals^{D \times D}$ are suitable covariance variance matrices, and we again write $\sta^{-n} \coloneqq \myVec(\sta^{-n})$. Specifically, 
\begin{align}
  \priorMean_{\particleagrad}(\sta^n) & = \priorMean_{\particlemgrad}(\sta^n)\\ %
  \priorVar_{\particleagrad} & = \priorVar_{\particlemgrad},\\ %
  \priorMeanAltAlt_{\particleagrad}(\sta^{-n}, \sta^n) & =  (\iMat + N\kalmanGain)^{-1}[(N+1) \bar{\sta} + \genericGrad(\sta^n) + N(\kalmanGain-\iMat)\priorMean],\\ 
  \priorVarAltAlt_{\particleagrad} & = \tfrac{\delta}{2}(\iMat + N\kalmanGain)^{-1},\\
  \priorMean_{\particleamala}(\sta^n) & = \priorMean_{\particleamala}(\sta^n),\\ %
  \priorVar_{\particlemala} & = \priorVar_{\particlemala}, \\%\tfrac{\delta}{2}\myBlockMat{N}(\iMat, \iMat) = \tfrac{\delta}{2} [\iMat_{ND} + \unitMat_{N \times N} \otimes \iMat].
  \priorMeanAltAlt_{\particleamala}(\sta^{-n}, \sta^n) & = \bar{\sta} + \tfrac{1}{N+1} \genericGrad(\sta^n),\\ 
  \priorVarAltAlt_{\particleamala} & =\tfrac{\delta}{2(N+1)}\iMat,
\end{align}
by Part~\ref{lem:auxiliary_lemma_1:conditional} of Lemma~\ref{lem:auxiliary_lemma_1} and Lemma~\ref{lem:inverse_of_special_block_matrix} from Appendix~\ref{app:sec:integrating_out_the_auxiliary_variables}. Of course, the standard \gls{CSMC} algorithm does not make use of the auxiliary variable $\aux$, so we extend the space to include $\aux$ with 
\begin{align}
  \priorMeanAltAlt_{\csmc}(\sta^{-n}, \sta^n) & = \priorMeanAltAlt_{\particleagrad}(\sta^{-n}, \sta^n), \label{eq:space_extension_for_csmc:1}\\ 
  \priorVarAltAlt_{\csmc} & = \priorVarAltAlt_{\particleagrad}. \label{eq:space_extension_for_csmc:2}
\end{align}

Key to our proofs will be the following bound which follows by the triangle inequality and a telescoping-sum decomposition (here: $\placeholderAlt$ is again a placeholder which takes values in $\{\csmc, \particlemala\}$ whilst we will always set $\placeholderAltAlt = \particleagrad$ and $q_{\placeholderAlt}^{-m}(\ccdot|\sta)$; and, unless otherwise stated, $q_{\placeholderAltAlt}^{-m}(\ccdot|\sta)$ denote the joint distributions on the space that includes the auxiliary variable $\aux$): 
\begin{align*}
\MoveEqLeft \lVert P_{\placeholderAlt}(\ccdot|\sta) - P_{\placeholderAltAlt}(\ccdot|\sta) \rVert_\tv\\
& \leq \lVert q_{\placeholderAlt}^{-0}(\ccdot|\sta) - q_{\placeholderAltAlt}^{-0}(\ccdot|\sta) \rVert_\tv\\*
& \quad + \sum_{l = 0}^N \int_{\spaceX^{N+2}} \delta_{\sta}(\diff \sta^0) q_{\placeholderAlt}^{-0} (\diff \sta^{-0} \times \diff \aux|\sta^0)\\*[-1.5ex]
& \qquad\qquad \times \lvert \Psi^l(\{h_{\placeholderAlt}^n(\sta^{0:N},\aux)\}_{n=1}^N) - \Psi^l(\{h_{\placeholderAltAlt}^n(\sta^{0:N},\aux)\}_{n=1}^N) \rvert\\*
& \leq \sqrt{\kl(q_{\placeholderAlt}^{-0}(\ccdot |\sta) \| q_{\placeholderAltAlt}^{-0}(\ccdot|\sta))}\\*[-1ex]
& \quad + \sum_{l = 0}^N [\Psi^l]_\lip \int_{\spaceX^{N+2}} \delta_{\sta}(\diff \sta^0) q_{\placeholderAlt}^{-0}(\diff \sta^{-0} \times \diff \aux|\sta^0) \sum_{n=1}^N \lvert h_{\placeholderAlt}^n(\sta^{0:N},\aux) - h_{\placeholderAltAlt}^n(\sta^{0:N},\aux) \rvert\\
& \leq 
 C \Bigl[\smash{\sqrt{D_{\placeholderAlt, \placeholderAltAlt}^0(\sta) + E_{\placeholderAlt, \placeholderAltAlt}^0(\sta)}} + \sum_{n=0}^N D_{\placeholderAlt, \placeholderAltAlt}^n(\sta) + \widetilde{E}_{\placeholderAlt, \placeholderAltAlt}^n(\sta) \Bigr].
 \label{eq:generic_tv_bound_auxiliary}
\end{align*}
Here, the penultimate line follows from Pinsker's inequality and the Lipschitz continuity of the selection function; $C \geq 0$ is some constant which may depend on these Lipschitz constants and on $N$ and $D$; $D_{\placeholderAlt, \placeholderAltAlt}^n(\sta)$ is defined exactly as in the marginal case \eqref{eq:d_quantity}. Furthermore, we have defined
\begin{align}
    E_{\placeholderAlt, \placeholderAltAlt}^n(\sta)
    & \coloneqq \int_{\spaceX^{N+1}} \delta_{\sta}(\diff \sta^0) q_{\placeholderAlt}^{-0}(\diff \sta^{-0} \times \diff \aux |\sta^0)  \lvert \log q_{\placeholderAlt}^{-n}(\aux| \sta^{-n}, \sta^n) - \log q_{\placeholderAltAlt}^{-n}(\aux| \sta^{-n},\sta^n) \rvert.
\end{align}
Finally, if $\placeholderAlt \neq \csmc$ and $\placeholderAltAlt \neq \csmc$, we we have defined
\begin{align}
 \widetilde{E}_{\particleamala, \particleagrad}^n(\sta)
 & \coloneqq E_{\particleamala, \particleagrad}^n(\sta),
\end{align}
whilst
\begin{align}
 \MoveEqLeft \widetilde{E}_{\csmc, \particleagrad}^n(\sta)\\
 & \coloneqq 
 \int_{\spaceX^{N+2}} \delta_{\sta}(\diff \sta^0) q_{\csmc}^{-0}(\diff \sta^{-0} \times \diff \aux|\sta^0) \\[-0.5ex]
   & \qquad \times \lvert \log q_{\particleagrad}^{-n}(\aux| \sta^{-n}, \sta^n) -\log q_{\particleagrad}^{-0}(\aux| \sta^{-0}, \sta^0) \rvert.
\end{align}

\subsection{Proofs of Part~2}

\begin{namedProof}[of Part~\ref{prop:convergence_for_highly_informative_prior:2} of Proposition~\ref{prop:convergence_for_highly_informative_prior}]
  Assume the same setting as in the proof of Part~\ref{prop:convergence_for_highly_informative_prior:1} of Proposition~\ref{prop:convergence_for_highly_informative_prior} with the same definition of $F_k$. 
  
  We proceed by controlling the terms in \eqref{eq:generic_tv_bound_auxiliary}. Note that $D_{\csmc, \particleagrad,k}^n = D_{\csmc, \particlemgrad,k}^n$. Hence, by the arguments from the proof of Part~\ref{prop:convergence_for_highly_informative_prior:1} of Proposition~\ref{prop:convergence_for_highly_informative_prior}, 
  \begin{align}
  \sup_{\sta \in F_k} D_{\csmc, \particleagrad,k}^n(\sta) \in \bo(\lambda_k^{(1-\varepsilon)/2}).
  \end{align}
  Additionally, due to \eqref{eq:space_extension_for_csmc:1}--\eqref{eq:space_extension_for_csmc:2}, $E_{\csmc, \particleagrad}^n(\sta) = 0$. Finally, using similar arguments as in the proofs for the `marginal' algorithm, we can verify that
  \begin{align}
  \sup_{\sta \in F_k} \widetilde{E}_{\csmc, \particleagrad,k}^n(\sta) \in \bo(\lambda_k^{(1-\varepsilon)/2}).
  \end{align}
  This completes the proof. \hfill \qedwhite
\end{namedProof}

\begin{namedProof}[of Part~\ref{prop:convergence_for_weakly_informative_prior:2} of Proposition~\ref{prop:convergence_for_weakly_informative_prior}]
  Assume the same setting as in the proof ofPart~\ref{prop:convergence_for_weakly_informative_prior:1} of Proposition~\ref{prop:convergence_for_weakly_informative_prior} with the same definition of $F_k$. 
  
  We proceed by controlling the terms in \eqref{eq:generic_tv_bound_auxiliary}. Note that $D_{\particleamala, \particleagrad,k}^n = D_{\particlemala, \particlemgrad,k}^n$. Hence, by the arguments from the proof Part~\ref{prop:convergence_for_weakly_informative_prior:1} of Proposition~\ref{prop:convergence_for_weakly_informative_prior}, 
  \begin{align}
  \sup_{\sta \in F_k} D_{\particleamala, \particleagrad,k}^n(\sta) \in \bo(\lambda_k^{-(1-\varepsilon)/2}).
  \end{align}
  Additionally, using similar arguments as in the proofs for the `marginal' algorithm, we can verify that
  \begin{align}
  \sup_{\sta \in F_k} E_{\particleamala, \particleagrad,k}^n(\sta) \in \bo(\lambda_k^{-(1-\varepsilon)}).
  \end{align}
  This completes the proof. \hfill \qedwhite
\end{namedProof}

\section{Step-size adaptation}
\label{app:sec:step_size_adaptation}

All our algorithms involve the calibration of several step sizes $\delta_t$, one for each time step. To calibrate these, we implement a routine that recursively increases or decreases $\delta_t$ if the running average of the acceptance rate $\alpha_t$ (i.e., the relative frequency with which $\sta_t$ is updated) is respectively above or below a pre-specified target acceptance rate (in our experiments, we picked this to be $\alpha^* = \SI{75}{\percent}$). The only exception to this lies in the twisted algorithms of Section~\ref{subsec:twisted_particleagrad} which we calibrate using a single step-size $\delta$ (so that $\delta = \delta_1 = \dotsc = \delta_T$), and for which the target relates to the overall acceptance rate averaged across time steps. The reason for this difference stems from the fact that the twisting causes the acceptance rate at time $s$ additionally depend on \emph{future} auxiliary variables $\aux_t$, and therefore the future step-size parameters $\delta_t$ (for $t > s$), thereby making the behaviour per time-step harder to control. In practice, our calibration of the twisted \gls{PARTICLEAGRAD} is therefore more similar to that of \gls{AGRAD} than that of our other algorithms. The adaptation procedure is summarised in the following algorithm.

\begin{framedAlgorithm}[step-size adaptation] \label{alg:adaptation}~
    \begin{enumerate}
        \item Initialise the trajectory $\sta_{1:T}[0]$, the initial step sizes $\delta_t[0]$ (for $t \in [T]$), the initial learning rate $\rho[0] = \frac{1}{2}$.
        \item Initialise the history of accepted time steps $A \coloneqq (A_{w,t}) \in \{0, 1\}^{W \times T}$, with $0$ everywhere.
        \item For $k = 1, \dotsc, K$, 
        \begin{enumerate}
            \item sample $\sta_{1:T}[k] \sim P(\ccdot | \sta_{1:T}[k-1])$, where $P$ denotes the Markov kernel induced by one of the algorithms discussed in this work with step sizes $\delta_{1:T}$ set equal to $\delta_{1:T}[k-1]$,
            \item roll the array $A$ by one: set $A_{2:\min\{W, k\},t} \coloneqq A_{1:\min\{W - 1, k - 1\}, t}$, and $A_{1,t} = \ind\{\sta_t[k] = \sta_t[k-1]\}$, for $t \in [T]$,
            \item compute $\alpha_{t} = \frac{1}{\min\{W, k\}} \sum_{w=1}^{\min\{W, k\}} A_{w,t}$, for $t \in [T]$,
            \item if $\lvert\alpha_{t} - \alpha^*\rvert < \sigma$ then keep $\delta_t[k] = \delta_t[k-1]$ unchanged;\\
            otherwise, set
            \begin{align}
                \delta_t[k] \coloneqq \delta_t[k-1] + \max\{k^{\gamma} \rho, \rho_{\min}\} (\alpha_t - \alpha) / \alpha^*.
            \end{align}
        \end{enumerate}
    \end{enumerate}
\end{framedAlgorithm}
In our experiments, we took $\sigma = \SI{5}{\percent}$, $K = \num{10000}$, $\delta_t[0] = 10^{-2}$, $W = \num{100}$, $\rho = \frac{1}{2}$, $\rho_{\min} = 10^{-3}$, $\gamma = -\frac{1}{2}$.

\section{Additional experimental results}
\label{app:sec:additional_experimental_results}

In this section, we provide additional simulation results for the multivariate stochastic volatility model experiments from Section~\ref{sec:simulations}.

\subsection{Calibrated step sizes and acceptance rates}
\label{app:subsec:calibrated_step_sizes_and_acceptance_rates}

\begin{figure}
  \centering
  \includegraphics[]{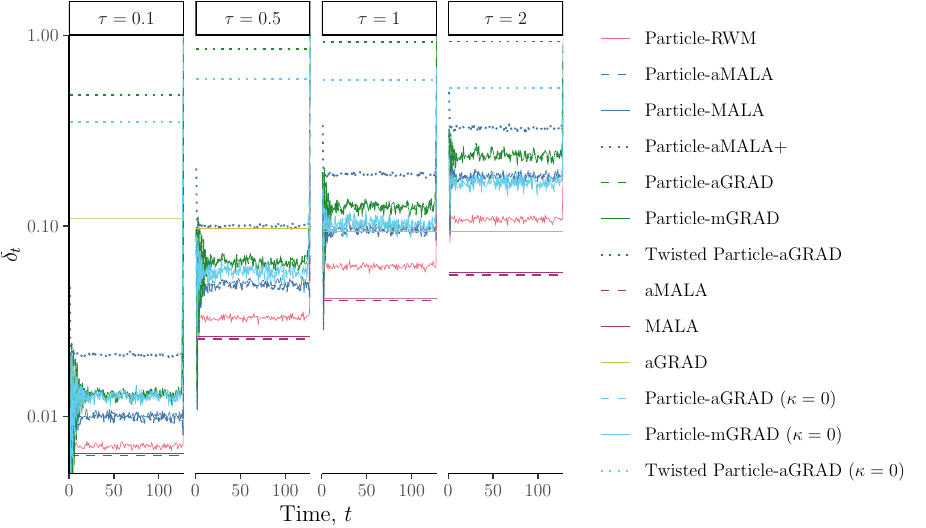}
  \caption{Adaptation of the step-size parameters $\delta_t$, averaged across all four chains and all five simulated data sets (per value of $\tau$) in the multivariate stochastic volatility model.}
  \label{fig:stoch_vol_delta}
\end{figure}

\begin{figure}
  \centering
  \includegraphics[]{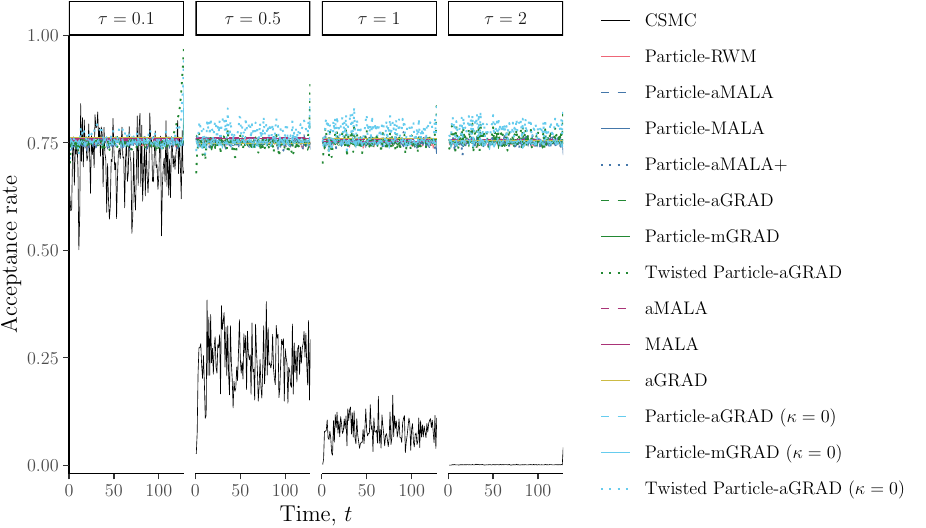}
  \caption{Acceptance rates (i.e., relative frequencies with which states are updated), averaged across all four chains and all five simulated data sets (per value of $\tau$) in the multivariate stochastic volatility model.}
  \label{fig:stoch_vol_acceptance_rate}
\end{figure}

Recall that the step sizes $\delta_t$ were calibrated to achieve an acceptance rate of \SI{75}{\percent}. Here, the `acceptance rate' at time $t$ refers to the relative frequency with which the state $\sta_t$ is updated. The calibrated step sizes are shown in Figure~\ref{fig:stoch_vol_delta}; and the corresponding acceptance rates are shown in Figure~\ref{fig:stoch_vol_acceptance_rate}.

The results are averaged over the four chains and five simulated data sets. We do not report \gls{CSMC} as it does not require calibration. All methods consistently resulted in acceptance rates close to the target \SI{75}{\percent}. Only the twisted \gls{PARTICLEAGRAD} algorithm showed more instability as the informativeness of the prior decreased: this is because, contrary to the methods, only a single step-size is used for all time steps, so calibrating for the informativeness of individual observations is not feasible. This seems to hint to the fact that the twisted \gls{PARTICLEAGRAD}, \emph{under our proposed calibration,} is less robust than alternatives to heterogeneous levels of informativeness.

\subsection{Breakdown of CSMC, aMALA and MALA}
\label{app:subsec:breakdown}

In this section, we illustrate the breakdown of \gls{CSMC}, \gls{AMALA} and \gls{MALA}. 

\begin{figure}
  \centering
  \includegraphics[]{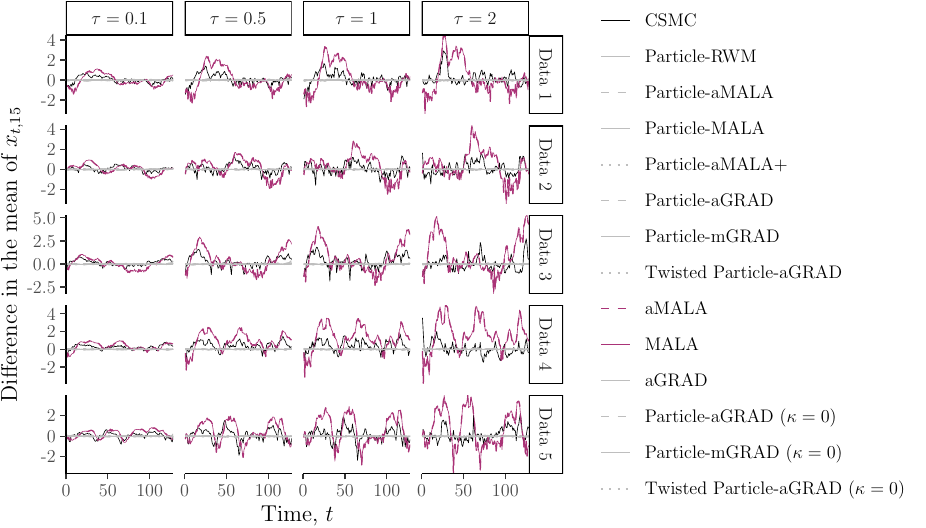}
  \caption{Estimated posterior mean of $x_{t, 15}$ minus the estimated posterior mean of $x_{t, 15}$ under the \gls{AGRAD} algorithm, averaged across all four chains for each of the five simulated data sets (per value of $\tau$) in the multivariate stochastic volatility model. The figure shows that the estimated posterior means of \gls{CSMC}, \gls{AMALA} and \gls{MALA} differ substantially from those of all the other algorithms.}
  \label{fig:stoch_vol_marginals}
\end{figure}

Firstly, Figure~\ref{fig:stoch_vol_marginals} illustrates that the estimates of the marginal posterior means of $x_{t,15}$ (the \num{15}th component of the state at time $t$) produced by \gls{CSMC}, \gls{AMALA} and \gls{MALA} differ substantially from those produced by all the other algorithms. We emphasise that the \num{15}th component was arbitrarily chosen as an example and is representative of the other components.

\begin{figure}
  \centering
  \includegraphics[]{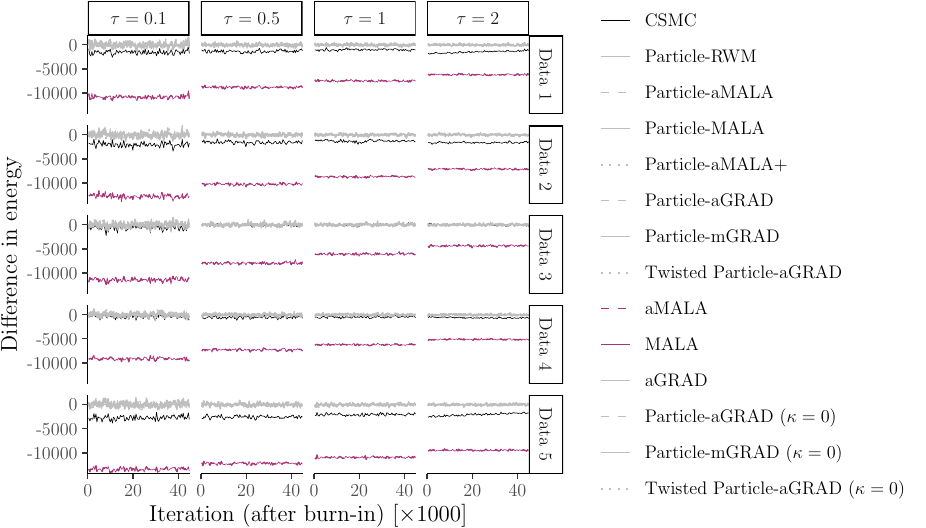}
  \caption{Energy (i.e., $\log \target_T(\sta_{1:T}[i]) + \mathrm{const}$, where $\sta_{1:T}[i]$ is the sample from the $i$th iteration after burn-in) minus the energy under the \gls{AGRAD} algorithm, averaged across all four chains for each of the five simulated data sets (per value of $\tau$) in the multivariate stochastic volatility model. The figure shows that the energy traces of \gls{CSMC}, \gls{AMALA} and \gls{MALA} differ substantially from those of all the other algorithms.}
  \label{fig:stoch_vol_energy_trace}
\end{figure}

Secondly, Figure~\ref{fig:stoch_vol_energy_trace} illustrates that the energy traces of \gls{CSMC}, \gls{AMALA} and \gls{MALA} differ substantially from those of all the other algorithms. Here, the energy is defined as $\log \target_T(\sta_{1:T}[i])$, where $\sta_{1:T}[i]$ is the sample from the $i$th iteration after burn-in. Such energy traces serve as a visual illustration of both stationarity and mixing speed: if the energy trace of a sampler differs too much from the others, or is not consistent across the independent Markov chains we used, the sampler is unlikely to perform correctly.

\subsection{Effective sample sizes}
\label{app:subsec:ess}

\begin{figure}
  \centering
  \includegraphics[]{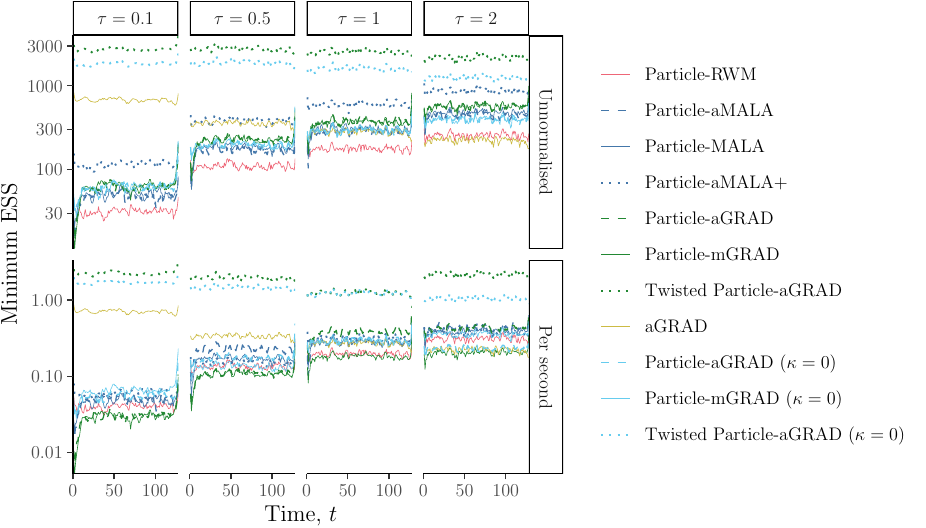}
  \caption{Minimum \gls{ESS} and \gls{ESS} per second averaged across all four chains and all five simulated data sets (per value of $\tau$) in the multivariate stochastic volatility model.}
  \label{fig:stoch_vol_ess_by_time_min}
\end{figure}

\begin{figure}
  \centering
  \includegraphics[]{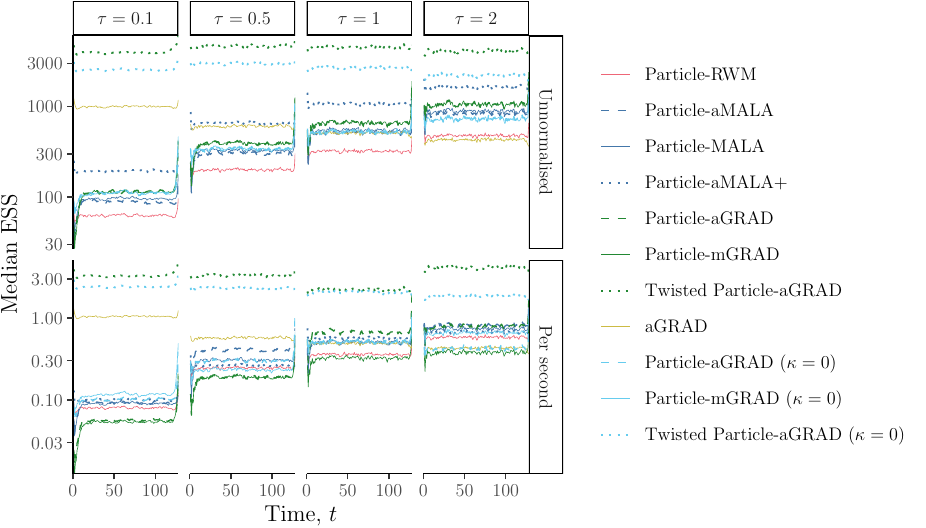}
  \caption{Medium \gls{ESS} and \gls{ESS} per second averaged across all four chains and all five simulated data sets (per value of $\tau$) in the multivariate stochastic volatility model.}
  \label{fig:stoch_vol_ess_by_time_med}
\end{figure}

\begin{figure}
  \centering
  \includegraphics[]{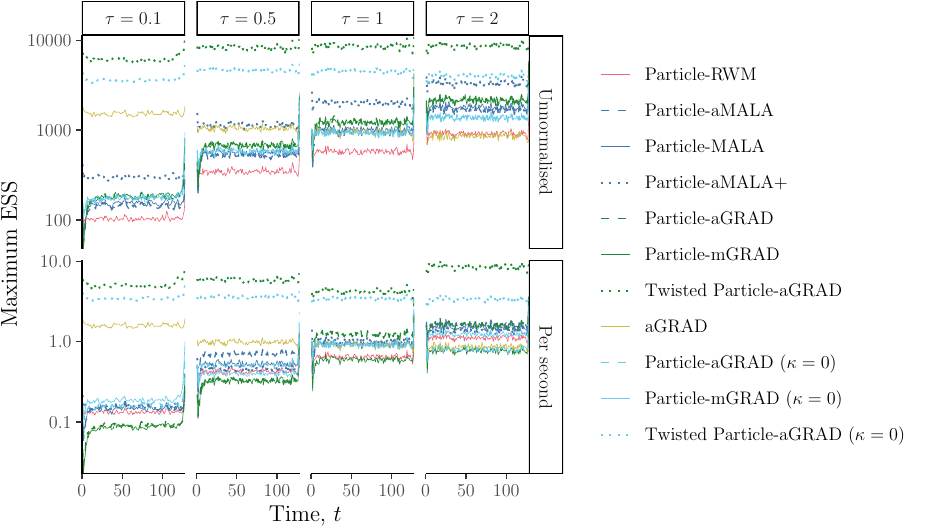}
  \caption{Maximum \gls{ESS} and \gls{ESS} per second averaged across all four chains and all five simulated data sets (per value of $\tau$) in the multivariate stochastic volatility model.}
  \label{fig:stoch_vol_ess_by_time_max}
\end{figure}

In this section, in Figures~\ref{fig:stoch_vol_ess_by_time_min}--\ref{fig:stoch_vol_ess_by_time_max} report the minimum, median and maximum \gls{ESS} and \gls{ESS} per second (averaged across all four chains and all five simulated data sets) individually for each time step $t = 1,\dotsc, T$.

\subsection{Autocorrelation}

\begin{figure}
  \centering
  \includegraphics[]{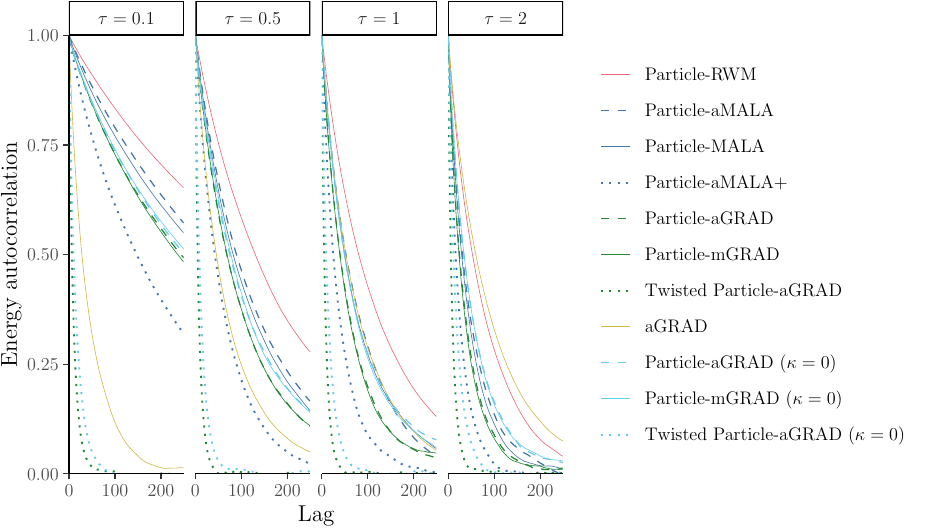}
  \caption{Autocorrelation of the energy from Figure~\ref{fig:stoch_vol_energy_trace} in the multivariate stochastic volatility model.}
  \label{fig:stoch_vol_energy_autocorrelation}
\end{figure}

Figure~\ref{fig:stoch_vol_energy_autocorrelation} shows the autocorrelation \citep[corrected using][]{Vehtari2021rank} of the energy from Figure~\ref{fig:stoch_vol_energy_trace}.
This serves as a visual confirmation of the statistical performance of the different algorithms considered under several prior dispersion regimes: as expected, the twisted \gls{PARTICLEAGRAD} dominates all other alternatives, while \gls{PARTICLEAMALAPLUS} dominates other alternatives, including \gls{AGRAD} as soon as the prior variance is large enough, followed by \gls{PARTICLEAGRAD}/\gls{PARTICLEAGRAD}, and then by \gls{PARTICLEAMALA}/\gls{PARTICLEMALA}, with \gls{PARTICLERWM} being the least efficient.

%% file: main.bbl
\begin{thebibliography}{}

\bibitem[Andrieu et~al., 2010]{andrieu2010particle}
Andrieu, C., Doucet, A., and Holenstein, R. (2010).
\newblock Particle {M}arkov chain {M}onte {C}arlo methods.
\newblock {\em Journal of the Royal Statistical Society: Series B (Statistical
  Methodology)}, 72(3):269--342.
\newblock With discussion.

\bibitem[Andrieu et~al., 2018]{andrieu2018uniform}
Andrieu, C., Lee, A., and Vihola, M. (2018).
\newblock Uniform ergodicity of the iterated conditional {SMC} and geometric
  ergodicity of particle {Gibbs} samplers.
\newblock {\em Bernoulli}, 24(2):842--872.

\bibitem[Andrieu and Vihola, 2016]{andrieu2016establishing}
Andrieu, C. and Vihola, M. (2016).
\newblock Establishing some order amongst exact approximations of {MCMC}s.
\newblock {\em Annals of Applied Probability}, 26(5):2661--2696.

\bibitem[Besag, 1994]{besag1994representations}
Besag, J.~E. (1994).
\newblock Contribution to the discussion on `{R}epresentations of knowledge in
  complex systems' by {G}renander, {U} and {M}iller, {M}.\ {I}.\@.
\newblock {\em Journal of the Royal Statistical Society: Series B (Statistical
  Methodology)}, 56(4):549--581.

\bibitem[Ceperley and Dewing, 1999]{ceperley1999penalty}
Ceperley, D.~M. and Dewing, M. (1999).
\newblock The penalty method for random walks with uncertain energies.
\newblock {\em The Journal of Chemical Physics}, 110(20):9812--9820.

\bibitem[Chopin and Singh, 2013]{chopin2013particle}
Chopin, N. and Singh, S.~S. (2013).
\newblock On particle {G}ibbs sampling.
\newblock {\em arXiv e-prints}, arXiv:1304.1887v1.

\bibitem[Corenflos et~al., 2022]{corenflos2022sequentialized}
Corenflos, A., Chopin, N., and S{\"a}rkk{\"a}, S. (2022).
\newblock De-sequentialized {M}onte {C}arlo: {A} parallel-in-time particle
  smoother.
\newblock {\em Journal of Machine Learning Research}, 23(283):1--39.

\bibitem[Corenflos and S{\"a}rkk{\"a}, 2023]{corenflos2023auxiliary}
Corenflos, A. and S{\"a}rkk{\"a}, S. (2023).
\newblock Auxiliary {MCMC} and particle {Gibbs} samplers for parallelisable
  inference in latent dynamical systems.
\newblock {\em arXiv preprint arXiv:2303.00301}.

\bibitem[Cotter et~al., 2013]{cotter2013crank}
Cotter, S.~L., Roberts, G.~O., Stuart, A.~M., and White, D. (2013).
\newblock {MCMC} methods for functions: {M}odifying old algorithms to make them
  faster.
\newblock {\em Statistical Science}, 28(3):424--446.

\bibitem[Fearnhead and Meligkotsidou, 2016]{fearnhead2016augmentation}
Fearnhead, P. and Meligkotsidou, L. (2016).
\newblock Augmentation schemes for particle {MCMC}.
\newblock {\em Statistics and Computing}, 26:1293--1306.

\bibitem[Finke, 2015]{finke2015extended}
Finke, A. (2015).
\newblock {\em On Extended State-Space Constructions for {M}onte {C}arlo
  Methods}.
\newblock PhD thesis, Department of Statistics, University of Warwick, UK.

\bibitem[{Finke} et~al., 2016]{finke2016embedded}
{Finke}, A., {Doucet}, A., and {Johansen}, A.~M. (2016).
\newblock {On embedded hidden Markov models and particle Markov chain Monte
  Carlo methods}.
\newblock {\em arXiv e-prints}, arXiv:1610.08962.

\bibitem[Finke and Thiery, 2023]{finke2023conditional}
Finke, A. and Thiery, A.~H. (2023).
\newblock {C}onditional sequential {M}onte {C}arlo in high dimensions.
\newblock {\em The Annals of Statistics}, 51(2):437--463.

\bibitem[Guarniero et~al., 2017]{guarniero2017iterated}
Guarniero, P., Johansen, A.~M., and Lee, A. (2017).
\newblock The iterated auxiliary particle filter.
\newblock {\em Journal of the American Statistical Association},
  112(520):1636--1647.

\bibitem[Hastings, 1970]{hastings1970monte}
Hastings, W.~K. (1970).
\newblock {M}onte {C}arlo sampling methods using {M}arkov chains and their
  applications.
\newblock {\em Biometrika}, 57(1):97--109.

\bibitem[Henderson and Searle, 1981]{henderson1981deriving}
Henderson, H.~V. and Searle, S.~R. (1981).
\newblock On deriving the inverse of a sum of matrices.
\newblock {\em SIAM Review}, 23(1):53--60.

\bibitem[Heng et~al., 2020]{heng2020controlledSMC}
Heng, J., Bishop, A.~N., Deligiannidis, G., and Doucet, A. (2020).
\newblock {Controlled sequential Monte Carlo}.
\newblock {\em The Annals of Statistics}, 48(5):2904 -- 2929.

\bibitem[Kalman, 1960]{kalman1960new}
Kalman, R.~E. (1960).
\newblock A new approach to linear filtering and prediction problems.
\newblock {\em Journal of Basic Engineering}, 82:35--45.

\bibitem[Karjalainen et~al., 2023]{karjalainen2023mixing}
Karjalainen, J., Lee, A., Singh, S.~S., and Vihola, M. (2023).
\newblock Mixing time of the conditional backward sampling particle filter.
\newblock {\em arXiv e-prints}, arXiv:2312.17572.

\bibitem[Karppinen et~al., 2023]{Karppinen2023bridge}
Karppinen, S., Singh, S.~S., and Vihola, M. (2023).
\newblock Conditional particle filters with bridge backward sampling.
\newblock {\em Journal of Computational and Graphical Statistics}, 0(0):1--15.

\bibitem[Karppinen and Vihola, 2021]{karppinen2021conditional}
Karppinen, S. and Vihola, M. (2021).
\newblock Conditional particle filters with diffuse initial distributions.
\newblock {\em Statistics and Computing}, 31:1--14.

\bibitem[Lee et~al., 2020]{lee2020coupled}
Lee, A., Singh, S.~S., and Vihola, M. (2020).
\newblock Coupled conditional backward sampling particle filter.
\newblock {\em Annals of Statistics}, 48(5):3066--3089.

\bibitem[Lindsten et~al., 2015]{lindsten2015uniform}
Lindsten, F., Douc, R., and Moulines, E. (2015).
\newblock Uniform ergodicity of the particle {G}ibbs sampler.
\newblock {\em Scandinavian Journal of Statistics}, 42(3):775--797.

\bibitem[Lindsten et~al., 2017]{lindsten2017divide}
Lindsten, F., Johansen, A.~M., Naesseth, C.~A., Kirkpatrick, B., Sch{\"o}n,
  T.~B., Aston, J.~A., and Bouchard-C{\^o}t{\'e}, A. (2017).
\newblock Divide-and-conquer with sequential {M}onte {C}arlo.
\newblock {\em Journal of Computational and Graphical Statistics},
  26(2):445--458.

\bibitem[Lindsten et~al., 2012]{lindsten2012ancestor}
Lindsten, F., Jordan, M.~I., and Sch\"{o}n, T.~B. (2012).
\newblock Ancestor sampling for particle {G}ibbs.
\newblock In {\em {P}roceedings of the 2012 {C}onference on {N}eural
  {I}nformation {P}rocessing {S}ystems}, Lake Tahoe, NV.

\bibitem[Liu, 1996]{liu1996peskun}
Liu, J.~S. (1996).
\newblock Peskun's theorem and a modified discrete-state {G}ibbs sampler.
\newblock {\em Biometrika}, 83(3):681--682.

\bibitem[Livingstone and Zanella, 2022]{livingstone2022barker}
Livingstone, S. and Zanella, G. (2022).
\newblock The {B}arker proposal: {C}ombining robustness and efficiency in
  gradient-based {MCMC}.
\newblock {\em Journal of the Royal Statistical Society: Series B (Statistical
  Methodology)}, 84(2):496--523.

\bibitem[Malory, 2021]{malory2021bayesian}
Malory, S. (2021).
\newblock {\em Bayesian inference for stochastic processes}.
\newblock PhD thesis, Lancaster University.

\bibitem[Metropolis et~al., 1953]{metropolis1953equation}
Metropolis, N., Rosenbluth, A.~W., Rosenbluth, M.~N., Teller, A.~H., and
  Teller, E. (1953).
\newblock Equation of state calculations by fast computing machines.
\newblock {\em Journal of Chemical Physics}, 21(6):1087--1092.

\bibitem[Murray et~al., 2013]{murray2013disturbance}
Murray, L.~M., Jones, E.~M., and Parslow, J. (2013).
\newblock On disturbance state-space models and the particle marginal
  {M}etropolis--{H}astings sampler.
\newblock {\em SIAM/ASA Journal on Uncertainty Quantification}, 1(1):494--521.

\bibitem[{Nicholls} et~al., 2012]{nicholls2012coupled}
{Nicholls}, G.~K., {Fox}, C., and {Muir Watt}, A. (2012).
\newblock {Coupled MCMC with a randomized acceptance probability}.
\newblock {\em arXiv e-prints}, arXiv:1205.6857.

\bibitem[Rhodes and Gutmann, 2022]{rhodes2022enhanced}
Rhodes, B. and Gutmann, M. (2022).
\newblock Enhanced gradient-based {MCMC} in discrete spaces.
\newblock {\em arXiv e-prints}, arXiv:2208.00040.

\bibitem[Roberts et~al., 1997]{roberts1997weak}
Roberts, G.~O., Gelman, A., and Gilks, W.~R. (1997).
\newblock Weak convergence and optimal scaling of random walk {M}etropolis
  algorithms.
\newblock {\em The Annals of Applied Probability}, 7(1):110--120.

\bibitem[Roberts and Rosenthal, 1998]{roberts1998optimal}
Roberts, G.~O. and Rosenthal, J.~S. (1998).
\newblock Optimal scaling of discrete approximations to {L}angevin diffusions.
\newblock {\em Journal of the Royal Statistical Society: Series B (Statistical
  Methodology)}, 60(1):255--268.

\bibitem[Roberts and Rosenthal, 2001]{roberts2001optimal}
Roberts, G.~O. and Rosenthal, J.~S. (2001).
\newblock Optimal scaling for various {M}etropolis--{H}astings algorithms.
\newblock {\em Statistical Science}, 16(4):351--367.

\bibitem[S{\"a}rkk{\"a} and Svensson, 2023]{sarkka2023bayesian}
S{\"a}rkk{\"a}, S. and Svensson, L. (2023).
\newblock {\em Bayesian filtering and smoothing}, volume~17.
\newblock Cambridge University Press.

\bibitem[Sherlock and Thiery, 2022]{sherlock2022discrete}
Sherlock, C. and Thiery, A.~H. (2022).
\newblock A discrete bouncy particle sampler.
\newblock {\em Biometrika}, 109(2):335--349.

\bibitem[Shestopaloff and Neal, 2018]{shestopaloff2018sampling}
Shestopaloff, A.~Y. and Neal, R.~M. (2018).
\newblock Sampling latent states for high-dimensional non-linear state space
  models with the embedded {HMM} method.
\newblock {\em Bayesian Analysis}, 13(3):797--822.

\bibitem[Singh et~al., 2017]{singh2017blocking}
Singh, S.~S., Lindsten, F., and Moulines, E. (2017).
\newblock Blocking strategies and stability of particle {G}ibbs samplers.
\newblock {\em Biometrika}, 104(4):953--969.

\bibitem[Titsias, 2011]{titsias2011riemann}
Titsias, M.~K. (2011).
\newblock Contribution to the discussion on `{R}iemann manifold {L}angevin and
  {H}amiltonian {M}onte {C}arlo methods' by {G}irolami, {M.}, and {C}alderhead,
  b.
\newblock {\em Journal of the Royal Statistical Society Series B: Statistical
  Methodology}, 73(2):123--214.

\bibitem[Titsias and Papaspiliopoulos, 2018]{titsias2018auxiliary}
Titsias, M.~K. and Papaspiliopoulos, O. (2018).
\newblock Auxiliary gradient-based sampling algorithms.
\newblock {\em Journal of the Royal Statistical Society: Series B (Statistical
  Methodology)}, 80(4):749--767.

\bibitem[Tjelmeland, 2004]{tjelmeland2004using}
Tjelmeland, H. (2004).
\newblock Using all {M}etropolis--{H}astings proposals to estimate mean values.
\newblock preprint 4/2004, Norwegian University of Science and Technology,
  Trondheim, Norway.

\bibitem[Vehtari et~al., 2021]{Vehtari2021rank}
Vehtari, A., Gelman, A., Simpson, D., Carpenter, B., and B{\"u}rkner, P.-C.
  (2021).
\newblock Rank-normalization, folding, and localization: {A}n improved
  {$\smash{\widehat{R}}$} for assessing convergence of {MCMC} (with
  discussion).
\newblock {\em Bayesian Analysis}, 16(2):667--718.

\bibitem[Vogrinc and Kendall, 2021]{vogrinc2021counterexamples}
Vogrinc, J. and Kendall, W.~S. (2021).
\newblock Counterexamples for optimal scaling of {M}etropolis--{H}astings
  chains with rough target densities.
\newblock {\em The Annals of Applied Probability}, 31(2):972--1019.

\bibitem[Whiteley, 2010]{whiteley2010particle}
Whiteley, N. (2010).
\newblock Contribution to the discussion on `{P}article {M}arkov chain {M}onte
  {C}arlo methods' by {A}ndrieu, {C}., {D}oucet, {A}., and {H}olenstein, {R}.
\newblock {\em Journal of the Royal Statistical Society: Series B (Statistical
  Methodology)}, 72(3):306--307.

\bibitem[Whiteley and Lee, 2014]{Whiteley2014twisted}
Whiteley, N. and Lee, A. (2014).
\newblock {Twisted particle filters}.
\newblock {\em The Annals of Statistics}, 42(1):115--141.

\bibitem[Zanella, 2020]{zanella2020discrete}
Zanella, G. (2020).
\newblock Informed proposals for local {MCMC} in discrete spaces.
\newblock {\em Journal of the American Statistical Association},
  115(530):852--865.

\end{thebibliography}
